\documentclass[final,5p,times,twocolumn]{elsarticle}
\usepackage{amsmath}

\usepackage{lineno,hyperref}
\modulolinenumbers[5]

\journal{New Astronomy}
\usepackage{natbib}
\usepackage{xspace}
\usepackage{xcolor}

\biboptions{authoryear}

\newcommand{\arcsec}{"}

\newcommand{\Ha}{\ifmmode {\rm H}\alpha \else H$\alpha$\fi\xspace}
\newcommand{\Hb}{\ifmmode {\rm H}\beta \else H$\beta$\fi\xspace}
\newcommand{\Hba}{\ifmmode {\rm H}\beta^{\prime} \else H$\beta^{\prime}$\fi\xspace}
\newcommand{\Hg}{\ifmmode {\rm H}\gamma \else H$\gamma$\fi\xspace}
\newcommand{\Hd}{\ifmmode {\rm H}\delta \else H$\delta$\fi\xspace}

\newcommand{\Hii}{\ifmmode \rm{H}\,\textsc{ii} \else H\,{\sc ii}\fi}
\newcommand{\Nii}{\ifmmode [\rm{N}\,\textsc{ii}]\lambda 6584 \else [N\,{\sc ii}]$\lambda$6584\fi\xspace}
\newcommand{\NNii}{\ifmmode [\rm{N}\,\textsc{ii}]\lambda 6548 \else [N\,{\sc ii}]$\lambda$6548\fi\xspace}
\newcommand{\Niip}{[N\,{\sc ii}]$\lambda\lambda$6548,6583\xspace}
\newcommand{\nii}{\ifmmode [\rm{N}\,\textsc{ii}] \else [N\,{\sc ii}]\fi\xspace}

\newcommand{\oi}{\ifmmode [\rm{O}\,\textsc{i}] \else [O\,{\sc i}]\fi\xspace}
\newcommand{\Oii}{[O\,{\sc ii}]$\lambda$3727}
\newcommand{\neiii}{\ifmmode [\rm{Ne}\,\textsc{iii}] \else [Ne\,{\sc iii}]\fi\xspace}

\newcommand{\heii}{\ifmmode [\rm{He}\,\textsc{ii}] \else [He\,{\sc ii}]\fi\xspace}
\newcommand{\hei}{\ifmmode [\rm{He}\,\textsc{i}] \else [He\,{\sc i}]\fi\xspace}
\newcommand{\oii}{\ifmmode [\rm{O}\,\textsc{ii}] \else [O\,{\sc ii}]\fi\xspace}

\newcommand{\Oiiip}{[O\,{\sc iii}]$\lambda\lambda$4958,5007\xspace}
\newcommand{\OOiii}{\ifmmode [\rm{O}\,\textsc{iii}]\lambda 4959 \else [O\,{\sc iii}]$\lambda$4959\fi\xspace}
\newcommand{\Oiii}{\ifmmode [\rm{O}\,\textsc{iii}]\lambda 5007 \else [O\,{\sc iii}]$\lambda$5007\fi\xspace}
\newcommand{\oiii}{\ifmmode [\rm{O}\,\textsc{iii}] \else [O\,{\sc iii}]\fi\xspace}
\newcommand{\Siip}{[S\,{\sc ii}]$\lambda\lambda$6731,6716}
\newcommand{\Sii}{\ifmmode [\rm{S}\,\textsc{ii}]\lambda 6731 \else [S\,{\sc ii}]$\lambda$6731\fi\xspace}
\newcommand{\SSii}{\ifmmode [\rm{S}\,\textsc{ii}]\lambda 6716 \else [S\,{\sc ii}]$\lambda$6716\fi\xspace}
\newcommand{\sii}{\ifmmode [\rm{S}\,\textsc{ii}] \else [S\,{\sc ii}]\fi\xspace}
\newcommand{\siii}{\ifmmode [\rm{S}\,\textsc{iii}] \else [S\,{\sc iii}]\fi\xspace}
\newcommand{\WHa}{\ifmmode \rm{W}_{{\rm H}\alpha} \else W${}_{{\rm H}\alpha}$\fi\xspace}
\newcommand{\WHacen}{\ifmmode \rm{W}_{{\rm H}\alpha}^{\rm cen} \else W${}_{{\rm H}\alpha}^{\rm cen}$\fi\xspace}
\newcommand{\WHaRe}{\ifmmode \rm{W}_{{\rm H}\alpha}^{\rm Re} \else W${}_{{\rm H}\alpha}^{\rm Re}$\fi\xspace}
\newcommand{\SFRHaSF}{\ifmmode {\rm SFR}_{\rm H\alpha} \else SFR${}_{\rm H\alpha}$\fi\xspace}
\newcommand{\age}{\ifmmode \mathcal{A}_\star \else $\mathcal{A}_\star$\fi\xspace}
\newcommand{\agej}{\ifmmode \mathcal{A}_{\star, j} \else $\mathcal{A}_{\star, j}$\fi\xspace}
\newcommand{\stmetL}{\ifmmode [{\rm Z}/{\rm H}]_{\rm L} \else [Z/H]$_{\rm L}$\fi\xspace}
\newcommand{\stmetM}{\ifmmode [{\rm Z}/{\rm H}]_{\rm M} \else [Z/H]$_{\rm M}$\fi\xspace}
\newcommand{\stageL}{\ifmmode \left<\log \age\right>_{\rm L} \else $\left<\log \age\right>_{\rm L}$\fi\xspace}
\newcommand{\stageM}{\ifmmode \left<\log \age\right>_{\rm M} \else $\left<\log \age\right>_{\rm M}$\fi\xspace}

\newcommand{\mlL}{$\Upsilon_{V}$}
\newcommand{\ml}{$\Upsilon$}

\newcommand{\dbq}[1]{``#1''}

\newcommand{\textt}[1]{{\tt #1}}

\def\pyf{\texttt{pyFIT3D}\xspace}
\def\pyp{\texttt{pyPipe3D}\xspace}

\DeclareRobustCommand{\ion}[2]{%
\relax\ifmmode
\ifx\testbx\f@series
{\mathbf{#1\,\mathsc{#2}}}\else
{\mathrm{#1\,\mathsc{#2}}}\fi
\else\textup{#1\,{\mdseries\textsc{#2}}}%
\fi}

\begin{document}

\begin{frontmatter}
\title{\pyf\ and \pyp\ - The new version of the Integral Field Spectroscopy data analysis pipeline}


\author[0000-0001-7231-7953]{Eduardo A. D. Lacerda}
\author[0000-0001-6444-9307]{S. F. S\'anchez}
\author[0000-0002-8931-2398]{A. Mej\'ia-Narv\'aez}
\author[0000-0002-2555-1074]{A. Camps-Fari\~na}
\author[0000-0002-9658-8886]{C. Espinosa-Ponce}
\author[0000-0003-2405-7258]{J. K. Barrera-Ballesteros}
\address[0000-0001-7231-7953]{Instituto de Astronom\'ia, Universidad Nacional Aut\'onoma de M\'exico, AP 70-264, CDMX 04510, M\'exico}
\author[0000-0002-9790-6313]{H. Ibarra-Medel}
\address[0000-0002-9790-6313]{University of Illinois Urbana-Champaign, Department of Astronomy, 1002 W Green St, Urbana, Illinois, 61801, USA}
\author[0000-0001-9226-9178]{A. Z. Lugo-Aranda}
\address[0000-0001-7231-7953]{Instituto de Astronom\'ia, Universidad Nacional Aut\'onoma de M\'exico, AP 70-264, CDMX 04510, M\'exico}

\begin{abstract}

We present a new version of the FIT3D and Pipe3D codes, two packages to derive properties of the stellar populations and the ionized emission lines from optical spectroscopy and integral field spectroscopy data respectively. The new codes have been fully transcribed to Python from the original Perl and C versions, modifying the algorithms when needed to make use of the unique capabilities of this language with the main goals of (1) respecting as much as possible the original philosophy of the algorithms, (2) maintaining a full compatibility with the original version in terms of the format of the required input and produced output files, and (3) improving the efficiency and accuracy of the algorithms, and solving known (and newly discovered) bugs.
The complete package is freely distributed, with an available repository online. \pyf and \pyp are fully tested with data of the most recent IFS data surveys and compilations (e.g. CALIFA, MaNGA, SAMI and AMUSING++), and confronted with simulations.  We describe here the code, its new implementation, its accuracy in recovering the parameters based on simulations, and a showcase of its implementation on a particular dataset.

\end{abstract}


\end{frontmatter}
%

\section{Introduction} \label{sec:intro}

The large volume of observed data available nowadays in every area of science demands high speed, precise and accurate analysis algorithms. Due to the availability of different data, the analysis packages are increasing their capability and coverage of usability cases. In the exploration of the nearby Universe ($z<0.1$), the scenario could not be different. Over the last couple of decades, the datasets evolve from a few number of surveys sampling a single value and/or a single spectrum for each galaxy, to an exploding amount of surveys with observations of millions of spectra covering the entire field-of-view of a huge amount of galaxies (2dFRGS: \citealt{Folkes.etal.1999}; SDSS: \citealt{York.etal.2000}; MaNGA: \citealt{Blanton.etal.2017}; CALIFA, \citealt{CALIFApresent}; SAMI: \citealt{Croom.etal.2012}; AMUSING: \citealt{Galbany.etal.2016}). As a consequence, the new datasets require dedicated pipelines for their data reduction and analysis (e.g. \texttt{LZIFU}, \citealt{ho16}; MaNGA DAP, \citealt{Westfall.etal.2019}; \texttt{GIST}, \citealt{Bittner.etal.2019}).

Pipe3D \citep[][hereafter S16b]{pipe3d_ii} is one of these tools, a dedicated pipeline to extract the properties of the stellar populations and emission lines in the optical spectra of Integral Field Spectroscopic data of galaxies. This pipeline uses as basic fitting algorithms the ones provided by FIT3D \citep[][hereafter S16a]{pipe3d}, a package that can explore the same properties for individual spectra, row-stacked multi-object spectra (or classical long-slit ones), or IFS datacubes. Both tools were coded in {\sc Perl} (with some routines coded in {\sc C}), making use of the numerical and fitting algorithms included in the Perl Data Language \citep[{\sc PDL}][]{PDL}. Pipe3D has been broadly used in the analysis of IFS data from individual galaxies and large datasets of different IFS galaxy surveys including CALIFA \citep[e.g.][]{laura16}, MaNGA \citep[][]{sanchez18}, MUSE \citep[e.g.][]{carlos20}, and SAMI \citep[e.g.][]{sanchez19}. Its capabilities to recover the properties of the stellar populations and emission lines have been contrasted with hydrodynamical simulations \citep[e.g.][]{guidi18,ibarra19,vandesande19}, and fully compared with the values recovered using other similar tools \citep[e.g.][]{dap,sanchez19}.

Despite of its well proved capabilities there are some drawbacks with the current implementation of the code. The most important one is that it is deeply attached to a coding language and a numerical package that are used only by a small fraction of the astronomical community. Furthermore, it uses PGPLOT\footnote{\url{https://sites.astro.caltech.edu/~tjp/pgplot/}}, a graphics library that is no longer officially supported, and again, with a small number of users nowadays. In addition, new improvements in basic algorithms that frequently appear for other languages (like Python) and numerical packages (like {\sc numpy} or {\sc scipy}) cannot be implemented. For all those reasons the code is difficult to be updated, upgraded and maintained. It is difficult to be installed in new versions of operative systems, and therefore, to be distributed.

In order to solve all those problems, update the code, and maintain it as competitive as possible, we embarked in the transcription of the  code to {\sc Python}, the coding language that is nowadays the one with
largest fraction of users among the astronomical community. We present in here the new code, describing its conceptual and coding philosophy, summarizing the main algorithms, and showing its capabilities in the analysis of the stellar populations and emission lines for both individual spectra and IFS data. The structure of the article is as follows: (i) Section \ref{sec:data} describes the dataset adopted as a showcase of the use of the code; (ii) In Sec. \ref{sec:philosophy} we describe the philosophy of the new code; (iii) Sec. \ref{sec:new_ver} comprises the description of the algorithms, with \ref{sec:new_ver:fitupd} describing {\sc pyFIT3D}, including the updates performed (Sec. \ref{sec:new_ver:fitupd}, a description of procedures to recover the non-linear parameters (velocity, velocity dispersion and dust attenuation, Sec. \ref{sec:new_ver:fitupd:nlfit}) and linear parameters (i.e., decomposition of the stellar population in the adopted single-stellar population library, Sec. \ref{sec:new_ver:fitupd:stpopsyn}), and \ref{sec:new_ver:pipeupd} describing the different steps included in {\sc pyPipe3D} to analyze an individual IFS datacube; (iv) the accuracy in the recovery of the properties of the new code is presented in Sec. \ref{sec:accur}, first contrasted against simulations for both the stellar population and emission line properties (Sec. \ref{sec:accur:stpop} and Sec. \ref{sec:accur:eml}, respectively), and then against real data (Sec. \ref{sec:accur:realdata}); (v) Finally, the conclusions of this study are presented in Sec. \ref{sec:summary}.




\section{Data}
\label{sec:data}

The data adopted in this study as a showcase of the code is obtained from the extended Calar Alto Legacy Integral Field Area survey sample (eCALIFA, \citealt{CALIFADR3, Galbany.etal.2018}). The survey was built with observations from the 3.5m telescope at the Calar Alto observatory, using PPAK Integral Field Unit \citep{Kelz.etal.2006} of the Potsdam Multi-Aperture Spectrograph \citep[PMAS,]{Roth.etal.2005}, with a covering factor of 60\% of a $74\arcsec \times 64\arcsec$ field-of-view (FoV) with 331 fibers of $2.7\arcsec$. A complete coverage of the FoV is achieved with a three position dithering scheme resulting in a spatial resolution (characterized by the point-spread function, PSF) with a full-width at half-maximum (FWHM) of $\sim$2.5\arcsec. The sample of galaxies covered by eCALIFA, corresponding to objects within a narrow range of redshifts around $z\sim$0.015, was primarily selected by diameter \citep{walcher:2014} in order to fit the optical extension of the galaxies within the FoV of the instrument. We use the V500 observation setup ($3745--7500$\AA, $\lambda/\Delta\lambda \sim 850$) for the dataset analyzed during this study, which guaranties the simultaneous covering of (i) the most relevant spectral features of the stellar populations in the optical regimes and (ii) the more relevant emission lines from [OII]$\lambda$3727 to [SII]$\lambda$6717,31 to explore the properties of the ionized gas. The observing strategy guarantees to sufficient signal-to-noise through the optical extension of the galaxies, making this dataset ideal for the purposes of testing the new version of the code. It is worth noticing that this code, like its predecessor (Pipe3D), has been already applied to other similar datasets, like MaNGA (Sánchez et al., in prep.) or data of IFS data of better spectral resolution (e.g., MUSE).

\section{Philosophy of the new code}
\label{sec:philosophy}

Contrary to the previous version of the code, that was mostly a single-coder package, the new version was developed by a group of people. This requires to adopt an inclusive and easy-to-write programming language. Following the massive growth of \texttt{python} users \citep[e.g.][]{python1995} in astronomy, hence the availability of modern, well documented, helpers and packages of analysis, we rewrite the code package entirely from \texttt{perl} \citep{perl} to \texttt{python} 3 \citep[][3.6 or higher]{python3}. Furthermore, to rewrite Pipe3D with such an inclusive programming language will encourage more people to participate in the development of the code at a lower level. At the end, we choose not to rename the pipeline and the spectral fitting tool, but to append a prefix {\it \dbq{py}} to original names.

One of the most important aspects of this code refactoring project is the new adopted programming philosophy: to provide modular and reusable standalone pieces of code (such as classes and modules) making easy the production of analysis, data-exploration and interactive\footnote{Such as jupyter-notebooks: \url{https://jupyter.org/}.} scripts.
Moreover, the new code is constructed based on well documented public libraries to facilitate users/developers to understand how the analysis is made. In addition to some standard \texttt{python} 3 library modules, we utilize numerical methods from \texttt{NumPy} \citep{numpy2006} and \texttt{SciPy} \citep{scipy2020}, the FITS reading package of \texttt{astropy} \citep[\texttt{astropy.io.fits};][]{astropy} together with some plot functionalities from \texttt{matplotlib} \citep{matplotlib} and \texttt{seaborn} \citep{seaborn}. Finally, the new package carries a complete code-based documentation and an online repository\footnote{\url{http://gitlab.com/pipe3d}} including handy tools to help the users dealing with the output files from the analysis and interactive scripts with various types of procedures taking advantage of the code re-usability.

\section{New code implementation}
\label{sec:new_ver}

As indicated before \pyf and \pyp are based on FIT3D/Pipe3D. We attain the detailed description along the following subsections for the parts that are needed to the understanding of the article and also for those which receive some update. We present the main changes revisiting some details of the spectral fitting code and introducing the updates to the code. Finally, we close this Section with the description of \pyp procedure for the analysis of an IFS datacube.

\subsection{pyFIT3D: The spectral fitting code update}
\label{sec:new_ver:fitupd}

\begin{figure*}[t]
    \centering
    \includegraphics[width=\textwidth]{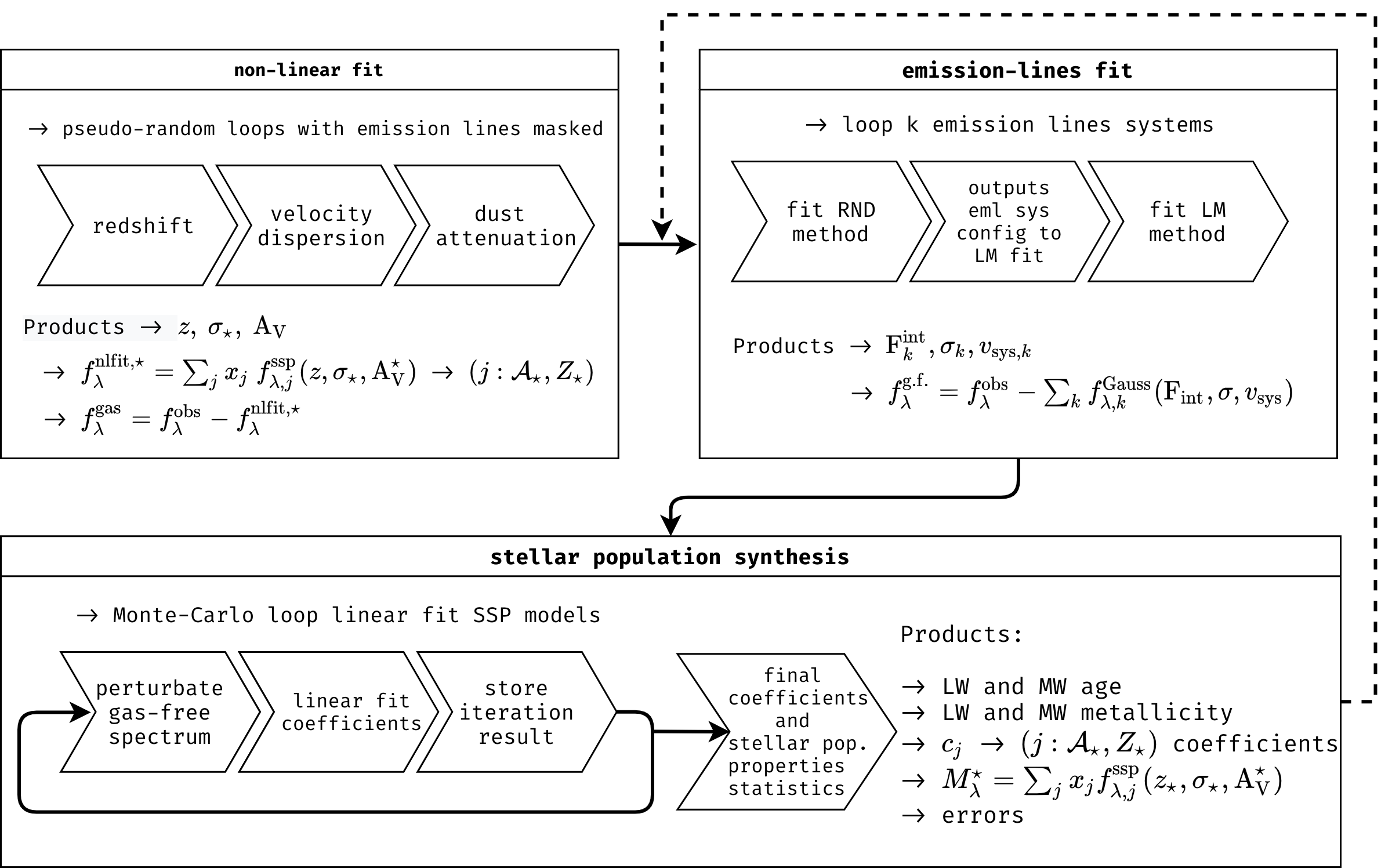}

    \caption{Flow-chart showing the three-step fitting process included in \pyf, together with the output products at every step: non-linear fit (top-left panel), where the systemic velocity/redshift ($z_\star$), line-of-sight velocity dispersion ($\sigma_\star$) and the dust attenuation (A$_{\rm V}^\star$) are estimated; emission lines fit (top-right panel), where it is created a model for the ionized gas emission lines spectrum; and finally the stellar population synthesis (bottom panel), where the gas-free resultant spectrum is then modeled creating the light population vector ($x_j$, i.e., the coefficients of the decomposition of the stellar spectrum in a base of synthetic stellar templates). At the end, the light-weighted (LW) and mass-weighted (MW) age and metallicity are estimated, based on this decomposition, together with associated errors.}
    \label{fig:anaspec_flow}
\end{figure*}
The spectra of a galaxy (or a region of a galaxy) are the result of the emitting light from stars and the ionized gas, kinematically shifted due to its dynamical stage and dust attenuated. The idea behind \pyf is to decouple the stellar continuum and the ionized emission spectrum in the observed spectral energy density (SED).
The fitting process works disentangling the stellar and the interstellar medium spectra without a direct assumption of the star-formation and chemical enrichment histories (SFH and ChEH). However, we assume that the individual spectrum comprises an unresolved stellar population. Thus, the physical aperture from which the spectra is obtained is large enough to encircle $\sim10^4$ stars, a number sufficient to fully sample an entire Mass Function \citep[without stochastic problems][]{Cervino.and.Luridiana.2004}. This way, the stellar population can be modeled by a linear combination of single-stellar populations (SSPs), each one corresponding to the different star-formation burst along the SFH. In comparison to the stellar continuum, we further assume that the nebular continuum is  negligible throughout the considered aperture. The flowchart presented in Figure \ref{fig:anaspec_flow} schematize the three-steps process of the spectral analysis described along this Section:
\begin{itemize}
    \item Non-linear fit step ({\it upper-left panel}), where it is derived the redshift (systemic velocity), the line-of-sight velocity dispersion (LOSVD) and the dust attenuation ($z_\star$, $\sigma_\star$ and A$_{\rm V}^\star$).
    \item Emission lines fit step ({\it upper-right panel}), where it is derived the main properties of a pre-defined set of emission lines.
    \item Stellar population synthesis step ({\it bottom panel}), where the stellar spectrum is decomposed in a set of SSPs, deriving its main properties (e.g., light-weighted ages and metallicities).
\end{itemize}
This sequence is essentially the same as the one adopted by the previous version of the code.

There are other two additional analysis that can be performed by FIT3D (and are an integral part of Pipe3D) which are not included originally in the formal loop of analysis, although they complement it:
\begin{itemize}
    \item Moment analysis: a non-parametric estimation of emission lines properties.
    \item Stellar-indices analysis: where it is measured a set of stellar indices, such as the Lick/IDS index system \citep[e.g.][]{Burstein.etal.1984, Worthey.1994}, and D4000 \citep{Bruzual.1983}.
\end{itemize}

Details on the procedures and their modifications are given over next subsections.


\subsubsection{Step one: non-linear fit}
\label{sec:new_ver:fitupd:nlfit}

As indicated before, the first group of analysis performed by \pyf is the determination of the stellar spectrum kinematic parameters ($z_\star$ and $\sigma_\star$) and the dust attenuation (A$_{\rm V}^\star$). During this step, the parameters are explored once at a time following the sequence shown in upper-left diagram of Figure \ref{fig:anaspec_flow}. For each explored parameter \pyf covers a range of values constraint between an user defined {\it minimum} and {\it maximum} values (defining an explored {\it interval}), following a random sampling with an average {\it step} (defined by the user too). An {\it input guess} value is also required, which use will be described later on. Optionally, setting up the {\it step} to zero makes the program to fix the value of this parameter to the {\it input guess} along the analysis, i.e. the parameter is not determined by \pyf. The kinematic parameters receive a second round of exploration as a refinement of the first one. In this turn, the {\it minimum, maximum} and {\it step} of exploration are automatically updated for a fine search around the best value found during the first round of the parameter determination.


To estimate the best fitted parameter, for each set of values ($z_\star$,$\sigma_\star$ and A$_{\rm V}^\star$) the program computes a model for the stellar spectrum (following the inversion procedure described below). For doing so, the adopted single/synthetic stellar population (SSP) library is configured to the candidate observed frame (i.e., redshifted and broadened with the explored values of the kinematic properties and dust attenuated based on the assumed extinction law) with a predefined set of emission lines (in the observed-frame) masked together with the strongest night-sky emission lines (in the rest-frame). For the broadening of the spectra due to the velocity dispersion it is assumed a Gaussian profile for the Line-of-sight velocity distribution (LOSVD). We acknowledge that this is just a first order approximation and that effects like the asymmetry of the profiles cannot be recovered using this approximation \citep[e.g.][]{Cappellari.etal.2011a}. However, for the typical signal-to-noise (S/N) of the current IFS surveys this modeling is sufficient and preferred than a more accurate, but more demanding, modeling of the stellar populations. In addition, we also include the option to input an instrumental dispersion (in \AA) to be considered during the broadening of the spectra.

\pyf includes the option to use two SSPs libraries, one for this step and another for the stellar population synthesis at the final part of the program (Section \ref{sec:new_ver:fitupd:stpopsyn}). We recommend the use of a simple set of SSPs for the non-linear fit step for two reasons: (1) speed-up the process (that is very time-consuming) and (2) avoid known degeneracies between the parameters (e.g. stellar metallicity and $\sigma_\star$). Then, the coefficients ($x_j$) of the linear combination of the SSP models which better reproduces the stellar continuum are derived through a recursive weighted least-squares (WLS) inverse matrix method,
\begin{equation}
    f_{\lambda}^{{\rm nlfit},\star} = \sum_j^{x_j > 0} x_j f_{\lambda,j}^{\rm ssp}(z_\star, \sigma_\star, {\rm A}_{\rm V}^\star) \rightarrow (j: \age,{\rm Z}_\star).
    \label{eq:nlfit_model}
\end{equation}
Where $f_{\lambda}^{{\rm nlfit},\star}$ is the modeled spectrum, $f_{\lambda,j}^{\rm ssp}(z_\star, \sigma_\star, {\rm A}_{\rm V}^\star)$ is the shifted, broadened and dust attenuated individual SSP template, where $j$ runs through the range of ages (\age) and metallicities (Z$_\star$) covered by the considered library. The recursive procedure applied iterates the WLS matrix inversion, rejecting those SSP templates within the library that produce negative coefficients, shrinking the SSP library to include only those ones for which a positive coefficient is derived. The iteration continues until there are no negative coefficients. The adopted merit function is the reduced $\chi^2$, calculated as
\begin{equation}
    \chi^2 = \frac{1}{n - m}\sum_\lambda^n \left(\frac{f_{\lambda}^{\rm obs} - f_{\lambda}^{{\rm nlfit},\star}}{\sigma_{\lambda}}\right)^2,
    \label{eq:nlfit_merit}
\end{equation}
where $n$ is the number of pixels of the observed spectrum, $m$ is the number of templates present in the adopted SSP templates library and $\sigma_\lambda$ is the error associated with the observed spectrum. In addition to a global input wavelength interval for the analysis performed by \pyf, we include the option to configure an exclusive interval for the derivation of the kinematic parameters ($z_\star$ and $\sigma_\star$). Although the lower wavelength interval for the determination of ${\rm A}_{\rm V}^\star$ would speed-up the process, it is preferred a large baseline of wavelength to provide with the best estimation of this parameter.
At the end, the best-fitted value for each parameter corresponds to that one for which the stellar population model produces the lower value of $\chi^2$. Since all three properties are necessary to model the stellar spectrum, if a property is yet to be determined, it adopts the {\it input guess} value.

\begin{figure*}[t]
    \centering
    \includegraphics[trim=50 0 50 50, clip, width=\textwidth]{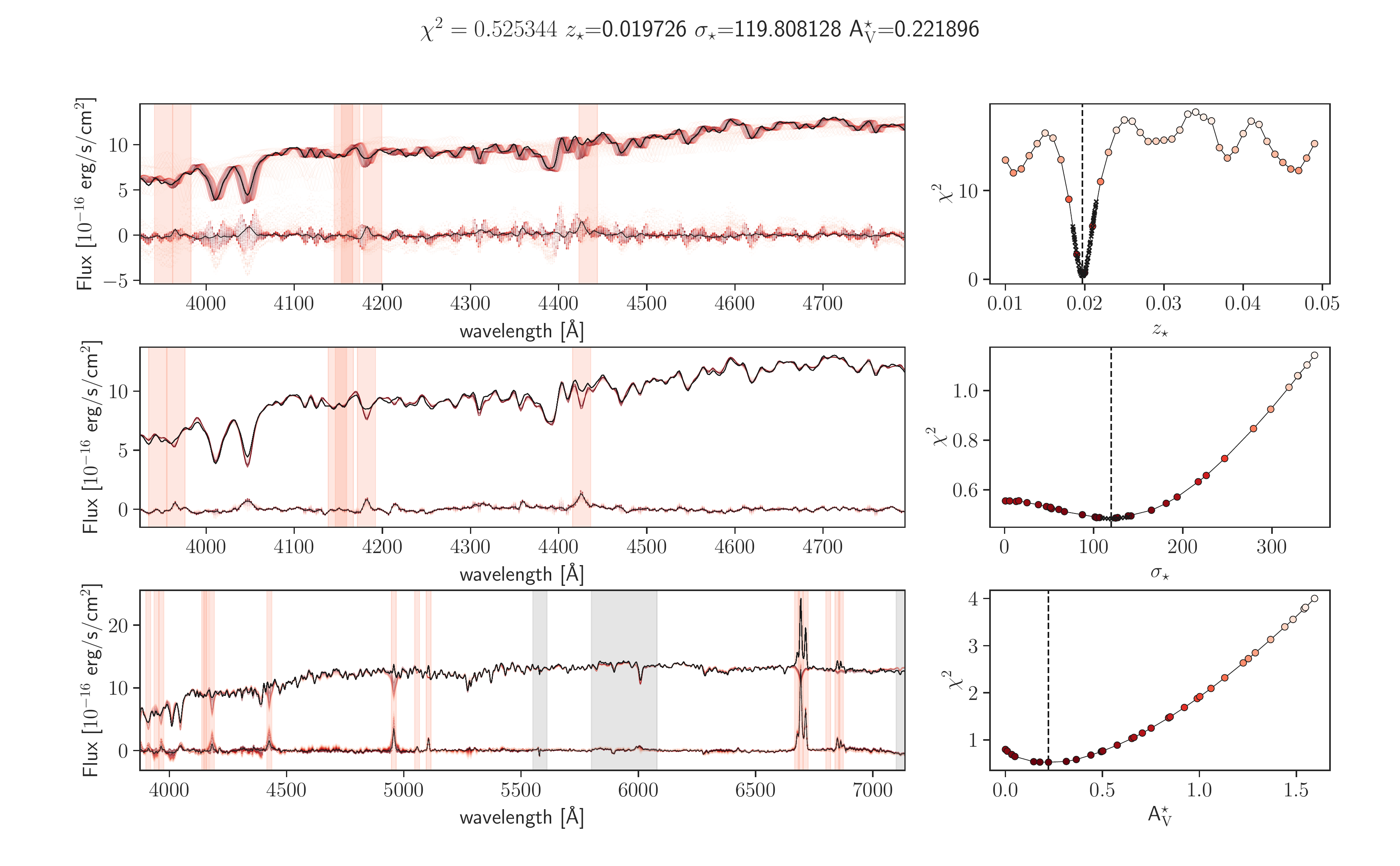}
    \caption{Showcase of the determination of the non-linear parameters for the central ($5\arcsec$ aperture) spectrum of galaxy NGC\,5947: Redshift ($z_\star$, upper-panels), line-of-sight velocity dispersion ($\sigma_\star$, middle-panels) and dust attenuation (A$_{\rm V}^\star$, lower-panels). The rows are ordered from top to bottom in the sequence that the three parameters are determined (as appears in Figure \ref{fig:anaspec_flow}). At each row, the left panel shows the observed spectrum (black line) and the different models evaluated during the parameter exploration with associated residuals (colors distributed from light- to dark-red following the range of explored values). The vertical shaded regions indicate the wavelength ranges masked during the process. The right panel shows the merit curve ($\chi^2$) as a function of each parameter. The colored dots follow the parameter value during the exploration of the first loop and, the black crosses, the second step exploration, a fine sweep around the best value found in the first loop. The best value found for each parameter is then indicated as a vertical dashed black line.}
    \label{fig:anaspec_nlfit_summary}
\end{figure*}

Figure \ref{fig:anaspec_nlfit_summary} illustrates this entire process for the central spectrum (5\arcsec) of the showcase dataset (NGC\,5947, observed by CALIFA). Details on the adopted parameters, masked wavelength intervals and configuration files for the current fitting procedure are given in Appendix \ref{appendix:showcase_config}. The final result depends on the particular parameters included in this configuration files, and in particular on the adopted SSP library \citep[see S16a and][for a discussion on the topic]{CF.etal.2014}. For this particular example, we adopted a sub-set of the \texttt{gsd156} stellar population library comprising just 12 templates (four ages and three metallicities).  The \texttt{gsd156} library \citep{CF.etal.2013}, comprises 156 SSP templates, that sample 39 ages (1 Myr to 14 Gyr, on an almost logarithmic scale), and 4 different metallicities (Z/Z$\odot$=0.2, 0.4, 1, and 1.5), adopting the Salpeter IMF \citep[][]{Salpeter:1955p3438}.
We must clarify that the code can run using any suitable SSP library, once transformed to the required input format (described in Appendix \ref{app:ssp}). All attempts of the modeled spectra (i.e., all the realizations of Eq. \ref{eq:nlfit_model}), residuals and the best model are shown along the first column following the sequence of determination from top to bottom. In each panel, vertical shaded regions represent the masked wavelengths during the procedure. The second column shows the merit curve for each explored parameter.

We highlight here the importance of a good choice of the {\it interval} explored in the determination of each parameter. The reason is that there may be local minima present in the parameter space exploration, as can be seen in the {\it upper-right panel} of Figure \ref{fig:anaspec_nlfit_summary}. For this, a selection of the {\it interval} of parameters based on the known properties of the observed object (e.g., redshift) is required, improving the velocity and reliability of the program. We will discuss that in Section \ref{sec:new_ver:pipeupd}.

\subsubsection{Step two: Emission lines fit}
\label{sec:new_ver:fitupd:elfit}

\begin{figure*}[t]
    \centering
    \includegraphics[trim=20 0 40 25, clip, width=\textwidth]{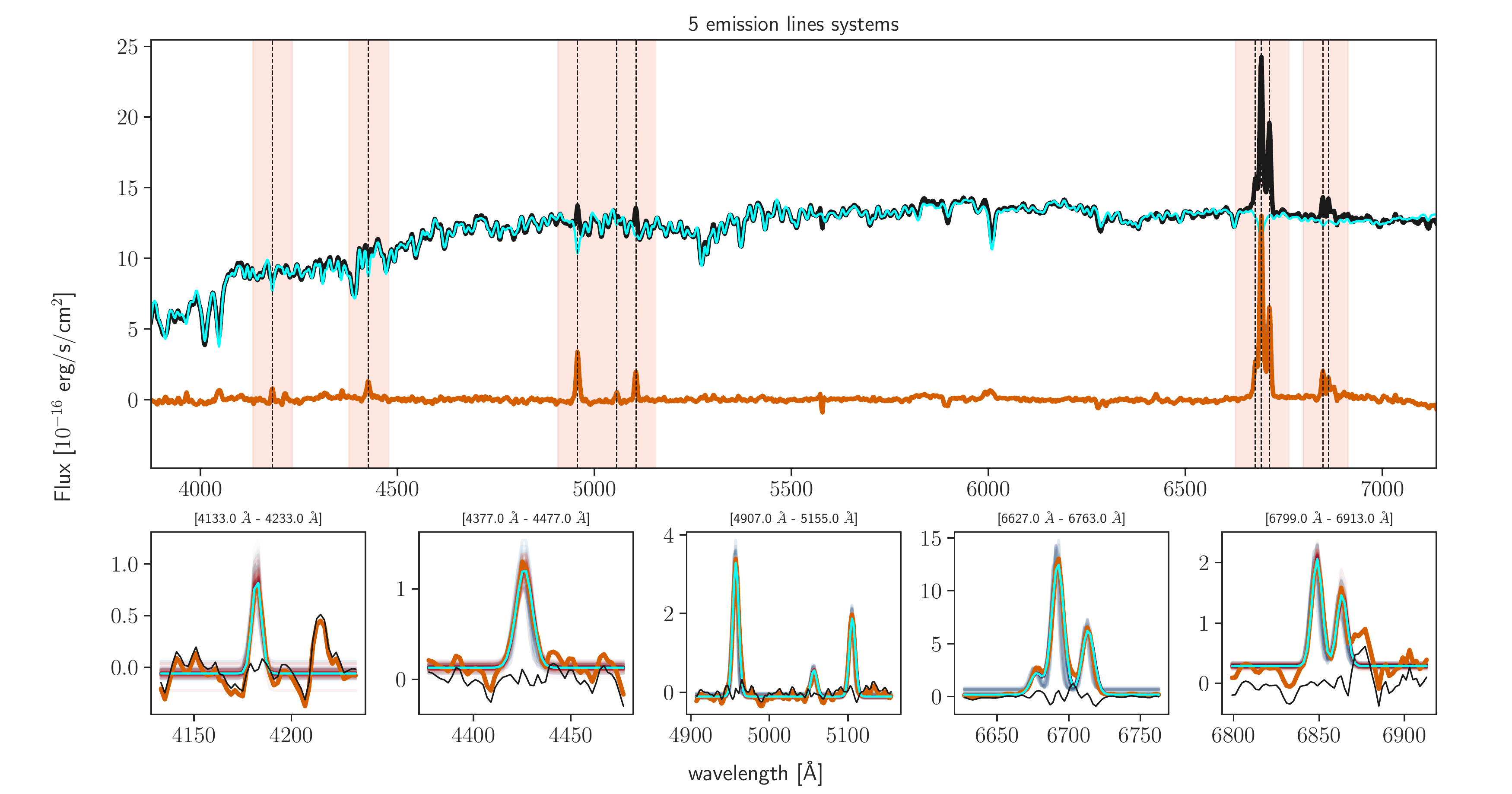}
    \caption{Example of the exploration of the properties of the ionized gas emission lines for the central spectrum of NGC\,5947. Top panel shows the observed spectrum (black) together with the best fitted model of the stellar population spectrum provided by the 1st non-linear step of the procedure (cyan). The residual spectrum, once subtracted the best model to the original spectrum, is shown in red. This spectrum comprises the ionized gas emission lines to be fitted (together with the residual of the stellar population fitting plus noise). Light-red shaded vertical regions highlights the intervals selected for fitting each emission lines {\it system}. Bottom panels shows a zoom on those wavelength intervals. In each panel the red line corresponds to the same residual spectrum shown in the top panel. This time, the cyan corresponds to the best fitted model for the set of emission lines in each wavelength regime and the residual of the fit is shown as a thin black line. The shaded area highlights all the models explored by the MC method implemented in the fitting procedure. \pyf derives the intensity, the velocity and the velocity dispersion of all implicated emission lines.}
    \label{fig:anaspec_ELfit_summary}
\end{figure*}

After the derivation of the best kinematic and the dust attenuation parameters the program produces a first model for the stellar spectrum based on the simple SSPs template (i.e., the one adopted for the non-linear fit step). This stellar spectrum model is then subtracted to the original one to produce a spectrum with just the information of the ionized gas emission lines (plus noise and residuals). The new process of the emission lines fit implemented in \pyf is made through a two-round procedure summarized at the upper-right diagram of Figure \ref{fig:anaspec_flow}. Which {\it systems}\footnote{We call {\it system} to a set of emission lines fitted at same time, i.e., included by the same configuration file.} the program will fit depends on the configuration made by the user\footnote{More details and examples on the configuration files are included in the \pyp webpage}. 

For the fit of the emission lines, we have implemented now two different methods that we call RND (for random) and LM (for Levenberg-–Marquardt), both modeling the emission lines as Gaussian profiles:
\begin{itemize}
    \item {\bf The RND method} mimics the one already implemented in the previous version of the code. The method is constructed based on a Monte-Carlo (MC) loop with a $\chi^2$ minimization scheme where all emission line models are created in a pseudo-random search of the free parameters. The process of the exploration of these parameters works in a similar way as the one performed by the non-linear fit step (i.e., the range of allowed values is fully covered using a pseudo-random exploration). Again, this is important to avoid local minima inherent to other fitting/minimization procedures (that may be faster in principal). At the end of each MC loop it is evaluated the best fitted $\chi^2$ and the intervals of exploration are narrowed around the best fitted parameters (following a similar process to a Markov chain MC inference). Once the program provides with a first estimation of the parameters that define each emission line, the algorithm performs a second fit with the {\bf LM method}.
    \item {\bf The LM method} is the implementation of the \textit{Levenberg--Marquardt} minimization, which has good precision and is very fast. However, this method, like many other minimization algorithms, may present issues with the accuracy due to the presence of local minima (e.g., right panels of Figure \ref{fig:anaspec_nlfit_summary}). Moreover, if the intervals for the exploration of each parameter are not well defined, the method could diverge. As indicated before, the RND method has a very robust accuracy. However, it is slower and its precision is affected on low S/N regime. The combination of both methods produce optimal results in terms of accuracy and precision. This is the reason why we include the second round of analysis in the emission lines fit step of \pyf. At the end, the final values for the explored parameters are derived using the LM method and the associated errors are estimated based on a MC iteration.
\end{itemize}

The Figure \ref{fig:anaspec_ELfit_summary} exemplifies this process for our example dataset. In this example we show the fit of five {\it systems} encompassing ten emission lines (bottom row, from left to right): \Hd; \Hg; \Oiiip\,+\,\Hb ; \Niip\,+\,\Ha; \Siip) through the {\bf RND+LM method}. The set of models for different emission lines are combined to derive an ionized gas emission spectrum ($\sum_k f^{\rm Gauss}_{\lambda,k}$). This is then subtracted from the observed one, producing a gas-free spectrum (plus residuals and noise), which will be used as the input spectrum for the stellar population synthesis.

\subsubsection{Step three: Stellar population analysis}
\label{sec:new_ver:fitupd:stpopsyn}

\begin{figure*}
    \centering
    \includegraphics[trim=10 0 10 10, clip, width=\textwidth]{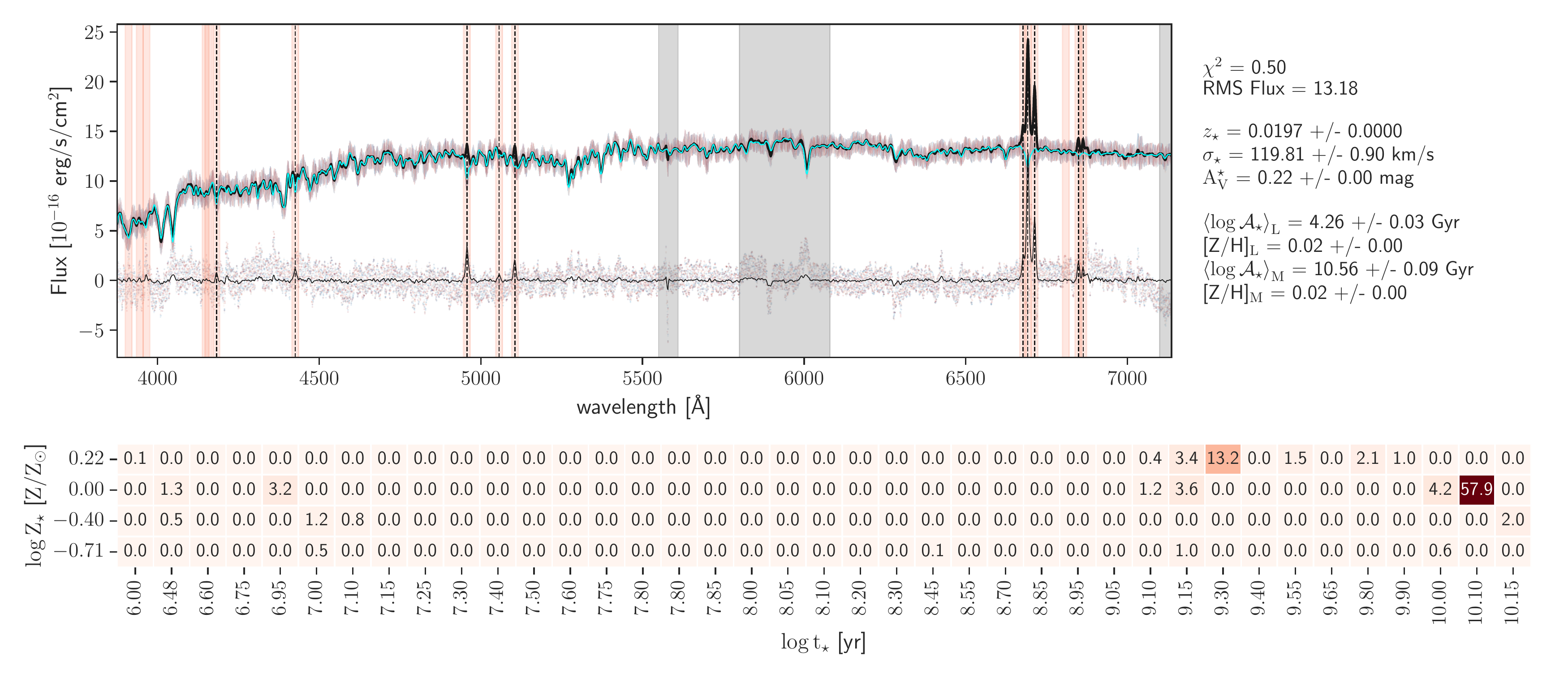}
    \caption{Example of the stellar population synthesis procedure for the central spectrum of NGC\,5947. The upper panel shows the spectral fit (left) together with the values of the main properties derived by the synthesis (right). The observed spectrum (dark-thick line) is overplotted with the best model for the stellar spectrum (cyan line) with the final residual spectrum draw as a black-thin line at the bottom. All the ${\rm F}^{k}_\lambda$ MC realizations are represented by the hatched gray area around the observed spectrum. The noise range (as in Eq. \ref{eq:MC_noise}) and the residual of the k-realizations have been amplified by a factor of five to be better appreciated. The vertical shaded regions correspond to the masked areas due to intrinsic emission lines (red) and telluric ones (gray). Vertical dashed lines highlight the fitted emission lines (represented in upper-right panel of Figure \ref{fig:anaspec_flow} and in Figure \ref{fig:anaspec_ELfit_summary}). The bottom panel shows the fraction of light in the selected wavelength range ($5500 \pm 45$\AA) that corresponds to each SSP within the template (i.e., the coefficients of the stellar decomposition), color coded in red.}
    \label{fig:anaspec_summary}
\end{figure*}

The last step of the fitting procedure included in \pyf comprises the synthesis of the stellar population. This process works by solving the linear decomposition of the gas-free spectrum ($f^{\rm g.f}_\lambda$) between the different SSP templates of the adopted library. The decomposition is done using a MC method over perturbed realizations of the input spectrum. First, all SSP templates are shifted to the observed frame and dust attenuated using the kinematics parameters and  ${\rm A}_{\rm V}^\star$ derived in the first step (Section \ref{sec:new_ver:fitupd:nlfit}).
Then, the average values of the set of $x_j$ parameters derived for each MC iteration are recovered as the final weights/coefficients of the stellar population model, with the standard deviation being adopted as the 1$\sigma$ error of each parameter. Figure \ref{fig:anaspec_summary} outlines this part of the procedure for our adopted example dataset. For this particular example we use the full \texttt{gsd156} stellar population library, that has been extensively used in many different studies \citep[e.g.~][]{perez13,rosa14,ibarra16,sanchez18,sanchez21}. Once again, we must remind that the code can run using any SSP library with the proper format.

Let F$^i_\lambda$ be the perturbed spectrum for the i$^{th}$ MC iteration, R$_\lambda^i$ the noise factor, corresponding to a random value following a (-1, 1) clipped Gaussian distribution, and $\sigma_\lambda$ being the noise spectrum:
\begin{equation}
    {\rm F}^i_\lambda = f^{\rm g.f}_\lambda + R_\lambda^i \sigma_\lambda.
    \label{eq:MC_noise}
\end{equation}
The left panel of Figure \ref{fig:anaspec_summary} shows the set of ${\rm F}^i_\lambda$ (gray lines) with the noise factor ($R_\lambda^i \sigma_\lambda$). In order to visualize better the differences between the $i$ iterations the noise has been multiplied by 5 (this factor was introduced just for visualization purposes, it is not included as such in the code). The reduced $\chi^2$ is calculated for each MC iteration as:
\begin{equation}
    \chi^2_i = \frac{1}{n - m} \sum_\lambda^n \left(\frac{{\rm F}^i_\lambda - M_{\star, \lambda}^i}{\sigma_{\lambda}}\right)^2,
\end{equation}
where $n$ is the number of pixels in the observed spectrum, $m$ is the number of SSP models and $M_{\star, \lambda}^i$ is the modeled spectrum at the i$^{th}$ MC realization,
\begin{equation}\label{eq:spec_mod}
    M_{\star, \lambda}^i = \sum_j^{x_j > 0} x_j^i f_{\lambda,j}^{\rm ssp}(z_\star, \sigma_\star, {\rm A}_{\rm V}^\star),
\end{equation}
and $x_j^i$ is the coefficient of the j$^{th}$ template of the adopted SSP library at the i$^{th}$ MC realization. The final modeled stellar spectrum and associated uncertainties are given by the mean,
\begin{equation}
    M_{\star, \lambda} = {\rm mean}(M^{1..i}_{\star, \lambda}),
\end{equation}
and standard deviation,
\begin{equation}
    \sigma(M_{\star, \lambda}) = {\rm stddev}(M^{1..i}_{\star, \lambda}),
\end{equation}
of the individual $i$ iterations, respectively. Finally, the coefficients of the best fitted multi-SSP model are given by the mean,
\begin{equation}
    x_j(\age, {\rm Z}_\star) = {\rm mean}[x_j^{1..i}(\age, {\rm Z}_\star)],
    \label{eq:coeffs}
\end{equation}
and their uncertainties by their standard deviation,
\begin{equation}
    \sigma[x_j(\age, {\rm Z}_\star)] = {\rm stddev}[x_j^{1..i}(\age, {\rm Z}_\star)].
    \label{eq:sigmacoeffs}
\end{equation}

We should note that these final coefficients provided by the algorithm correspond to the fraction of light that each SSP template contributes to the observed spectrum at a certain wavelength,
\begin{equation}\label{eq:coeff}
    c_j = \frac{x_j}{f^{\rm ssp}_{\lambda_{\rm norm},j}}.
\end{equation}
For the optical wavelength range $\lambda_{\rm norm}$ is usually selected to be 5500 \AA.

Once we derive the coefficients of the decomposition of the stellar population in the SSP templates, it is possible to derive different average properties. In particular, \pyf estimates the light-weighted (LW) stellar age,
\begin{equation}
    \left<\log \age\right>_{\rm L} = \sum_j^m c_j \log \agej,
    \label{eq:LWAge}
\end{equation}
and metallicity,
\begin{equation}
    \stmetL = \sum_j^m c_j \log {\rm Z}_{\star, j}.
    \label{eq:LWMet}
\end{equation}
The light-weighted averages highlight the contribution of the young stellar populations, i.e., recent star forming events, since they contribute more to the light relative to their old counterparts. On the other hand, mass averaged values are less sensitive to the presence of young stellar populations, highlighting the contribution of the old stellar ones. Mass-weighted values are also calculated defining the mass fraction contributed by each SSP model as
\begin{equation}\label{eq:coeff_mass}
    \mu_j = c_j \Upsilon_{\lambda_{\rm norm},j},
\end{equation}
where $\Upsilon_{\lambda_{\rm norm},j}$ is the stellar mass-to-light (i.e. M/L) associated with the $j^{th}$ SSP model. So, the mass-weighted (MW) mean stellar age and metallicity are calculated as
\begin{equation}
    \left< \log \age \right>_{\rm M} = \frac{\sum_j^m \mu_j \log \agej}{\sum_j^m \mu_j}
    \label{eq:MWAge}
\end{equation}
and
\begin{equation}
    \stmetM = \frac{\sum_j^m \mu_j \log {\rm Z}_{\star, j}}{\sum_j^m \mu_j}.
    \label{eq:MWMet}
\end{equation}

\subsubsection{Moment analysis}
\label{sec:new_ver:fitupd:momana}

After the description of the three steps present on the default procedure of analysis of \pyf (Figure \ref{fig:anaspec_flow}), we will move to revisit a second method of analysis of the ionized gas spectrum implemented on \pyf. We present in Section \ref{sec:new_ver:fitupd:elfit} the emission line parameters derived through the mixed {\bf RND+LM method} procedure. However, there are more emission lines than the strongest and frequently observed (and/or analyzed) ones, like those presented in the aforementioned Section. The parameters of the emission lines distributed by \pyp comprise those for more than 50 emission lines and the most majority of them weak ones. It is not practical to perform a Gaussian fit for all of them because it would be very time consuming. Considering this, we implement a different scheme to extract the main properties of these emission lines.

This method is based on a direct estimation of the flux intensity, velocity, velocity dispersion and equivalent width (EW) of the emission lines from the moment analysis of the ionized spectrum. For the operation of the method it is needed (i) a list of central wavelengths of emission lines to be fitted (in the rest-frame), (ii) the ionized gas spectrum (once subtracted the stellar population model), (iii) the stellar population model spectrum and (iv) an error spectrum. In addition, an input guess estimation of the gas velocity (in km/s) and velocity dispersion (including the instrumental dispersion, in \AA) are also required. For this initial guess we adopt the output of the \Ha emission analysis with the method described in Section \ref{sec:new_ver:fitupd:elfit}.

For each emission line, the algorithm defines a wavelength range at which it is going to perform the analysis. This range is centered in the wavelength of the emission line provided by the list (described before), shifted by the input guess velocity, and it covers a wavelength interval of $\pm$7.5 times the input velocity dispersion around this central wavelength. In this wavelength range the code calculates the first three statistical moments, that corresponds to the flux intensity, the central velocity and the velocity dispersion of the considered emission line. This moment estimation is performed weighting the fluxes using a Gaussian Kernel that adopted as width the input velocity dispersion. In this way, the possible contributions of adjacent (or even blended) emission lines are minimized, giving, at the same time, more weight to the regions of the emission lines with higher fluxes (and signal-to-noise values). Furthermore, the EW of the emission line is calculated by dividing the estimated integrated flux by the flux density of the underlying continuum. For doing so, the program estimates the flux density of the continuum from the stellar population model spectrum as the average flux density inside two wide (30\,\AA) side wavelength bands centered at $\pm\ 60$\,\AA\ from the central one. The size of the bands is chosen to mitigate contribution from stellar absorption features, although this is not perfect and some effects are always impossible to avoid. Finally, the process is iterated for each emission line following a MC loop to derive the errors of the four estimated parameters.


\subsubsection{Stellar-indices analysis}
\label{sec:new_ver:fitupd:indices}

The second additional analysis performed by \pyf different than the modeling of the stellar population and the emission lines is the estimation of the equivalent widths of a predefined set of stellar indices. This is part of a classical technique to derive properties of stellar populations in galaxies from the measurement of determined absorption line strength indices, such as the Lick/IDS system \citep[e.g.][]{Burstein.etal.1984, Faber.etal.1985, Worthey.1994}. Those indices are sensitive to the age and metallicity of the stellar populations and provide model-independent information complementary to that described in Section \ref{sec:new_ver:fitupd:stpopsyn} (fit of the stellar spectrum with multi-SSP templates).

The adopted algorithm of measurement follows the prescription (formulas and steps) implemented in {\sc indexf} \citep{Cardiel.etal.2003}. Briefly, the algorithm estimates the EW for each of the stellar indices from the stellar population model spectrum using the same procedure to derive the EW of the emission lines described in Section \ref{sec:new_ver:fitupd:stpopsyn}. In order to perform the analysis over a set of MC realizations, the program also considers the residuals from the stellar population synthesis as a guess of the noise pattern.

The current version of the code includes the measurement of eight stellar indices (\Hd, H$\delta$mod, \Hg, \Hb, Mg$b$, Fe5270, Fe5335 and D4000). We note that D4000 is not an EW, but a color parameter, that corresponds to the ratio between the flux density observed at wavelengths bluer and redder than the 4000\AA. In this particular case we use the definition based on the flux density in wavelengths (D4000$_\lambda$) instead of the most frequently adopted one that uses the flux density in frequencies (D4000$_\nu$) for this parameter. This parameter is frequently defined as B4000 and it corresponds to D4000$_\nu$/1.1619 \citep[e.g.][]{gorgas99}. The list of adopted bandwidths for each parameter is indicated in S16b, Table 2. For the calculation of the equivalent widths (and color parameters), the bandwidths are shifted to the observed-frame using the velocity estimated from the analysis of the kinematics described in Section \ref{sec:new_ver:fitupd:nlfit}. At the end, the program derives the standard deviation of the measurements of each index along the MC simulation.

\subsection{pyPipe3D: revisiting the procedure}
\label{sec:new_ver:pipeupd}

So far we have described the treatment performed to analyze a single spectrum using \pyf. Based on this fitting tool we develop a pipeline to provide with the spatially resolved properties of the stellar populations and emission lines from an IFS cube. This tool, named Pipe3D, was previously described in S16b. Here, we implemented the new \texttt{python} version of the code, and, based on the aforementioned philosophy described in Section \ref{sec:philosophy}, we create a single script, \pyp, that performs all the steps of this pipeline. To describe this new script we revisit briefly the steps implemented in this pipeline highlighting the differences and updates with respect to the previous version.

Pipe3D was thought to be a generalist tool, able to analyze data from any IFS galaxy survey (or individual observations). However, each survey (and instrument reduction package) provides the data in a slightly different format. In order to apply Pipe3D (and \pyp), it is needed to do some pre-processing on the data, which is not part of the pipeline itself. The goal of this pre-processing is to adapt any IFS data to the format expected by the pipeline. At the end, this format is a FITS file with at least two extensions (three preferred). The first extension should be a cube containing in each spaxel the observed spectrum (i.e., the flux intensity), with a regular linear step in the z-axis sampling the wavelength range. The spectra are assumed to be corrected for Galactic extinction. The other two extensions should have the same format (i.e., the same 3D world coordinate system, WCS). The second extension contains the 1$\sigma$ level of the error, estimated by the reduction process. Finally, the third extension should contain a 3D mask, including all the possible defects of the CCDs, cosmic rays masks, etc.

\begin{figure*}[t]
    \includegraphics[trim=00 0 5 0, clip, width=\textwidth]{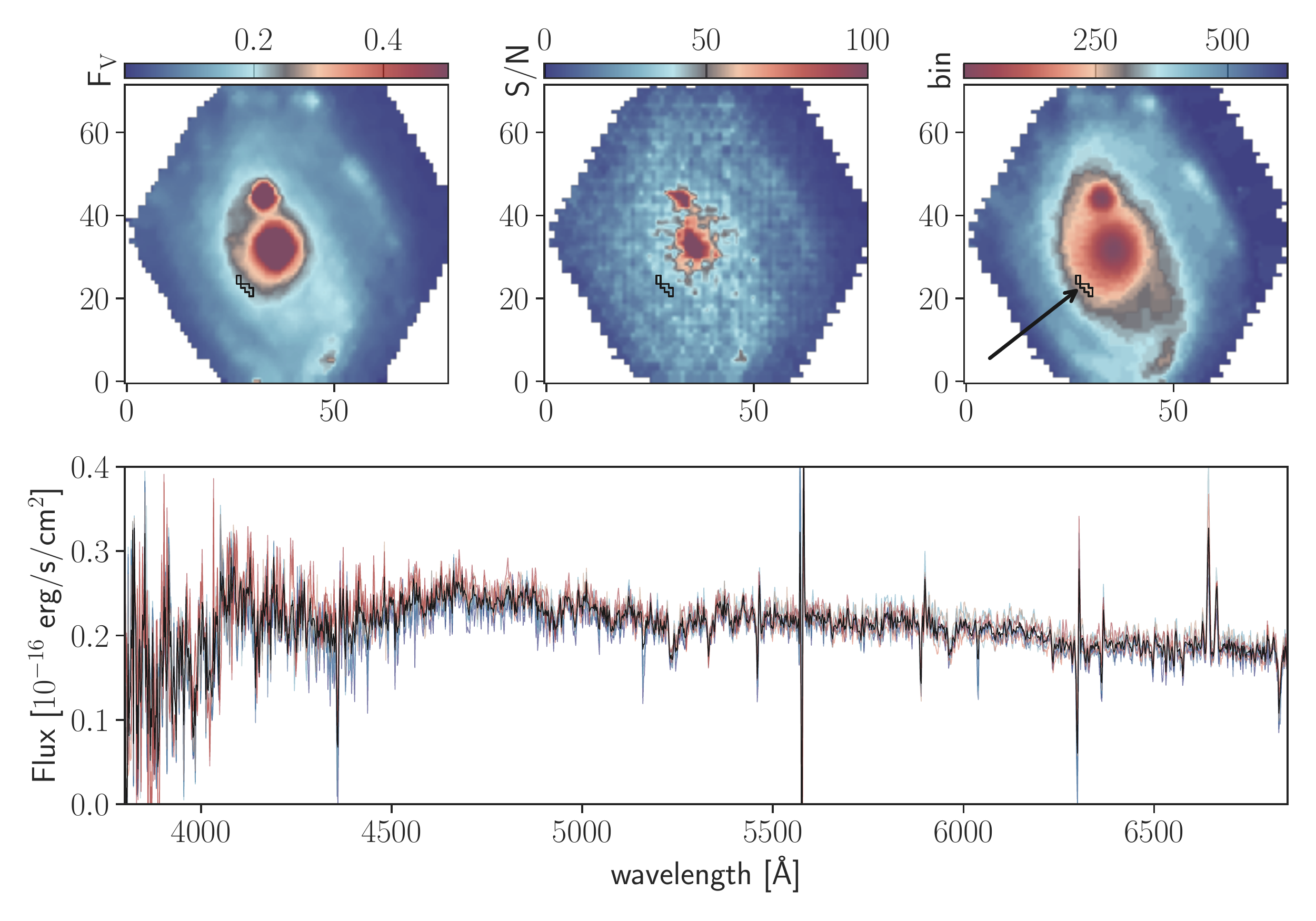}
    \caption{Example of the CS binning performed by \pyf for the IFS datacube of the galaxy NGC\,2916, observed by CALIFA. {\it Top panels:} From left to right, (i) map of the mean observed flux intensity with the wavelength range 5590-5680\,\AA, (ii) S/N map at this wavelength interval and (iii) the spatial bin (tessella) identification number (from 0 to 614 in this case). An arbitrary binned region (184) at approximately one effective radius of the center is highlighted on all three maps. The bottom panel shows the average spectrum (black) together with the individual spectra (red) within this tessella.}
    \label{fig:CSbin}
\end{figure*}

The analysis performed by \pyp begins with the extraction and the fitting of the central spectrum ($5\arcsec$ aperture) of the observed source, using the analysis described from Section \ref{sec:new_ver:fitupd:nlfit} to \ref{sec:new_ver:fitupd:stpopsyn}. This first analysis helps the subsequent ones, recovering the redshift and velocity dispersion of the central region and allowing to automatically select the guess and range of values to explore for each non-linear parameter (Section \ref{sec:new_ver:fitupd:nlfit}).

The surface brightness of galaxies decreases as a function of the galactocentric distance, with the S/N following the same behavior, degrading the reliability of the analysis of the stellar continuum in the outer regions (e.g. \citealt{Cappellari.and.Copin.2003, CF.etal.2013, CF.etal.2014}; S16a). \pyp overcomes this problem with a process of spatial binning called Continuum Segmentation binning, or CS-binning. This process, fully described in S16b, combines the criteria adopted in other segmentation/binning schemes, the isophotal \citep{Papaderos.etal.2002, Papaderos.etal.2013} and the Voronoi binning \citep{Cappellari.and.Copin.2003} schemes. The first method aggregates adjacent spaxels only if they belong to the same isophote (i.e., if their flux intensities differ less than a certain pre-defined percentage) and their distance is lower than a maximum value. By construction, this procedure preserves at a certain stage the shape of the original light distribution. However it does not guarantee to reach an optical S/N after binning, and the selection of the parameters is somehow arbitrary. On the other hand, the Voronoi binning uses as a unique criterium to aggregate adjacent spaxels the S/N. The procedure provides with the minimum number of voxels that increases the S/N above a desired threshold defined by the user. In this way, it aggregates spaxels without any consideration to their physical properties (e.g. it can aggregate arm and inter-arm regions in the same tessella/spatial bin).

The CS-binning combines both criteria in a procedure that uses both a goal S/N and a maximum difference in the flux intensity to aggregate adjacent spaxels. In summary, the procedure starts by selecting the un-binned spaxel with the largest flux intensity in a certain band (5590--5680\,\AA). If this spaxel already fulfills the requirement in S/N (i.e., if it is larger than the desired goal), the spaxel is selected as a bin itself. If not, the algorithm looks for all the adjacent spaxels for which the flux intensity differs less than a certain fractional threshold with respect to this flux intensity of the considered spaxel. If this criteria is fulfilled then each adjacent spaxel is aggregated to the bin, and the S/N and average flux are reevaluated. If the S/N is now above the threshold, the bin is defined and the procedure starts looking for a new bin (i.e., looking for the highest flux spaxel within the unbinned ones). If not, further adjacent spaxels are incorporated into the tessella until one of the two criteria are not fulfilled. The full procedure is finished when no un-binned spaxels are found above a threshold in S/N (normally 1$\sigma$). Then, the code provides with (i) a RSS with each row comprising the average spectrum within each of the final tessellas, (ii) a map of the fractional contribution of each spaxel to this spectrum (the so-called dezonification map \citealt{CF.etal.2013}; S16b) and (iii) a segmentation map in which is stored the unique ID of each tessella, for a future identification of the average spectra with their location in the sky.

Figure \ref{fig:CSbin} exemplifies this process for the CALIFA V500-datacube of galaxy NGC\,2916. The three maps are: the flux intensity of V-band; the S/N map at this band and the CS-binned map. The average spectrum and the spectra of the binned spaxels of tessella 184 are shown at the bottom panel (this tessella was arbitrary chosen within those located near one effective radius from the center of the galaxy). This binned region is highlighted in all three maps. The S/N map shows the patchy pattern due to the three-pointing dithering scheme adopted by CALIFA (and other IFS surveys, like MaNGA or SAMI)\footnote{The final datacubes in these surveys are the result of combining at least three dithered pointings with fibers larger than the final selected spaxel. For this reason certain spaxels comprise the information of three fibers, while adjacent spaxels comprise the information of only two or just one of the original fibers. Thus, for the same flux intensity the S/N is not homogeneous, maintaining the triangular structure of the adopted dithering scheme.}. The smooth distribution of the CS-binned map (top-right panel in Figure \ref{fig:CSbin}) shows that this spatial binning essentially removes this effect, conserving the shape of the light distribution of the galaxy. Our experiments have shown that when using other binning schemes, in particular the Voronoi binning, the light distribution of the galaxy is not preserved as well as using the adopted CS-binning. We need to recall the reader that this former binning was developed to explore the central regions of early type galaxies, which usually have a smooth light distribution without significant features \citep{Cappellari.and.Copin.2003}. S16b already performed a direct comparison of the three aforementioned spatial binning procedures for the same dataset confirming this result.

At this point, \pyp run the \pyf algorithm (described in Sections from \ref{sec:new_ver:fitupd:nlfit} to \ref{sec:new_ver:fitupd:stpopsyn}) for each individual spectrum of the CS-binned dataset. This process provides with all the parameters summarized in Section \ref{sec:new_ver:fitupd} at each location within the FoV of the IFS data. This way, for each parameter is generated a map (2D distribution), in which each derived parameter is associated with the location of the CS-bin defined by the segmentation map described before. Finally, all the maps with the distribution of properties are integrated into a set of cubes (3D arrays) for more convenient and simple distribution. These cubes will be described later on. In addition, \pyp provides with an additional cube with the spectral information provided by \pyf, comprising (i) the original spectra, (ii) the best model for the stellar populations, (iii) the best model combining both the stellar populations and the emission lines models, and (iv) three residual spectra: (a) once subtracted the combined model, (b) once subtracted the best stellar population model (i.e., gas spectra), and (c) once subtracted the model spectra for the emission lines (i.e., emission-line cleaned spectra). This product has a cube structure since it comprises at each slice the RSS including these six spectra for each tessella, with the z-axis running for the number of spatial bins.

The \pyp procedure continues with an analysis of the stellar indices, following the procedure described in Sec. \ref{sec:new_ver:fitupd:indices}. This exploration was performed for each individual spectrum corresponding to each tessella once subtracted the best model for the emission lines, removing their possible contribution/pollution to the estimation of the stellar indices. Like in the case of the previously described procedure the result of this analysis is a set of maps, one for each analyzed index (and another one for its corresponding error), with a spatial structure that follows the applied tessellation (i.e., all spaxels within the same spatial bin have the same estimated stellar index, by construction). Following the same storing philosophy all those maps are integrated into a single dataproducts cube for a more simple distribution of the results.


Finally, \pyp re-analyze of the emission lines adopting two different procedures, the parametric fitting described in Sec.  \ref{sec:new_ver:fitupd:elfit}, and the moment analysis described in Sec. \ref{sec:new_ver:fitupd:momana}. Contrary to the previous explorations described above these two new analysis are performed spaxel-by-spaxel, not using the CS-binned RSS spectra. The reason behind that is that the spatial distribution of the continuum, that was the basis of the adopted tessallation procedure, does not necessarily corresponds to that of the ionized gas \citep[which may present both a smooth component associated with the continuum and clumpy of filamentary structures][]{sanchez21}. Prior to any analysis it is constructed a GAS-pure datacube, i.e., a cube in which the contribution of the stellar population is removed. For doing so the best fitted stellar population model provided by the \pyf algorithms for each tessella is re-scaled to the flux intensity of each spaxel within this tessella using the dezonification map. A further polynomial correction is applied to remove the possible color effects between the model (resulting from the fitting to the average spectra) and the individual spectrum. This matched stellar-population model spectrum is then removed to the original one creating a spectrum that contains the emission by the ionized gas (i.e., the emission lines), plus noise and residuals. This GAS-pure cube is also provided as a dataproduct of the \pyp\ analysis.

The final analysis explores the strongest emission lines in the optical wavelength range (\Oii, \Hb, \OOiii, \Oiii, \NNii, \Ha, \Nii, \SSii, \Sii, in the current implementation of the code) using the parametric fitting (assuming a Gaussian profile for each emission line), and the full set of strong a weak emission lines described before, for the non-parametric moment analysis, spaxel by spaxel using the GAS-pure datacube. Once more, like in the previous cases, each of these procedures provides with a map with the spatial distribution for each of the derived properties (flux, velocity...) and each of the emission lines, and their corresponding errors. For convenience, once more, all those maps are rearranged into two different dataproduct cubes, one for each of the procedures.


In summary, \pyp provides with five different dataproduct cubes, that can be stored as individual FITS files, or as extensions of the same Pipe3D FITS file. Those dataproducts are tagged as:
\begin{itemize}
    \item {\bf \texttt{SSP}}: Main parameters derived from the analysis of the stellar populations, including the LW, and MW ages, metallicities, dust attenuation and stellar kinematics properties, derived for each tessella;
    \item {\bf \texttt{SFH}}: Weights of the decomposition of the stellar population for the adopted SSP templates library (e.g., values in Eq. \ref{eq:coeff}), derived for each tessella. It can be used to derive the spatial resolved star-formation and chemical enrichment histories of the galaxies and the LW and MW properties included in the {\bf SSP} dataproducts;
    \item {\bf \texttt{INDICES}}: Set of stellar absorption indices derived for each tessella once subtracted the emission line contribution;
    \item {\bf \texttt{ELINES}}: Flux intensities for the strongest emission lines in the optical wavelength range, together with the kinematics properties of \Ha, derived based on a Gaussian fitting of each emission line, spaxel-by-spaxel;
    \item {\bf \texttt{FLUX\_ELINES}}: Main parameters of a set of strong and weak emission lines ($\sim$50) derived using a weighted momentum analysis based on the kinematics of \Ha, derived spaxel-by-spaxel. It includes the flux intensity, equivalent width, velocity and velocity dispersion, and the corresponding errors for the different analyzed emission lines.
\end{itemize}

\section{Accuracy of the fitting code}
\label{sec:accur}

As indicated before, the full code has been transcribed and completely re-coded to \texttt{python} with a set of modifications. Therefore, it is required to evaluate the accuracy and the precision of all the recovered parameters. In order to do so  we run the code on a set of simulated and real spectra.




%
\subsection{Testing the stellar population analysis}
\label{sec:accur:stpop}

We generate a set of 2000 realistic simulated stellar spectra by adopting the set of parametric SFH and ChEH recently published by \citet{MejiaNarvaez.etal.2020}, Appendix B. These evolutionary histories were designed to reproduce the wide range of M/L, $\age$ and metallicity distribution functions observed in the most recent IFS galaxy surveys such as CALIFA or MaNGA \citep[e.g.][]{sanchez18,Lacerda.etal.2020}.
The considered SFH and ChEH were transformed to age and metallicity distribution functions in stellar masses, re-sampled to the grid of values covered by the adopted \textt{gsd156} SSP library. Once we estimate the amount of stellar mass that corresponds to each SSP template (e.g., values described in Eq. \ref{eq:coeff_mass}), this fraction is transformed to light based on the corresponding $\Upsilon_{\lambda_{\rm norm}}$ (i.e., coefficients described in Eq. \ref{eq:coeff}).
Then, applying Eq. \ref{eq:spec_mod}, a simulated stellar spectrum is created  corresponding to a particular SFH and ChEH. Finally, the set of simulated spectra are (i) re-sampled to 2\,\AA/pix\footnote{to mimic the sampling of the CALIFA V500 dataset}, (ii) shifted to a specific redshift; (iii) convolved with a Gaussian function of width $\sigma_\star$ to replicate the LoSVD of the stars and (iv) attenuated by the dust extinction value obtained applying the extinction law by \citet{Cardelli.etal.1989}. Hence, for each simulation we adopt a set of non-linear parameters randomly selected from a flat distribution within pre-defined intervals: (i) from 0.005 to 0.05 for $z_\star$; (ii) from 75 and 350 km/s for the $\sigma_\star$; and (iii) from 0 to 1.6 mag for the A$_{\rm V}^\star$. These parameters were selected to emulate the typical values for nearby galaxies.

Finally, we add white noise to the simulated spectra following a Gaussian distribution in order to simulate the range of S/N values reported in the most recent IFS galaxy surveys (1-$\sigma$ of the input have S/N between 50 and 100, and all simulated data with S/N values ranging between 20 and 300). We do not include other types of noise pattern in the simulated spectra, limiting our exploration to situations in which problems like spectrophotometric calibration, errors in sky subtraction, CCD defects, etc, are negligible. Furthermore, no emission lines have been included in these simulations. However, it is well known that a well-defined set of masks for the emission lines \cite[e.g.][]{Stencel.1977, Matsuoka.etal.2007, Schroder.etal.2009} is essential for a good determination of the stellar parameters in real data.

We adopted the same configuration files, initial {\it guess} parameters and {\it intervals} for the analysis of all the simulated spectra. For the initial {\it guess} we use the average values covered by our simulations and as {\it intervals} the simulated range of values. For these simulations we use the same stellar population library to simulate the spectra and to model them. Thus,
the \texttt{gsd156} library is adopted for the analysis of the stellar populations and the \texttt{gsd12} for the non-linear step. So far we want to explore how well the code reproduces well-known and controlled parameters. For a discussion on the effects of fitting a simulation generated with a particular stellar library using other libraries we refer the user to \citet{CF.etal.2014}.

\subsubsection{Recovery of simulated parameters}
\label{sec:accur:stpop:recov}

\begin{figure*}[t]
    \includegraphics[width=0.33\textwidth]{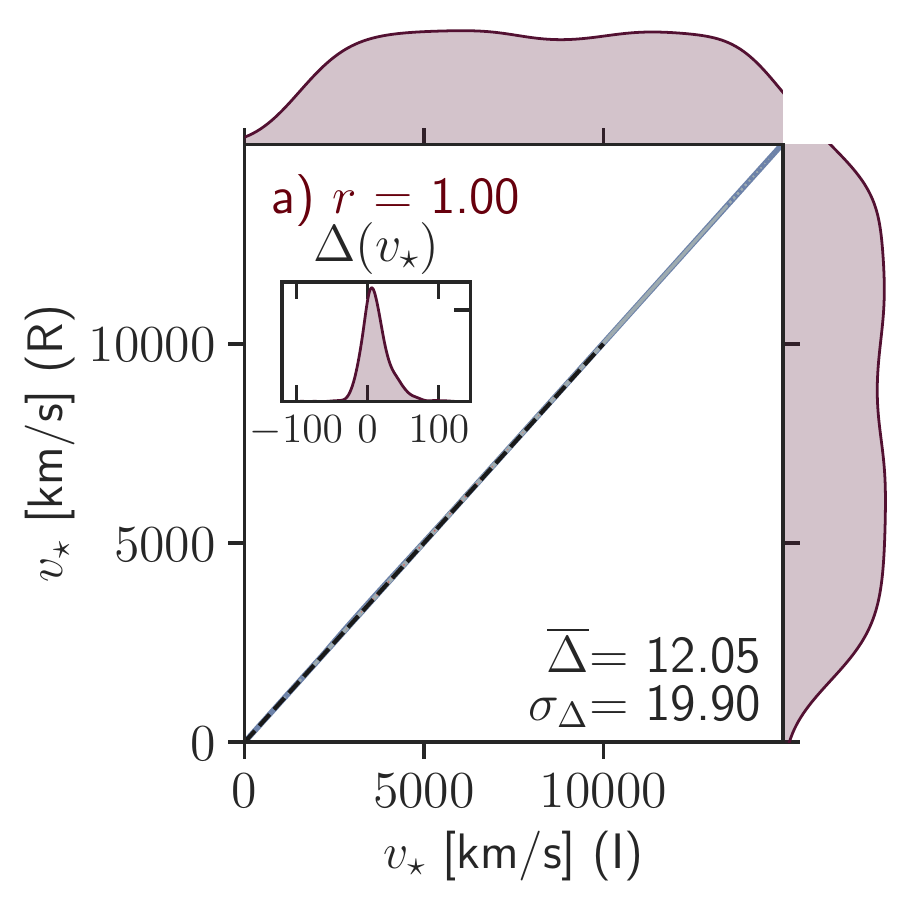}
    \includegraphics[width=0.33\textwidth]{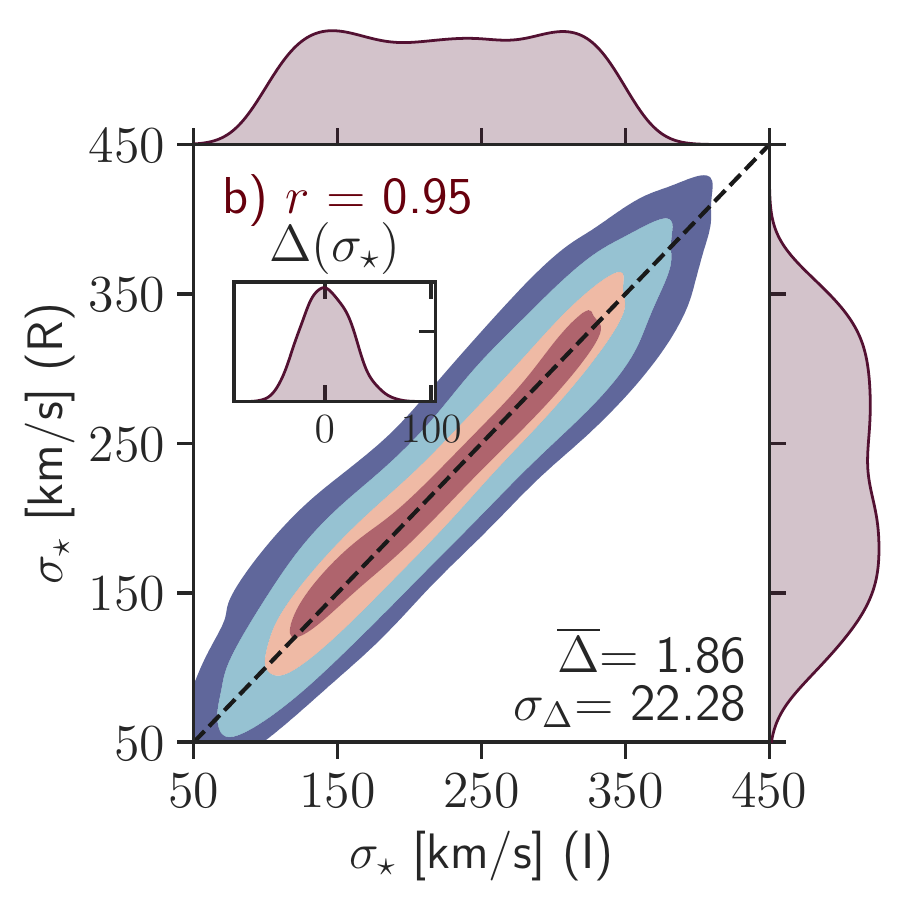}
    \includegraphics[width=0.33\textwidth]{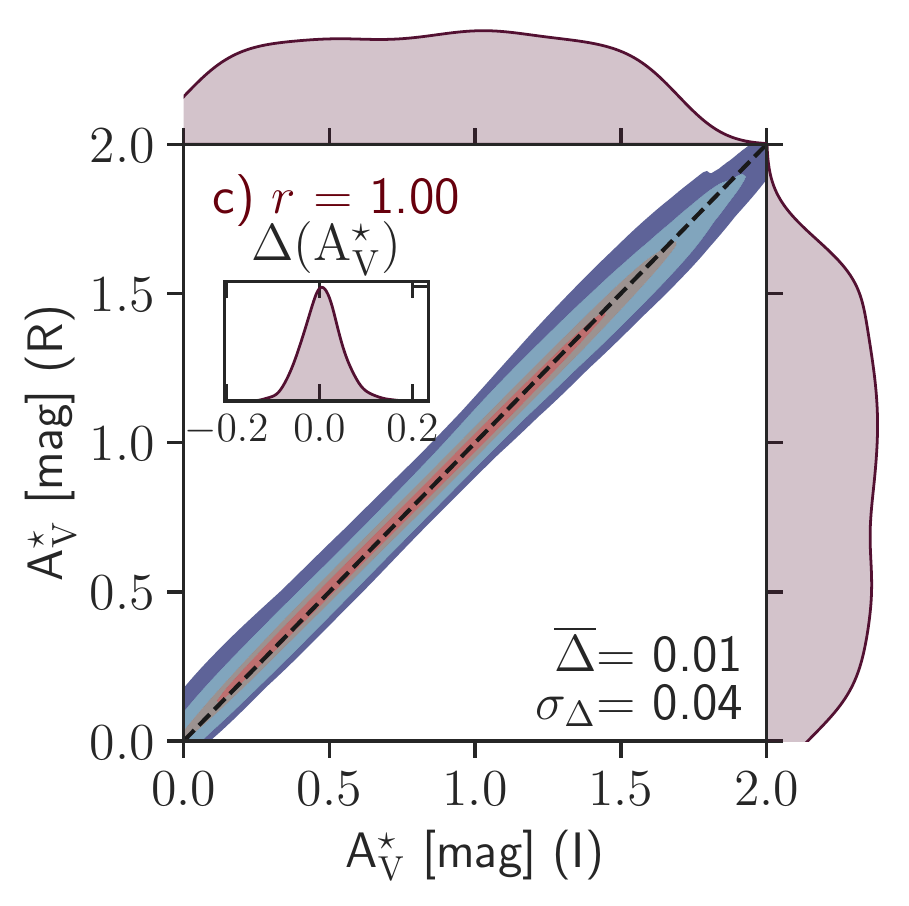} \\
    \includegraphics[width=0.33\textwidth]{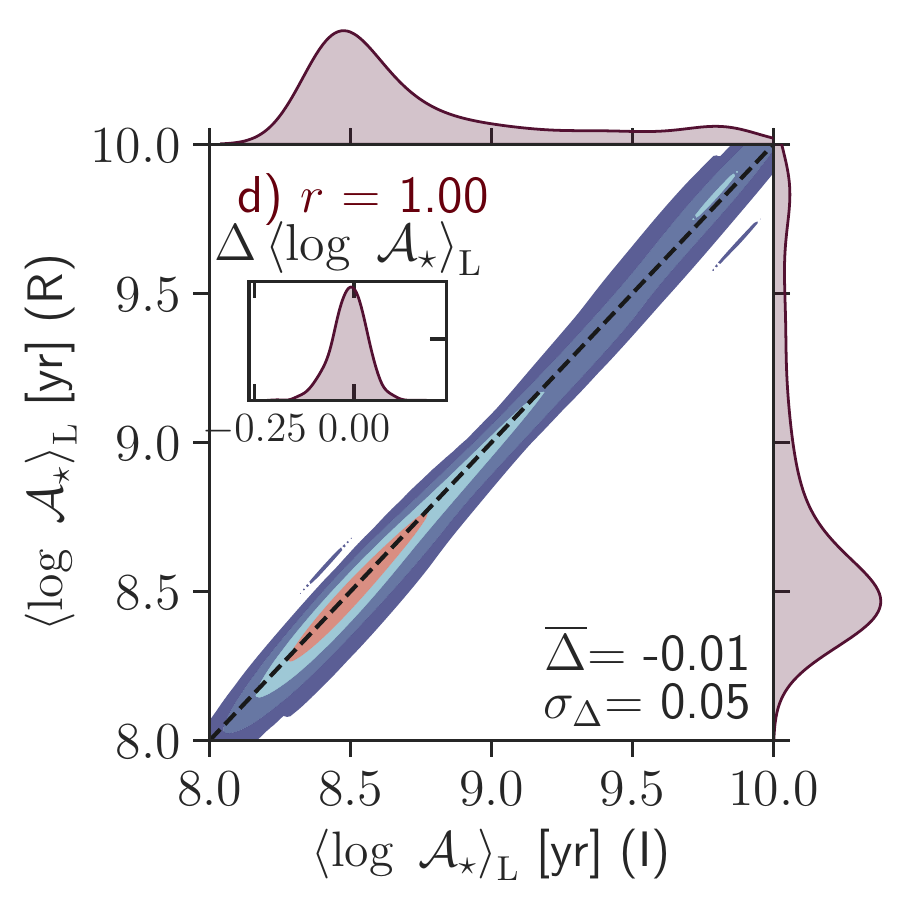}
    \includegraphics[width=0.33\textwidth]{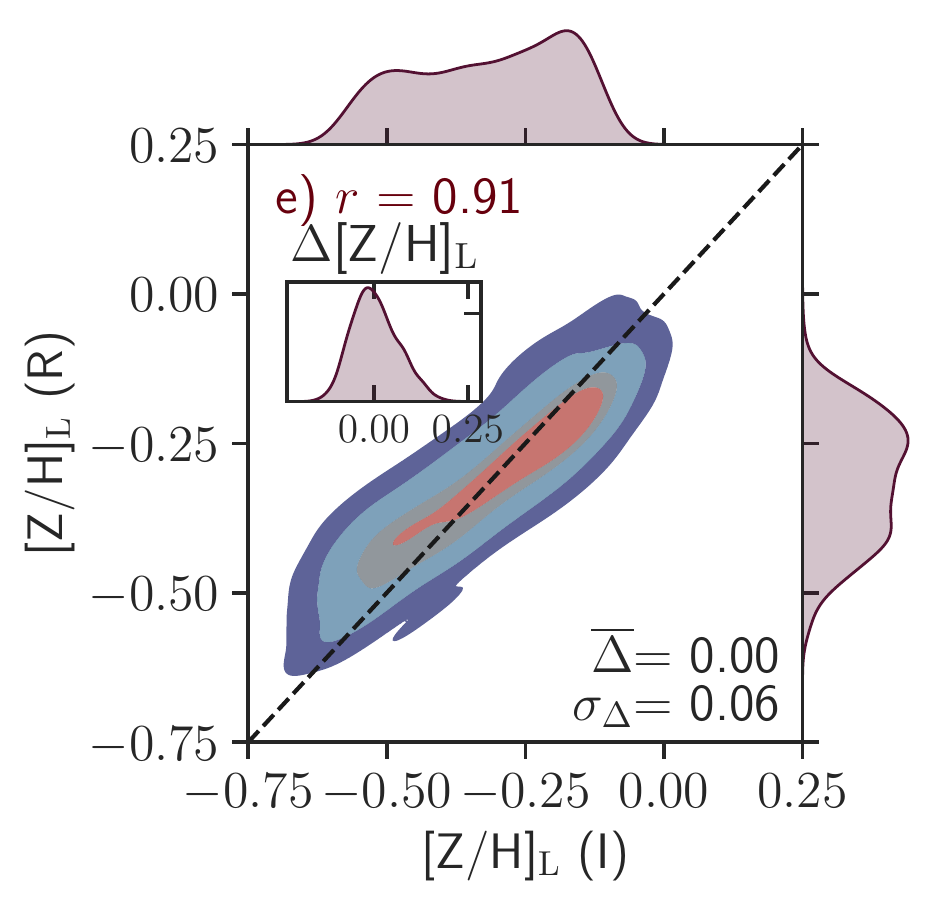}
    \includegraphics[width=0.33\textwidth]{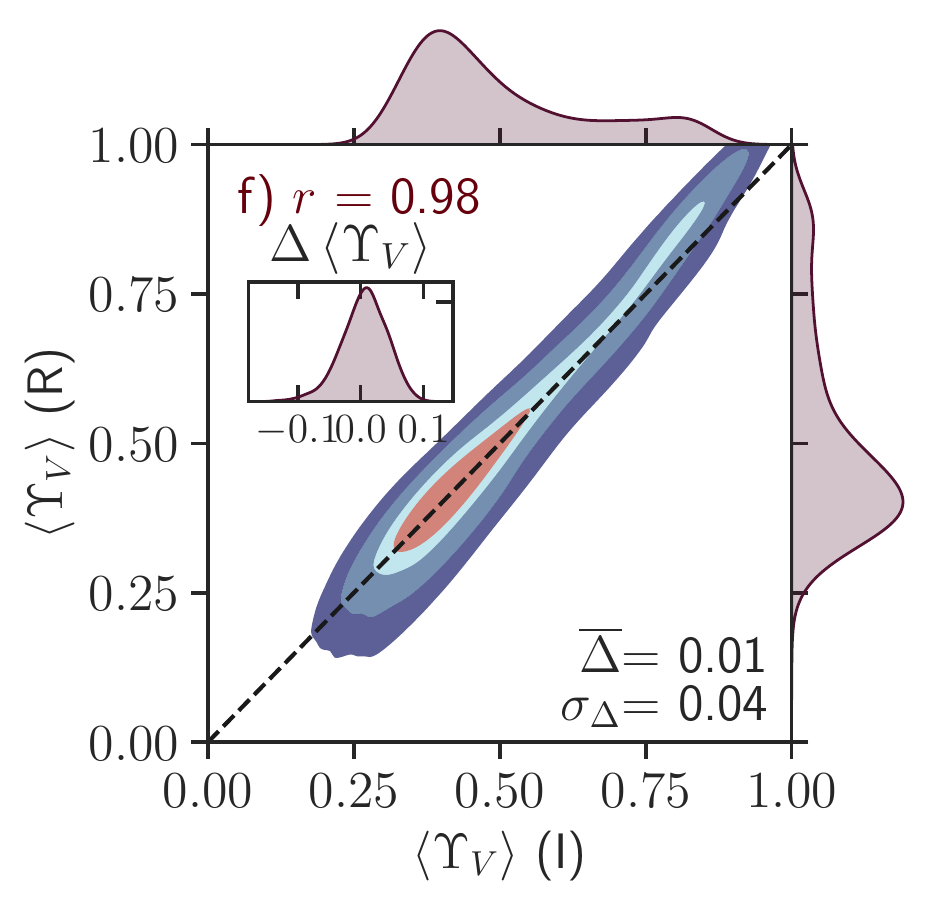}

    \caption{Stellar parameters recovered by \pyf (R) as a function of the input ones (I) for a set of 2000 simulations with an average S/N$\sim$60. Upper panels correspond, from left to right, to the non-linear parameters: redshift ($z_\star$), line-of-sight velocity dispersion ($\sigma_\star$) and A$_{\rm V}^\star$. Lower panels correspond, from left to right, to the parameters recovered by the stellar population analysis: LW age (\stageL), LW metallicity (\stmetL) and the stellar mass-to-light ratio (\mlL). Each panel shows the Pearson correlation coefficient ($r$), the mean and the standard deviation for the distribution, with each contour encircling a 1, 2, 4 and 6 standard deviations of the kernel density estimate (KDE) of points. We warn the reader that KDE method introduces extrapolation effects in the distributions. Most of data points used to generate distributions are within 2-$\sigma$. The projected KDE of each parameter has been included at the upper and right borders of each panel. The inset in each panel shows the KDE of the difference between both parameters, i.e., $\Delta$(par)}
    \label{fig:stpop_simul}
\end{figure*}


Figure \ref{fig:stpop_simul} shows the comparison between the input (I) and recovered (R) values for the following stellar parameters: $z_\star$, $\sigma_\star$, A$_{\rm V}^\star$, \stageL (Eq. \ref{eq:LWAge}), \stmetL (Eq. \ref{eq:LWMet}) and the stellar mass-to-light ratio (\mlL), based on the simulations described before. The offset between the input and recovered parameters and the corresponding standard deviations have been included in each panel of the figure. It is clearly appreciated that the all parameters are very well recovered.

Among the non-linear parameters, the redshift is the one recovered with the most precision and accuracy by \pyf. This is due to the procedure applied on the determination of the non-linear parameters . We have to remember that there is no curve fit of the parameter, i.e. the method relies on the lowest value of Eq. \ref{eq:nlfit_merit}, which is primarily affected by the match between the central wavelengths of the stellar absorption lines. On the other hand, the $\sigma_\star$ is the non-linear parameter recovered with less accuracy ($\sim$2 km/s) and precision ($\sim$22 km/s). Although \pyf has not been developed optimized for kinematic analysis of the stellar population, the values are recovered with great accuracy and without any significant bias related to the input values. However, the precision is somehow affected. The extraction of the LOSVD is a degenerated problem and both the adopted SSP library and the LOSVD profile chosen (in \pyf case, a Gaussian one) may affect it \citep[e.g.][]{Bender.1990, Statler.1995, Cappellari.and.Emsellem.2004, Ocvirk.etal.2006, FalconBarroso.etal.2011,FalconBarroso.etal.2021}.
For the last parameter determined during the non-linear fit step, the dust attenuation, there is a good agreement between input and recovered values. However, the determination of A$_{\rm V}^\star$ is another degenerated problem, which affects and is affected by the determination of other parameters. We will discuss this matter forward along this Section. %

\begin{figure*}
    \includegraphics[width=\textwidth]{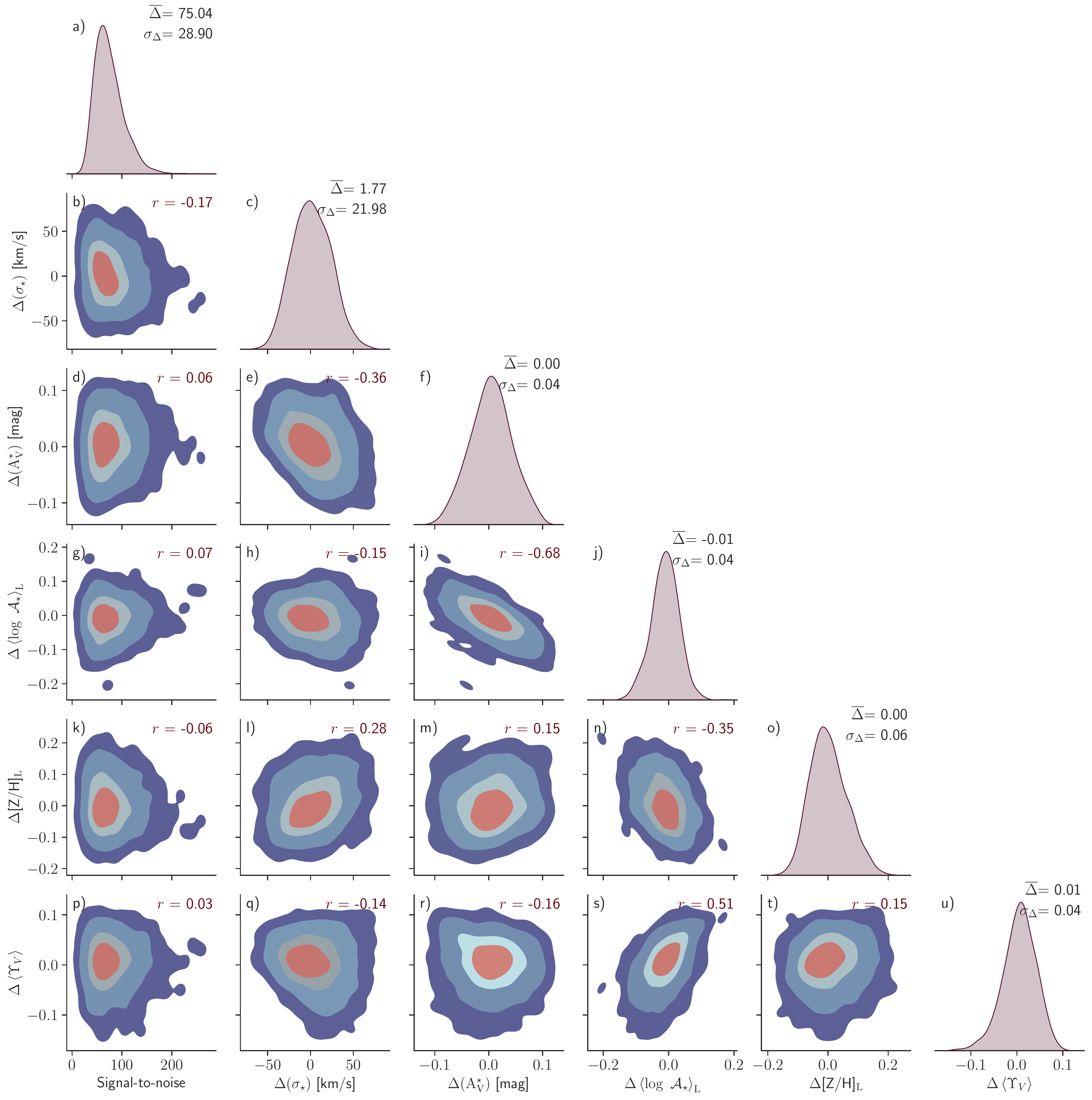}
    \caption{Pair-plot comparing the differences between the recovered (R) and input (I) parameters for the different explored parameters shown in Figure \ref{fig:stpop_simul}($\Delta$par) and their distributions along the input S/N of each simulation run. This plot illustrates the possible interdependence between the accuracy in the derivation of the different parameters. All panels include the Pearson ($r$) correlation coefficient for the corresponding KDE of the distribution at the upper-left corner. Each contour encircles a 1, 2, 4 and 6 standard deviations of the density of points.}
    \label{fig:stpop_resid_simul}
\end{figure*}

The remaining stellar parameters are then derived using the non-linear parameters already determined and fixed, following the procedure described in Section \ref{sec:new_ver:fitupd:stpopsyn}. From this analysis we recover the coefficients of the decomposition of the stellar population in the considered SSP library, from which we derive average properties, both LW and MW. 
Along the bottom row of Figure \ref{fig:stpop_simul} we evaluate the capability of \pyf to recover these properties. The LW mean age is the best recovered stellar property, with an accuracy of $\sim$-0.01 dex and a precision of $\sim$0.05 dex. On the other hand, the metallicity is recovered slightly worse, with no appreciable offset and a precision of $\sim$0.06 dex, on average. Contrary to the age, the metallicity presents a negative bias (recover lower values) at the high metallicity range, and a positive bias (recover higher values) at low metallicity. Usually the age is the main source of variance between the SSP templates with the metallicity responsible for higher order effects \citep[e.g.][]{Ronen.1999}. Thus, the age is the parameter that it is a priori easier to distinguish between different stellar populations, what explain why it is better recovered than the metallicity. For the stellar \mlL we see a good agreement between the input and recovered values, with an average accuracy of $\sim$0.01 dex and a precision of $\sim$0.04 dex. Like in the case of the metallicity we appreciate a positive offset (larger recovered values), in the range of larger \mlL values ($>$0.75 dex). However, this bias is never larger than a few percent.

\begin{table}[htb!]
    \label{tab:stpop_stats_SN}
    \caption{Stellar population simulation: statistics of the residuals for a typical S/N of $\sim$60.}
    \centering
    \begin{tabular}{cc}
    ${\Delta}$(par) & mean $\pm$ stdev \\
    \hline
    $v_\star$ & 13.71 $\pm$ 21.23 km/s \\
    $\sigma_\star$ & 4.94 $\pm$ 20.32 km/s \\
    A$_{\rm V}^\star$ & 0.01 $\pm$ 0.04 mag \\
    \stageL & -0.01 $\pm$ 0.04 dex \\
    \stmetL & 0.01 $\pm$ 0.07 dex \\
    \mlL & 0.01 $\pm$ 0.04 dex \\
    \hline
    \end{tabular}
    \\ Results of the mean $\pm$ the standard deviation of the distributions. \\
\end{table}


\subsubsection{Interdependence on the ability of estimating the parameters}
\label{sec:accur:stpop:bias}

Until now we characterized the capability of \pyf to recover a set of input parameters. Now we proceed to characterize how the accuracy and precision in the determination of one parameter affect other ones, i.e., the bias in the derivation of those parameters. Figure \ref{fig:stpop_resid_simul} shows the pair-plot for the distributions of the difference between the recovered and input values ($\Delta$par = par$_{\rm R}$ - par$_{\rm I}$), for the set of parameters shown in Figure \ref{fig:stpop_simul} (discussed in the previous Section). In addition, we include a first column with the input S/N of each run, illustrating the precision and accuracy of the fit accordingly to the input goodness of data, i.e., the relative residual offsets as a function of the S/N. In each panel is represented a pair of parameters, with the diagonal representing the corresponding kernel density estimate for the matching parameter.
The median and standard deviations of the distributions included in this diagonal are the same as the ones shown in Figure \ref{fig:stpop_simul}, being included only for completeness.
The plots involving $\Delta(z)$ have been deliberately removed since the redshift presents almost no bias, according to the results shown in the previous Section, and for this particular case it is appreciated no correlation at all with any of the explored $\Delta par$.


Examining the first column of Figure \ref{fig:stpop_resid_simul} we see no correlation between the residuals and the input S/N, what tunes the code accuracy. We also note that the results precision is improving with the increasing of the input S/N. We show in Table \ref{tab:stpop_stats_SN} the statistics (mean and standard deviation) of the explored residuals for a typical S/N of $\sim$60. Conversely, there are clear correlations that highlight the interdependence in the estimation of the different explored parameters, in agreement with the results found in the literature using different stellar synthesis codes \citep[e.g.][]{CF.etal.2005, SanchezBlazquez.etal.2011, CF.etal.2014}. The strongest correlation is found between the $\Delta$A$_{\rm V}^\star$ and $\Delta\stageL$ ($r=-$0.71), indicating that an over-estimation of the dust attenuation produces an under-estimation of the LW age of the stellar population (and the other way around). This degeneracy is somehow expected since the dust attenuation and the age produce similar effects in the shape of the spectra (i.e., making them redder). Despite of its strength we need to highlight that the effect involves a maximum variation of $\sim$0.1 dex in the \age for an error of $\sim$0.2 mag in A$_{\rm V}$. The second stronger correlation is that of the $\Delta\stageL$ with the \mlL ($r=$0.51), that illustrates the direct connection between both parameters. As indicated before, the strongest variance in the SSPs is due to the \age, changing essentially the \ml ratio. Since older stellar populations have larger \ml ratios than younger ones, the observed positive relationship between the two residuals is naturally understood.

\begin{figure*}[htb!]
    \includegraphics[width=0.333\textwidth]{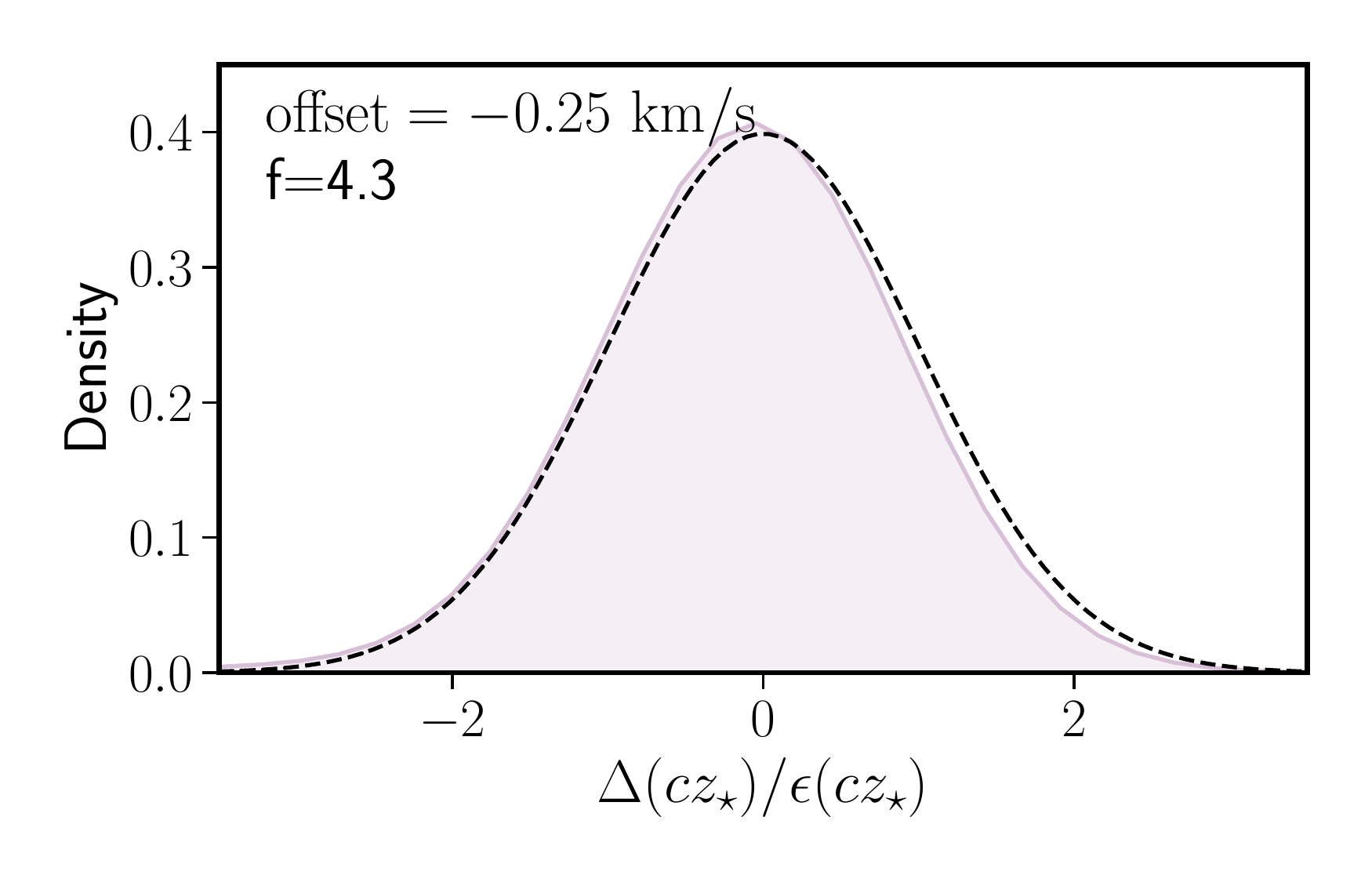}\includegraphics[width=0.333\textwidth]{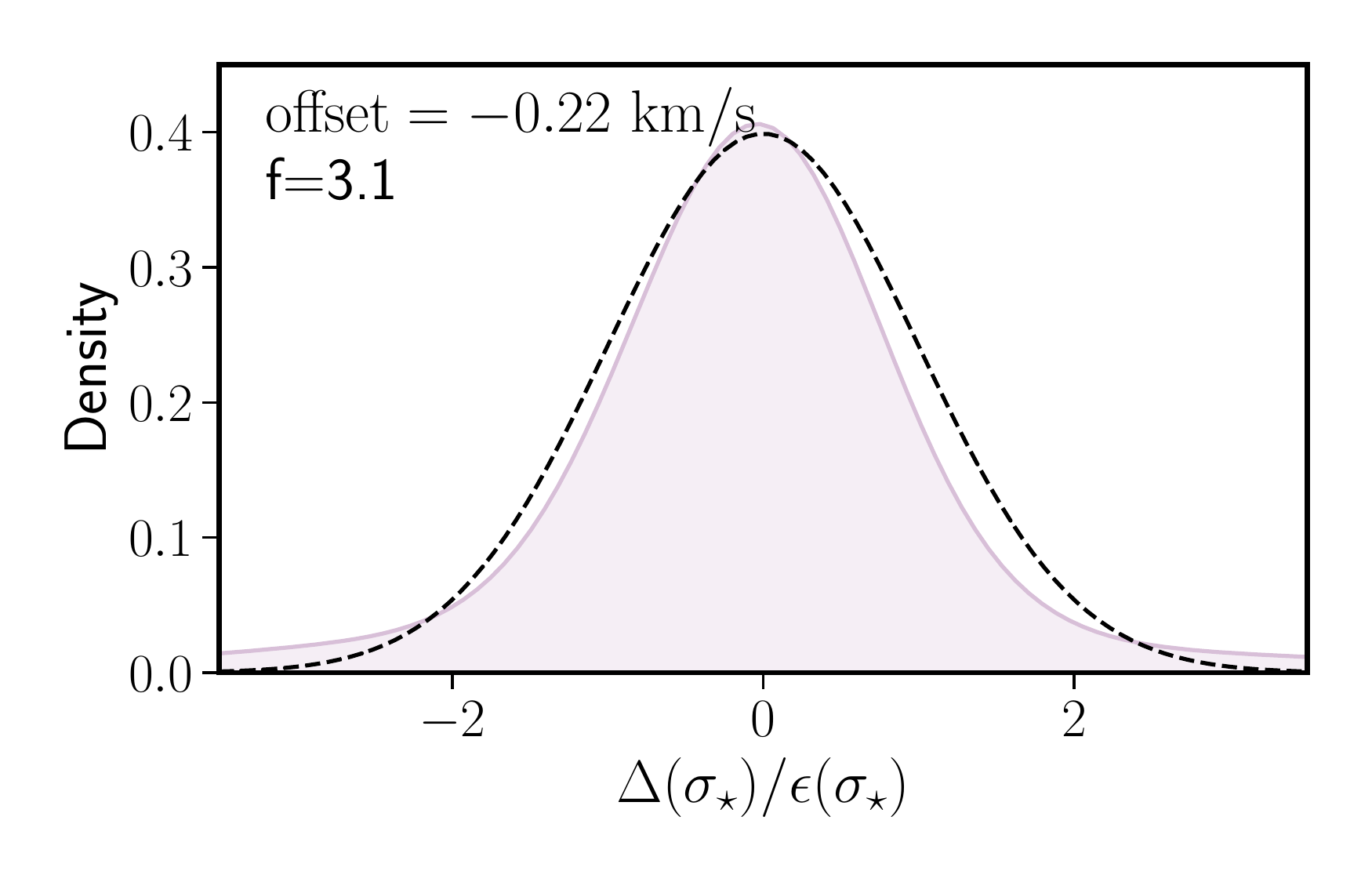}\includegraphics[width=0.333\textwidth]{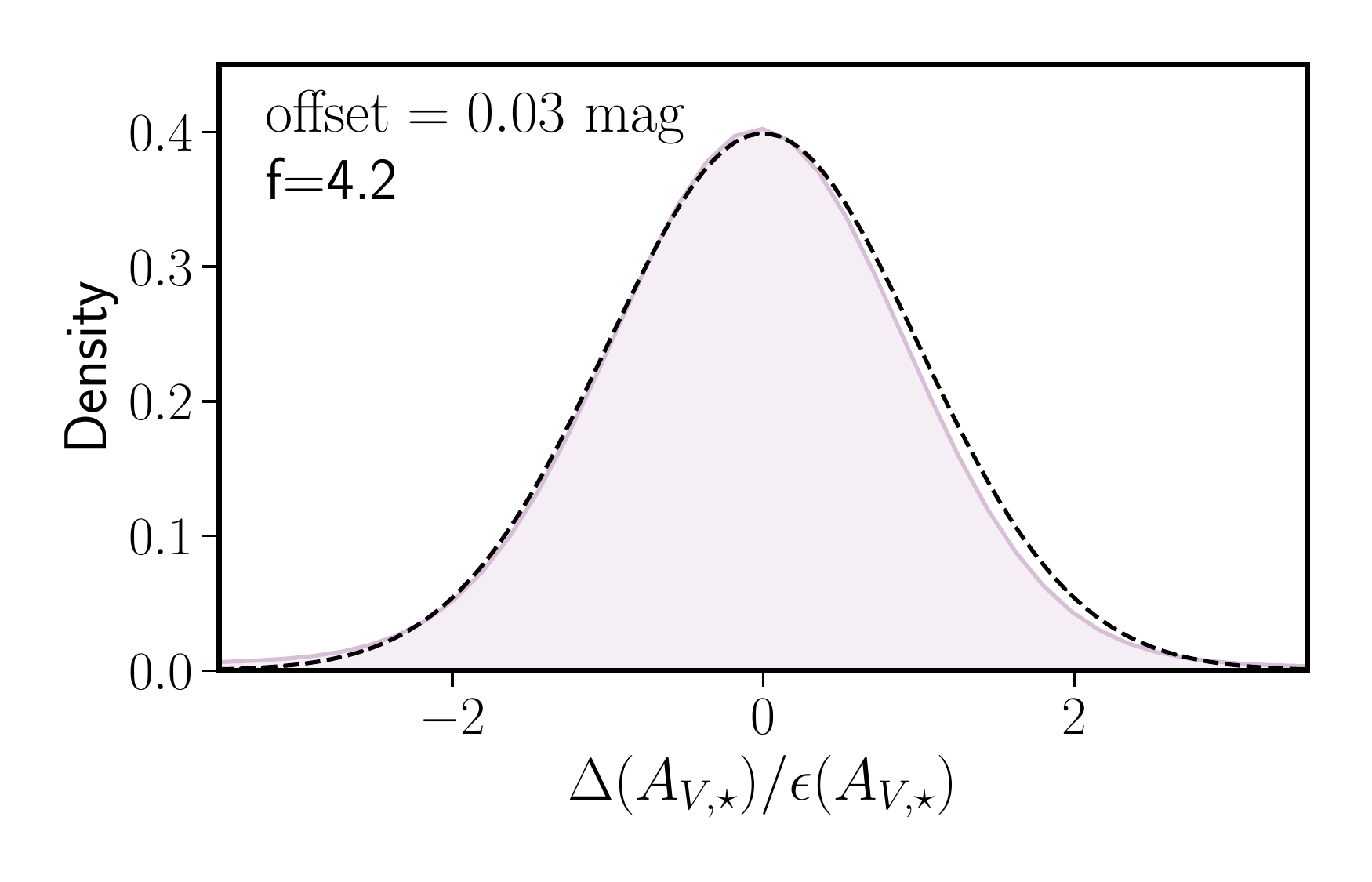}
    \includegraphics[width=0.333\textwidth]{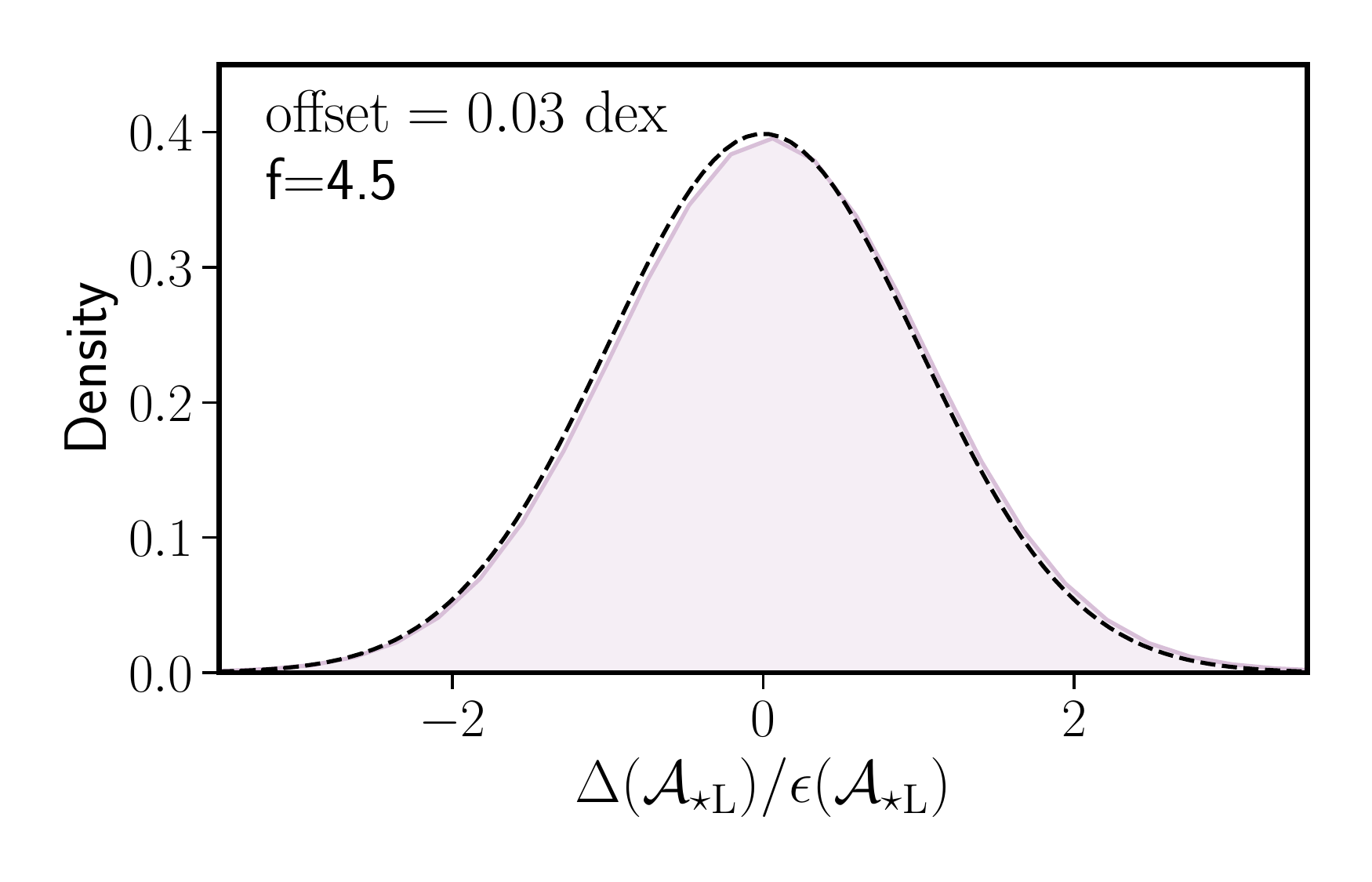}\includegraphics[width=0.333\textwidth]{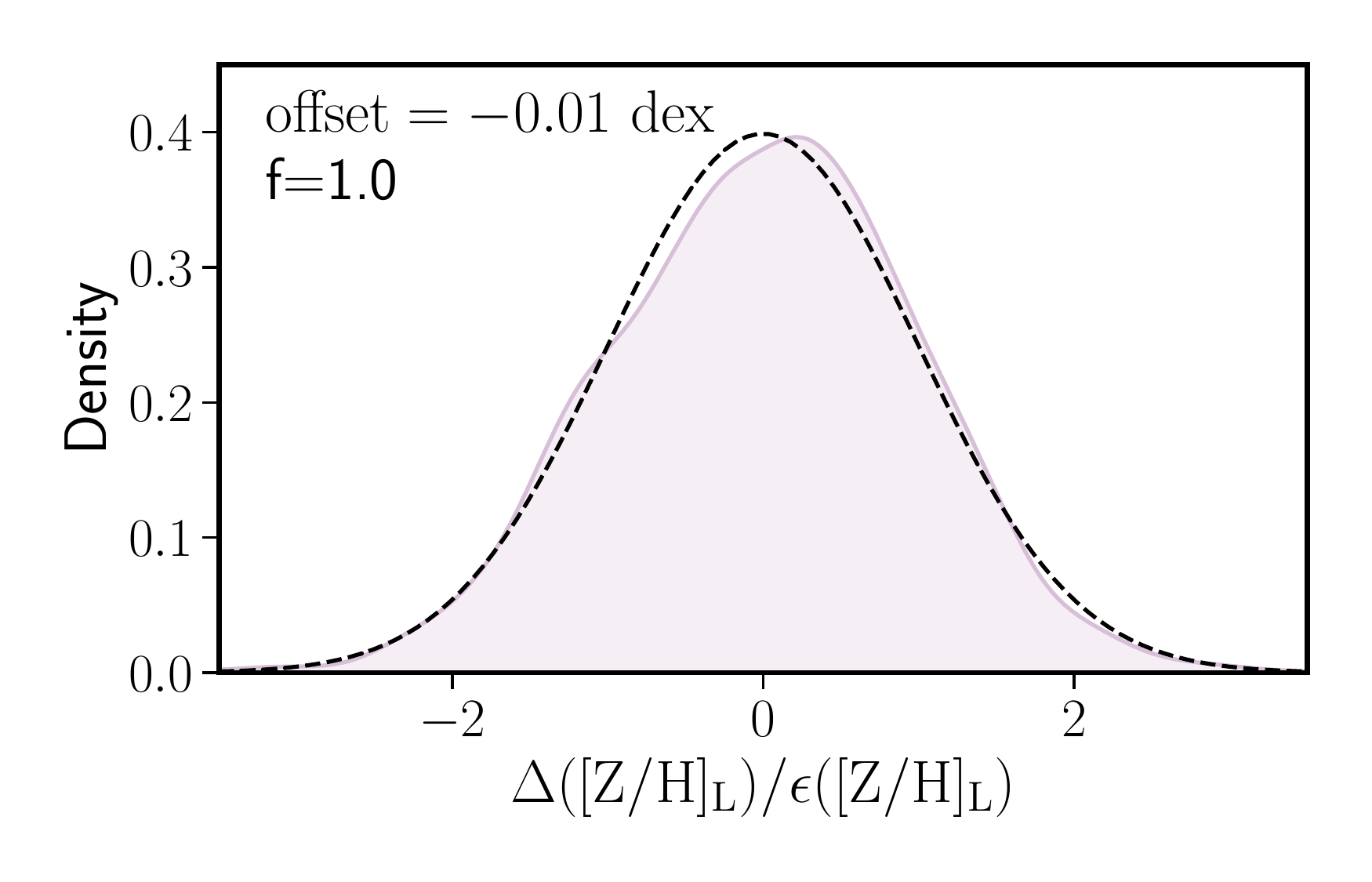}\includegraphics[width=0.333\textwidth]{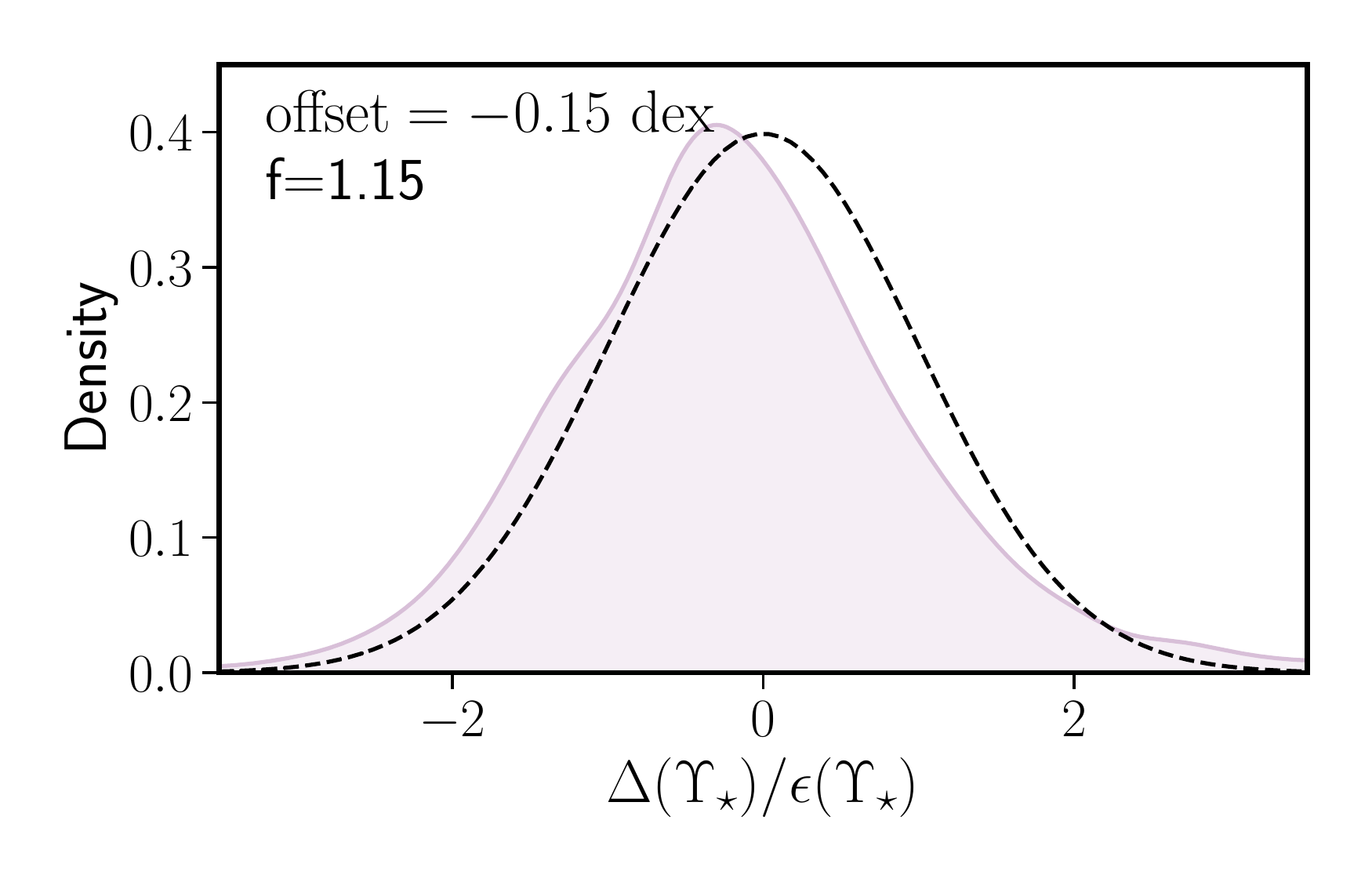}
    \caption{Density distribution (shaded curve) of the uncertainty in the derivation of the explored stellar parameters, $\Delta$(par) (i.e., the estimated value minus the input/simulated one), divided by the error estimated by \pyf, $\epsilon$(par). Top-panels show the distribution for the non-linear parameters ($z_\star$, $\sigma_\star$ and  A$_{\rm V}^\star$), while bottom panels the mean properties of the stellar populations (\stageL, \stmetL and \ml). The dotted line in each panel corresponds to a Gaussian function centered in zero with a standard deviation of one, representing the expected distribution if the estimated errors were fully representative of the real uncertainties in an statistically way.
    $\epsilon$(par) has been multiplied by a factor $f$ (shown in each panel) to match the observed and expected distributions as much as possible. For comparison purposes we also added the offset in $\Delta$(par), in the original units of each parameter, to illustrate the nature of the observed shift in some panels (i.e., a systematic bias in the derivation of the parameters, not related with the estimation of the errors).}
    \label{fig:DS_star}
\end{figure*}

The third strongest correlation ($r=-$0.37) corresponds to the well known degeneracy between age and metallicity: an increase (decrease) in the fraction of old and metal rich stars have a similar effect, i.e. the spectra becomes redder (bluer). Furthermore, intrinsic uncertainties in the stellar evolution models, in the understanding of the stellar evolution and stellar contributions for the light in different wavelength ranges depending on the age and metallicity contributes to this degeneracy too \citep[e.g.][]{Worthey.1994, CF.etal.2005, SanchezBlazquez.etal.2011}. The effect in the residuals of both parameters is of the same order as the one reported by \citep{CF.etal.2014} when using the STARLIGHT code for a similar SSP library. In that exploration, it was found that the use of different libraries may affect this degeneracy, but without completely removing it.

An almost similar correlation is found between the $\Delta$ $\sigma_\star$ and $\Delta$ A$_{\rm V}^\star$ ($r=-$0.35), in the sense than an over-estimation of the velocity dispersion produce an under-estimation of the dust-attenuation. Since a change in the velocity dispersion does not introduce a direct change in the shape of the observed spectrum (the primary property sensitive to the dust attenuation), this degeneracy should be induced by other interdependence among the explored parameters. The more suitable candidate is the effect that the velocity dispersion has on the derivation of the stellar metallicity: a deeper absorption feature can be obtained by either increasing the velocity dispersion or the metallicity. This leads to a correlation between the bias/errors when trying to derive both parameters at the same time \citep[e.g.][]{koleva08,SanchezBlazquez.etal.2011}. We indeed find a very weak trend between $\Delta$ $\sigma_\star$ and $\Delta \stmetL$ ($r=$0.28), an effect that via the stronger age-metallicity and age-dust degeneracy may induce the observed trend.

All the remaining parameters do not present any clear trend among their residuals, what suggests that their derivation is essentially independent: (i) $\sigma_\star$ vs \mlL and A$_{\rm V}^\star$; (ii)  A$_{\rm V}^\star$ vs \mlL; and (iii) \stmetL vs \mlL.

\subsubsection{Accuracy of the estimated uncertainties}
\label{sec:accur:stpop:uncert}



One of the goals of \pyf is to give a good estimation of the uncertainties for each derived parameter. The stellar population decomposition in \pyf comprises an intrinsic MC method over perturbed realizations of the input spectrum. This process enable us to estimate the uncertainties of the estimated coefficients of the stellar decomposition in the considered SSP library (Equation \ref{eq:sigmacoeffs}). These uncertainties are propagated to estimate the errors in the averaged properties of the stellar populations (such as the LW and MW ages and metallicities). On the other hand, the uncertainties estimated in the exploration of non-linear parameters (Sec. \ref{sec:new_ver:fitupd:nlfit}) are derived from the analysis of the $\chi^2$ curves (Figure \ref{fig:anaspec_nlfit_summary}, right columns), assuming as a minimum error the average sampling of the explored range. This latter estimation is valid when the parameters are independent one each-other and the errors have a normal Gaussian distribution, what it is not true in many cases (as discussed in the previous Section). In general, there is no guarantee a priori that the estimated errors are representative of the real ones.

In order to determine how the errors estimated by \pyf for each parameter, $\epsilon$(par), are representative of the real ones we compare them with the value of the offsets between the output and input values, $\Delta({\rm par})$, for our simulated dataset. 
As indicated before, in the case of an accurate estimation of the parameters (see Sec. \ref{sec:accur:stpop:recov}) this residual is a good estimation of the precision and therefore a realistic tracer of the error. We should note that uncertainties and errors should match only in an statistical sense. A direct comparison between both parameters is presented in Appendix \ref{app:rel}. Figure \ref{fig:DS_star} shows the distribution of the ratio between both parameters, where the error is scaled by a factor $f$ (i.e., $\Delta({\rm par})/(f\epsilon({\rm par}))$. In the ideal case in which our estimated errors are a good representation of the real uncertainties these distributions should follow a perfect Gaussian function (shown in Figure \ref{fig:DS_star}), centered in zero (with any offset measuring the accuracy of the estimation of the error), and with a standard deviation of one (with any difference measuring the precision of the estimation of the error). The mean value of  $\Delta({\rm par})$ is shown in each panel, showing that \pyf provides an accurate estimation of the parameters, as discussed in the previous Section. In addition we include in each figure the factor $f$ introduced to match the observed and expected distributions, i.e., the correction that has to be applied to the reported errors to be fully representative of the real uncertainties. It is noticed that \pyf underestimates the errors by a factor $\sim$3--4 for several parameters ($z_\star$,$\sigma_\star$, A$_{\rm V,\star}$ and \stageL)), although in other cases the error is of the same order of the uncertainty (\stmetL and  \mlL).

\begin{figure*}
    \includegraphics[width=\textwidth]{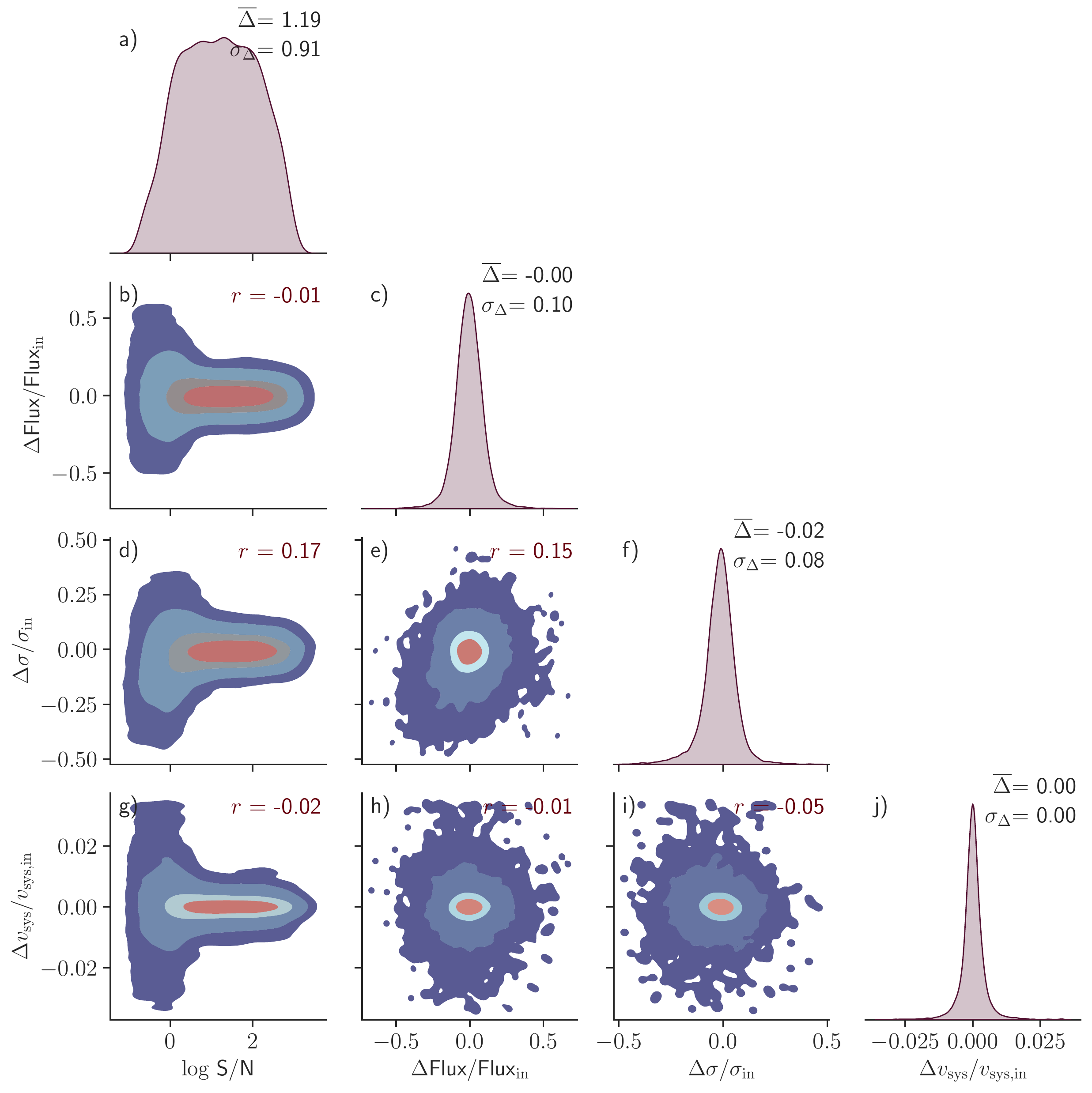}
    \caption{
    Pair-plot with the relative residuals for each explored properties of the \Niip\,+\,\Ha simulated emission lines and their peak flux-intensity S/N using the new double step fitting method (RND + LM) implemented in \pyf. Contours, density histograms, and legends in each panel follow the same scheme presented on Figure \ref{fig:stpop_resid_simul}.
    }
    \label{fig:elines_pairplot}
\end{figure*}

We are not completely aware of the nature of the reported differences between the estimated and the real errors for certain parameters. However, we consider that a combination of reasons may produce them. The most relevant one is that the estimation of the errors relies on the assumption that the derivation of the parameters is fully independent, and that the errors are propagated linearly from the uncertainties. Both assumptions are not fully valid, based on the results described in Sec. \ref{sec:accur:stpop:bias}, where we reported clear correlations in the residuals between the estimated and input values for different parameters. 
The results of this analysis force us to make a revision on the currently adopted procedure to propagate the errors. However, although the nominal errors reported by the code present clear offsets, by applying the reported correction factors the code is able to provide with reliable and representative values of the real uncertainties.


\subsection {Recovering the properties of the emission lines}
\label{sec:accur:eml}

Along this Section we estimate the accuracy and precision on the recovery of the properties of the emission lines based on the fitting procedure implemented in \pyf. For doing so, we simulate 10000 spectra including the \Niip\,+\,\Ha emission line system (i.e. three different emission lines), fixing their integrated flux intensities to 22, 66 and 100 respectively (in arbitrary units). Then, for each simulation we consider a different velocity and velocity dispersion for the entire system (i.e., same kinematic properties for the three lines), randomly chosen within the following predefined ranges: $5500 - 6500$ km/s and $2.5 - 6.5$\,\AA\ ($\sigma_v \approx 100 - 400$\,km/s) respectively. The simulated spectra covers the wavelength range between $6500 - 7000$\,\AA (rest-frame), repeating the sampling of 2\,\AA/pix employed in the stellar analysis simulation. Each emission line is simulated adopting a Gaussian profile, with the corresponding flux intensities listed before, shifted by the systemic velocity and broadened by the velocity dispersion corresponding to each simulation.
Subsequently, we add different levels of noise, assuming a Poissonian distribution, to cover a wide range of S/N ratios between $\sim$0.1 and $\sim$1500. We define here the S/N as the ratio between the peak flux-intensity of the emission line and the 1$\sigma$ level of the simulated white noise. For a conversion to the conventional S/N of the integrated flux, a factor of 3 should be applied to the reported S/N values (S/N$_{\rm int} \approx 3\,$S/N$_{\rm peak}$). Finally, all simulated spectra were fitted using the {\bf RND+LM} procedure described in Section \ref{sec:new_ver:fitupd:elfit}. In addition, we repeat the fit to the simulated spectra using only the {\bf RND} method in order to test the improvements from the introduction of a second step in the emission lines fit procedure employed by \pyf in comparison with the previous version of FIT3D.

For this simulation we use the same input configuration for all rounds (see Appendix \ref{appendix:showcase_config} for more details on the adopted configuration). The values for the {\it guess} ({\it interval}) adopted for the systemic velocity and the velocity dispersion were 6000 ($5500--6500$) km/s and 2.7 ($2.5--6.5$)\,\AA\ respectively.
However, the {\it intervals} are rewritten by \pyf taking into account the shift introduced in the simulated spectrum due to the systemic velocity. All emission lines in the modeled system are set to have the same kinematic parameters (systemic velocity and velocity dispersion). In addition, we set a physical link between the integrated flux of each \Niip line as $_{\NNii} = (1/3)$F$_{\Nii}$. On the other hand, no link was imposed between $_{\Nii}$ and $_{\Ha}$, i.e. both are free parameters on the simulation. By not fixing these values, the simulation can cover different sources of ionization (e.g. AGNs, \Hii\ regions, post-AGBs; \citealt{Kewley.etal.2001a, Kauffmann.etal.2003c, Stasinska.etal.2006a, CF.etal.2010}).

\begin{table}
    \label{tab:eml_stats_SN}
    \caption{Emission lines simulation: statistics of the relative residuals segregated by S/N}
    \begin{tabular}{cccc}
      S/N & ${\Delta}$Flux/Flux$_{\rm in}$ & ${\Delta}\sigma/\sigma_{\rm in}$ & 10${\times \Delta}v_{\rm sys}/v_{\rm sys, in}^1$\\
    \hline
     $<$ 3 & 0.00 $\pm$ 0.16 & -0.04 $\pm$ 0.12 & -0.01 $\pm$ 0.16 \\
     $3-10$ & -0.01 $\pm$ 0.09 & -0.01 $\pm$ 0.06 & 0.00 $\pm$ 0.07 \\
     $10-100$ & -0.01 $\pm$ 0.08 & -0.01 $\pm$ 0.05 & 0.00 $\pm$ 0.04 \\
      $>$ 100 & -0.01 $\pm$ 0.07 & -0.01 $\pm$ 0.04 & 0.00 $\pm$ 0.03 \\
      \hline
    \end{tabular}
    Results of the mean $\pm$ the standard deviation of the distributions. \\
    (1) ${\Delta}v_{\rm sys}$ statistics is multiplied$^2$ by 10 to show it in the same scale of the other two parameters. \\
    (2) We did not include the multiplication factor to the data points during the statistical calculation avoiding possible systematic bias.
\end{table}
\begin{figure*}
    \includegraphics[trim=0 4 0 5, clip, width=\textwidth]{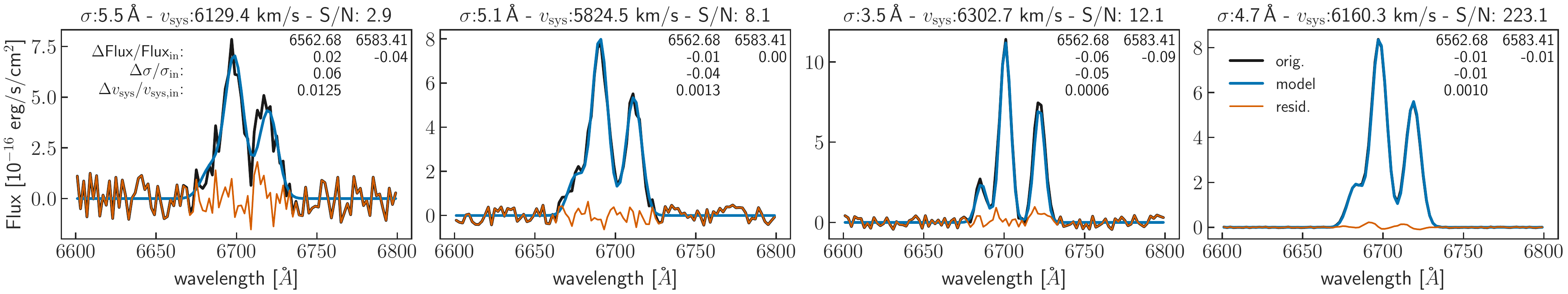}
    \caption{Four examples of the simulated spectra analysis, one for each S/N interval with statistics presented in Table \ref{tab:eml_stats_SN}. At each panel we show the original spectrum (black), the modeled spectrum (blue) and the residual spectrum (red). We also show the input parameters, peak S/N and the relative residuals of the simulation analysis.}
    \label{fig:tab2plots}
\end{figure*}
\begin{figure}[htb!]
    \includegraphics[trim=0 25 0 0, clip, width=0.45\textwidth]{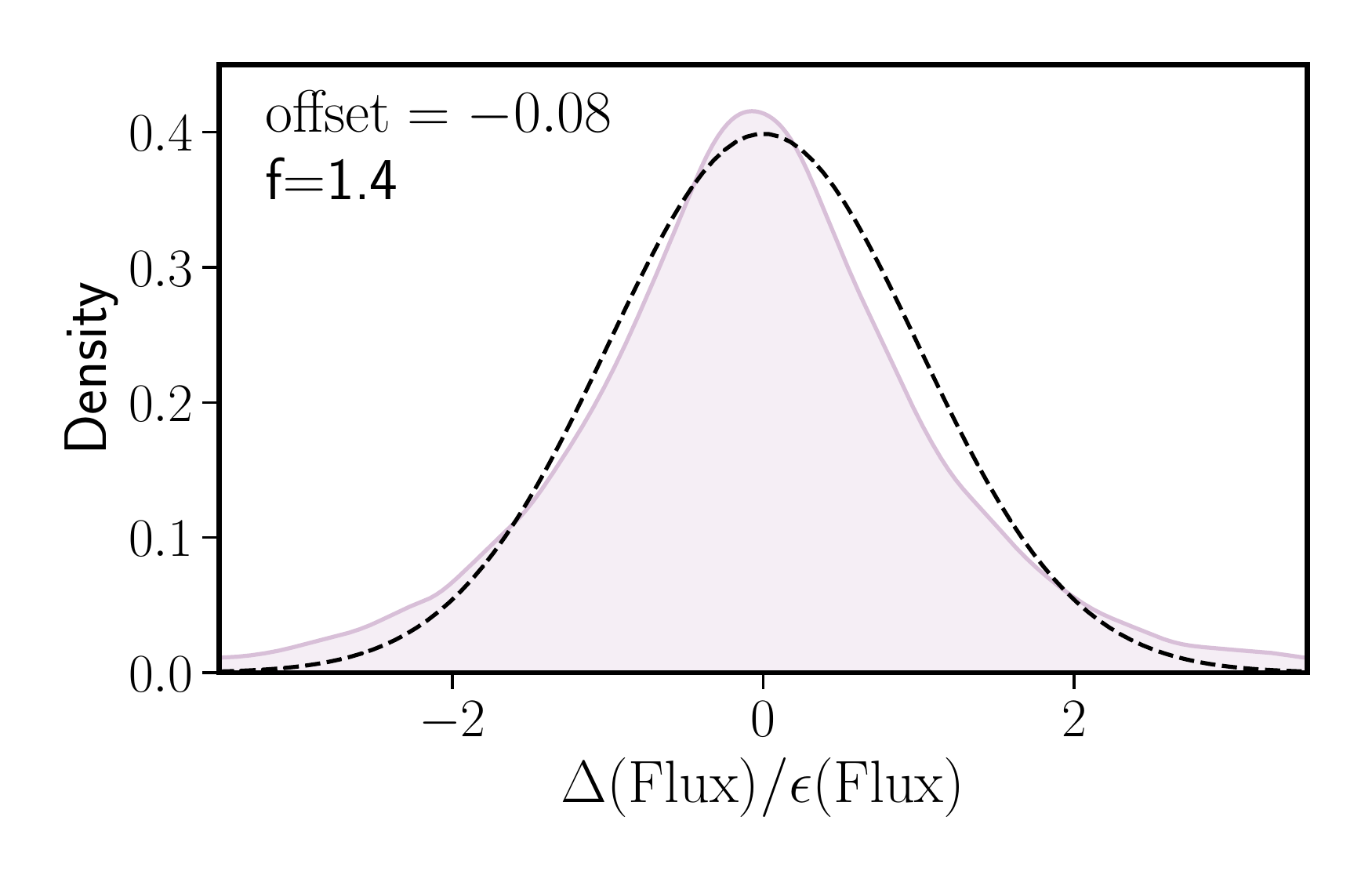}
    \includegraphics[trim=0 25 0 0, clip,width=0.45\textwidth]{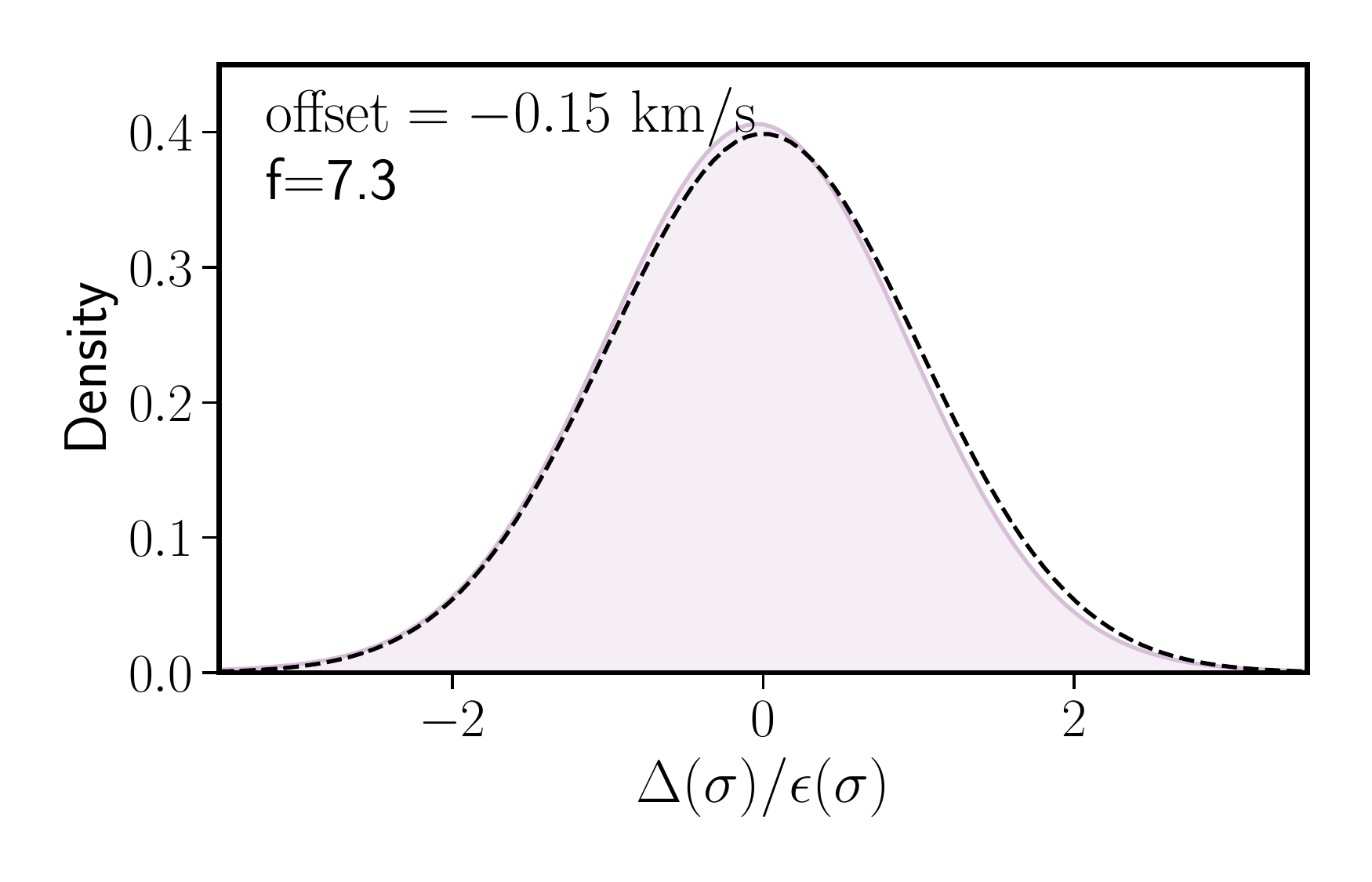}
    \includegraphics[trim=0 25 0 0, clip,width=0.45\textwidth]{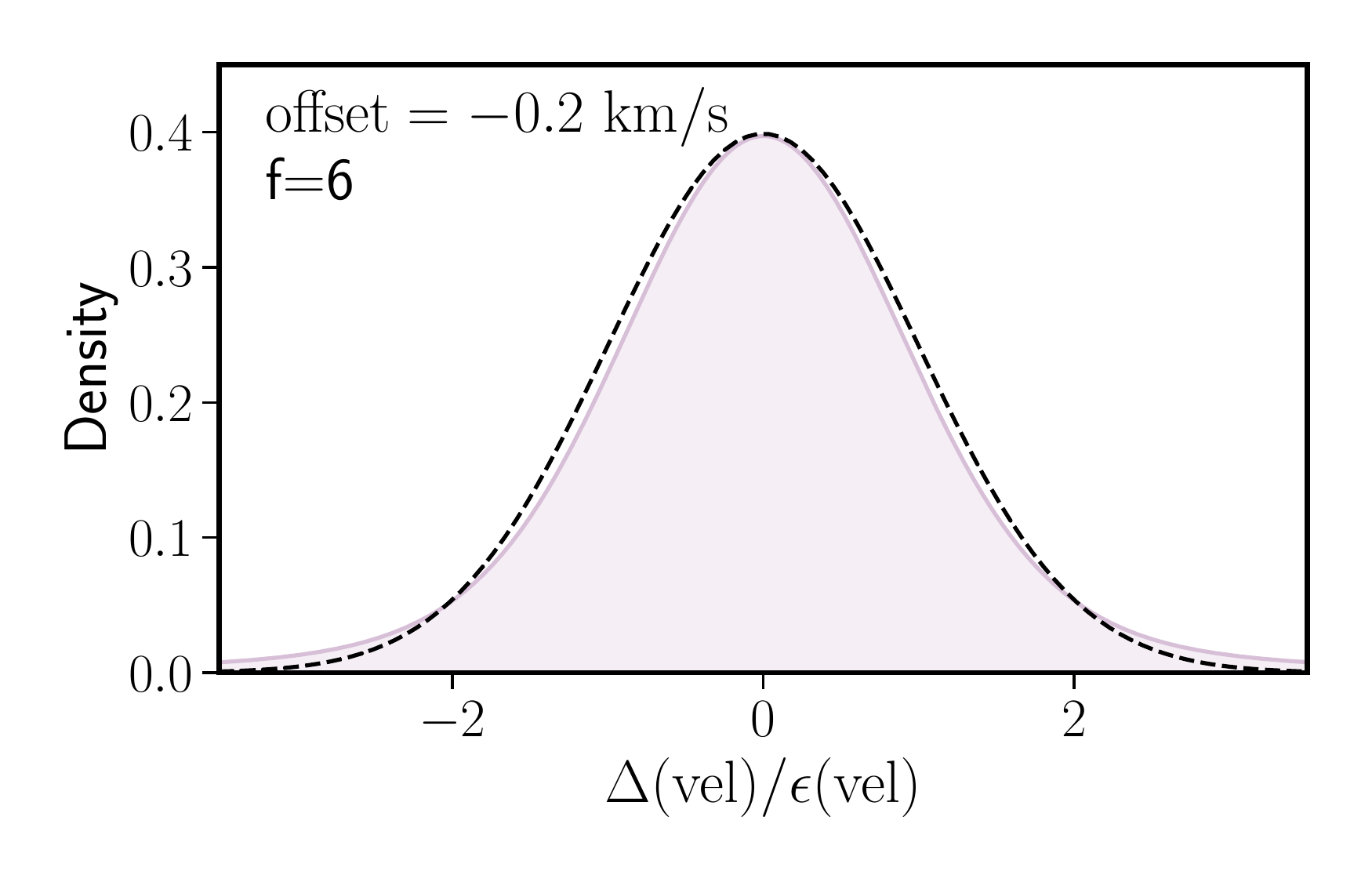}
    \caption{Similar figure as Figure \ref{fig:DS_star} for the properties derived from the analysis of the simulated emission lines. From top-to-bottom we show the distribution of  $\Delta$(par)/$\epsilon$(par) for the flux intensity, the velocity dispersion, and the systemic velocity.}
    \label{fig:DS_eml}
\end{figure}

\subsubsection{Bias on the estimation of the parameters for the emission lines}
\label{sec:accur:eml:recbias}

Figure \ref{fig:elines_pairplot} gives an overview of the results of this exploration showing a pair-plot with the relative residuals ($\Delta$par/par$_{\rm in}$) of the different parameters of the simulated emission lines system together with the input S/N of each simulation run. Table \ref{tab:eml_stats_SN} summarizes the results showing the mean and the standard deviation of the distributions of the relative residual offsets for four different S/N ranges. Figure \ref{fig:tab2plots} shows four examples of the analysis (one for each S/N range of Table \ref{tab:eml_stats_SN}).

Exploring the first column of Figure \ref{fig:elines_pairplot} we see no correlation between the input S/N and the relative residuals (i.e., accuracy of the results), with a clear improvement of the precision of the fitting with the S/N, confirmed by the residuals at each S/N range at Figure \ref{fig:tab2plots}. Even when the S/N is lower than 3, where the systematic effects dominate (like errors in the stellar population subtraction and noise patterns not considered in this simulation), the precision on the flux intensity is not worse than $\approx$16\% ($\approx$12\% for the velocity dispersion). Furthermore, \pyf provides with an almost constant precision and accuracy above S/N $>$ 10 for all parameters. However, the precision for those parameters does not get better than $\approx$4\% even in the high S/N cases considered in this simulation. In the case of the systemic velocity, the program does recover this parameter with a precision and an accuracy $\approx$10\% worse than the other two.
Although the accuracy on the flux and the velocity determination is less affected when compared to the sigma one, the uncertainties begin to suffer a considerable increase at the very low S/N regime (S/N $<$ 1), with offsets on the recovered values around 20\% on the flux and the velocity, and above 15\% for the sigma determination.

The estimation of the integrated flux depends directly on the velocity dispersion, so, in principal, any bias in the determination of one could affect both the accuracy and the precision of the determination of the other. However, we see no correlation between the relative residual offsets, i.e. the determination of the parameters do not present any interdependence bias. The bulk of the distribution rests within $\pm 0.1 \Delta$par/par$_{\rm in}$ in both axis for all three parameters (panels $e, h$ and $i$ of Figure \ref{fig:elines_pairplot}). 

\begin{figure}
    \includegraphics[trim=10 15 0 10, clip, width=\columnwidth]{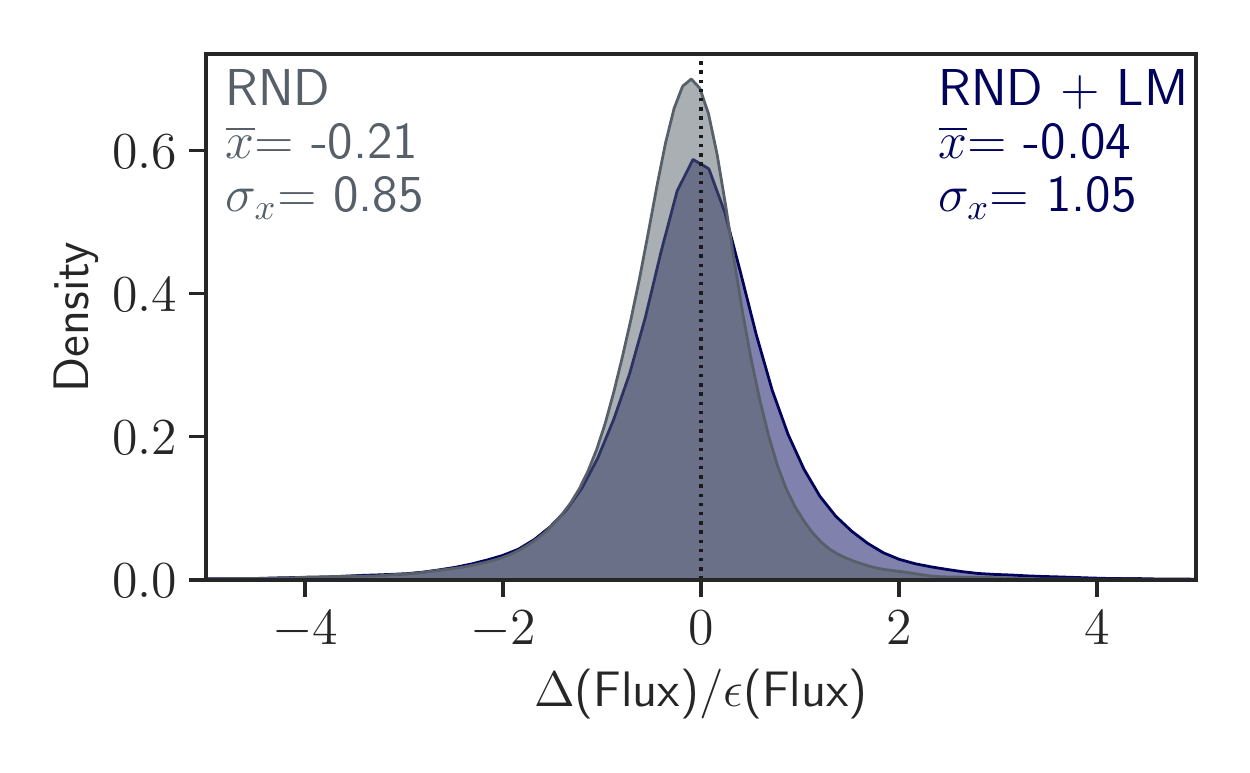}
    \caption{
    Density distributions of the uncertainty in the derivation of the flux, $\Delta$(Flux), divided by the error estimated by \pyf, $\epsilon$(Flux), for two rounds of fittings of 10000 simulated spectra comprising the \Niip\,+\,\Ha emission lines system. For the first round (light-red) it was adopted the {\bf RND method}, while for the second one (dark-red), it was adopted the combined {\bf RND+LM methods}.
    Upper-left and right corners present the mean and standard deviation of each distribution.
    }
    \label{fig:elines_residuals_histograms}
\end{figure}

Repeating the experiment using only the fit with the {\bf RND method} increases the values of the standard deviations of the relative residuals with S/N $> 3$, that in this case shows similar distributions as the simulations with S/N $< 3$ (in agreement with the results presented in S16 using the former version of the code). This tell us that the inclusion of a second round of fit with the {\bf LM method} increases the precision of the estimated parameters. Other notable result is that the correlation coefficient between $\Delta\sigma/\sigma_{\rm in}$ and $\Delta$Flux/Flux$_{\rm in}$ increases from 0.15 to 0.26. Although this is a mild increase, it is a significant change. This time we could conclude that the inclusion of the {\bf LM method} as a second step of the emission line fit process also helps to decrease the Flux$ - \sigma$ bias, already found for the previous version of the code (S16).

\begin{figure}[htb!]
    \centering
    \includegraphics[width=0.49\columnwidth]{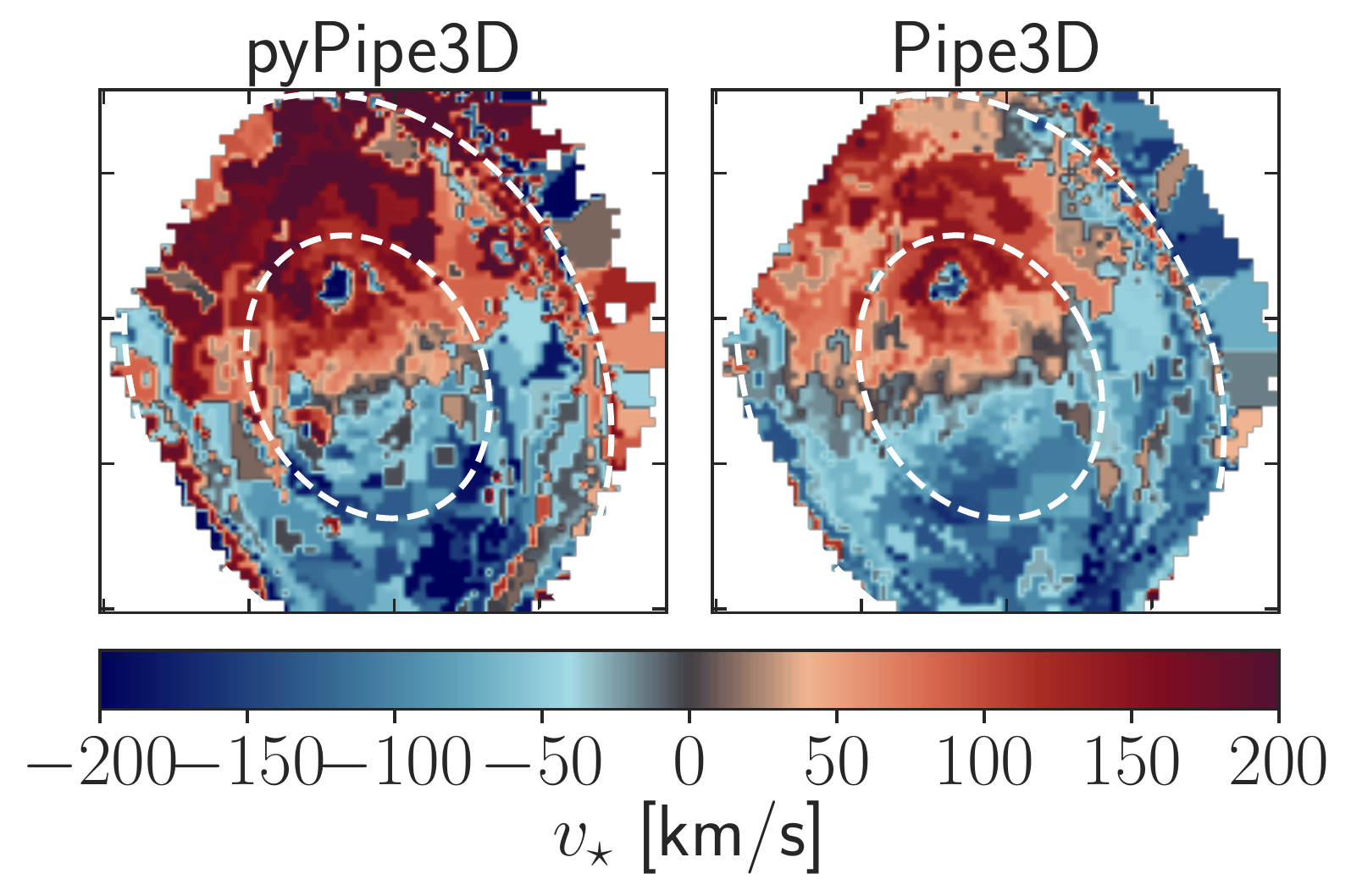}
    \includegraphics[width=0.49\columnwidth]{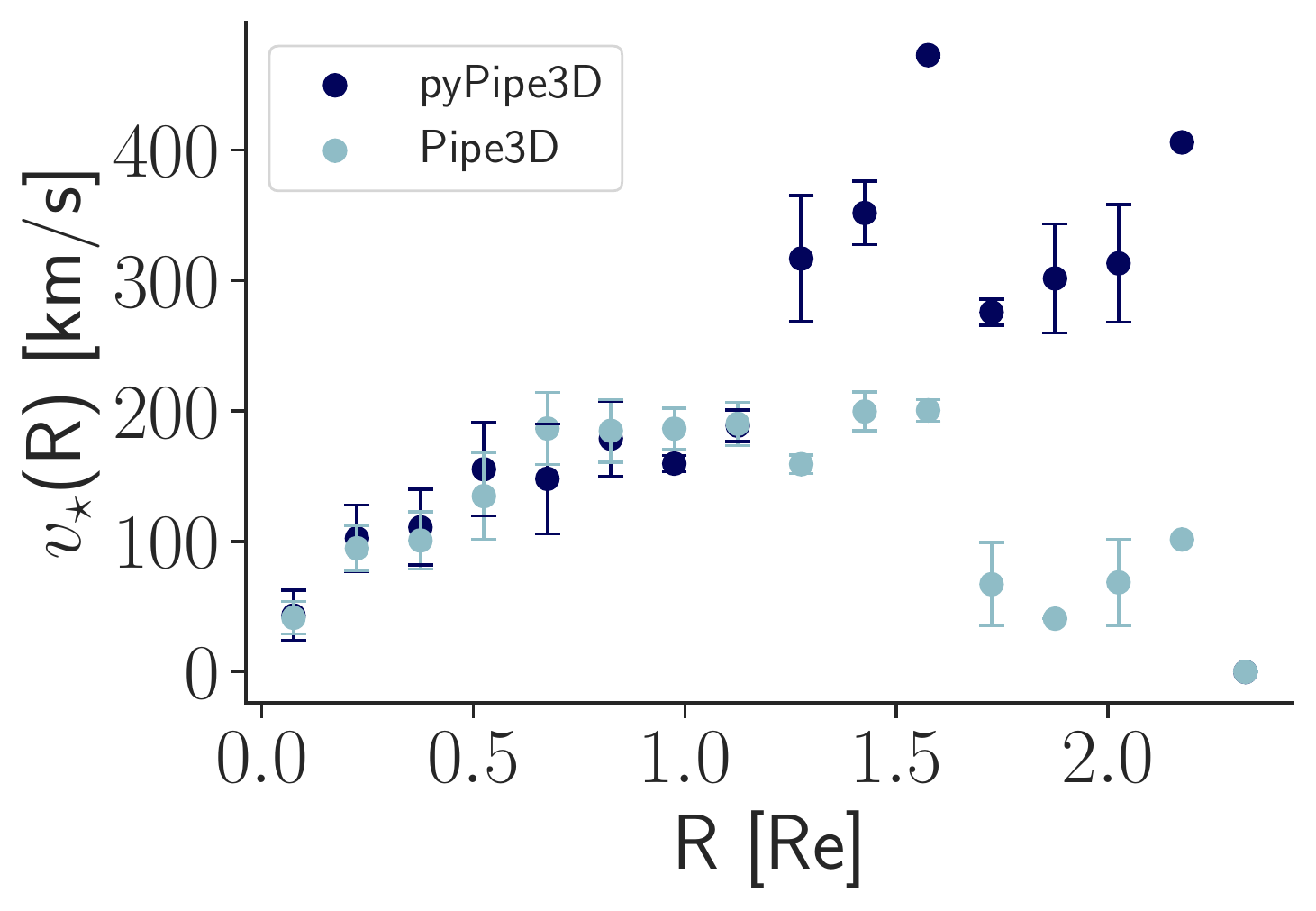} \\
    \includegraphics[width=0.49\columnwidth]{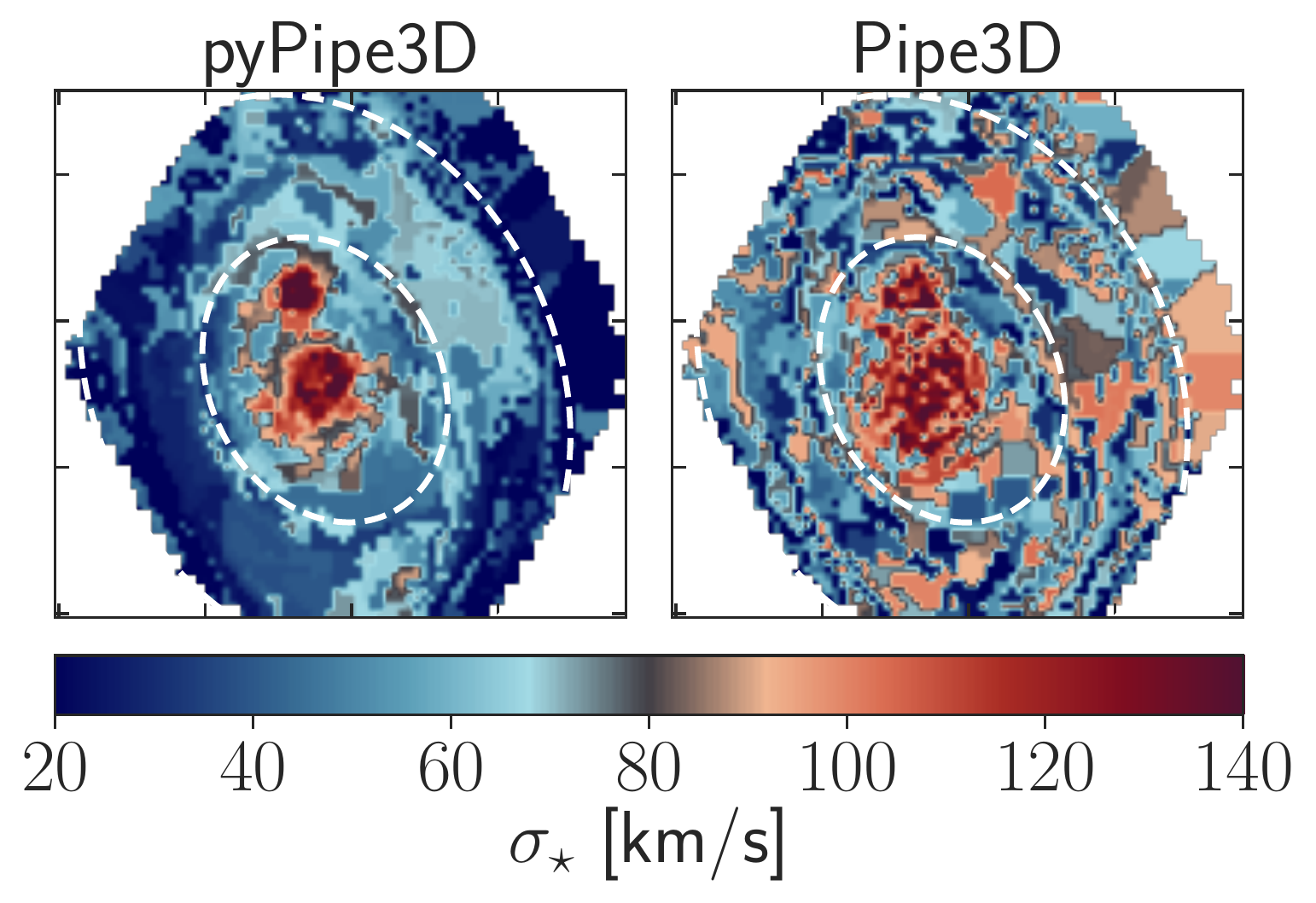}
    \includegraphics[width=0.49\columnwidth]{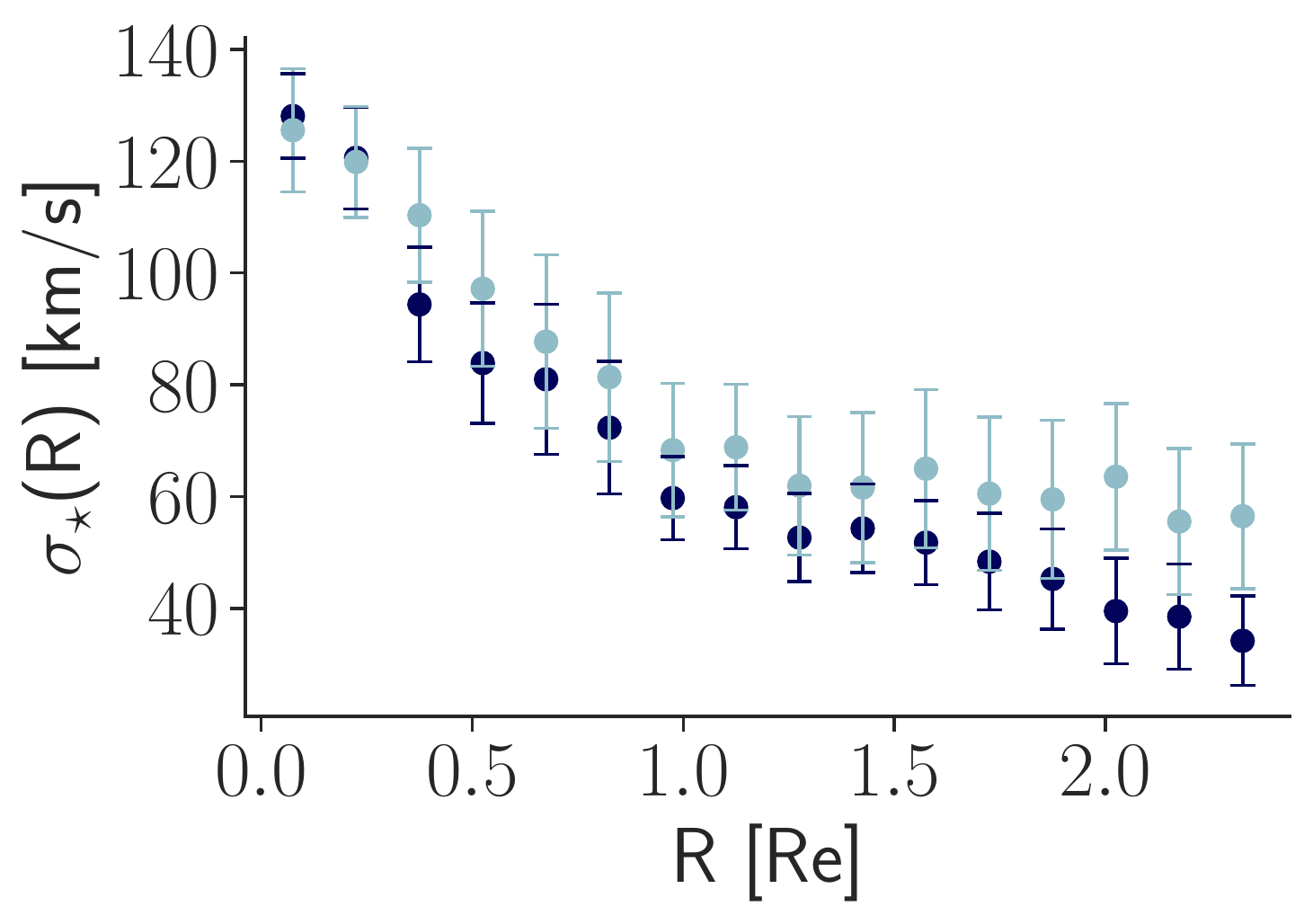} \\
    \includegraphics[width=0.49\columnwidth]{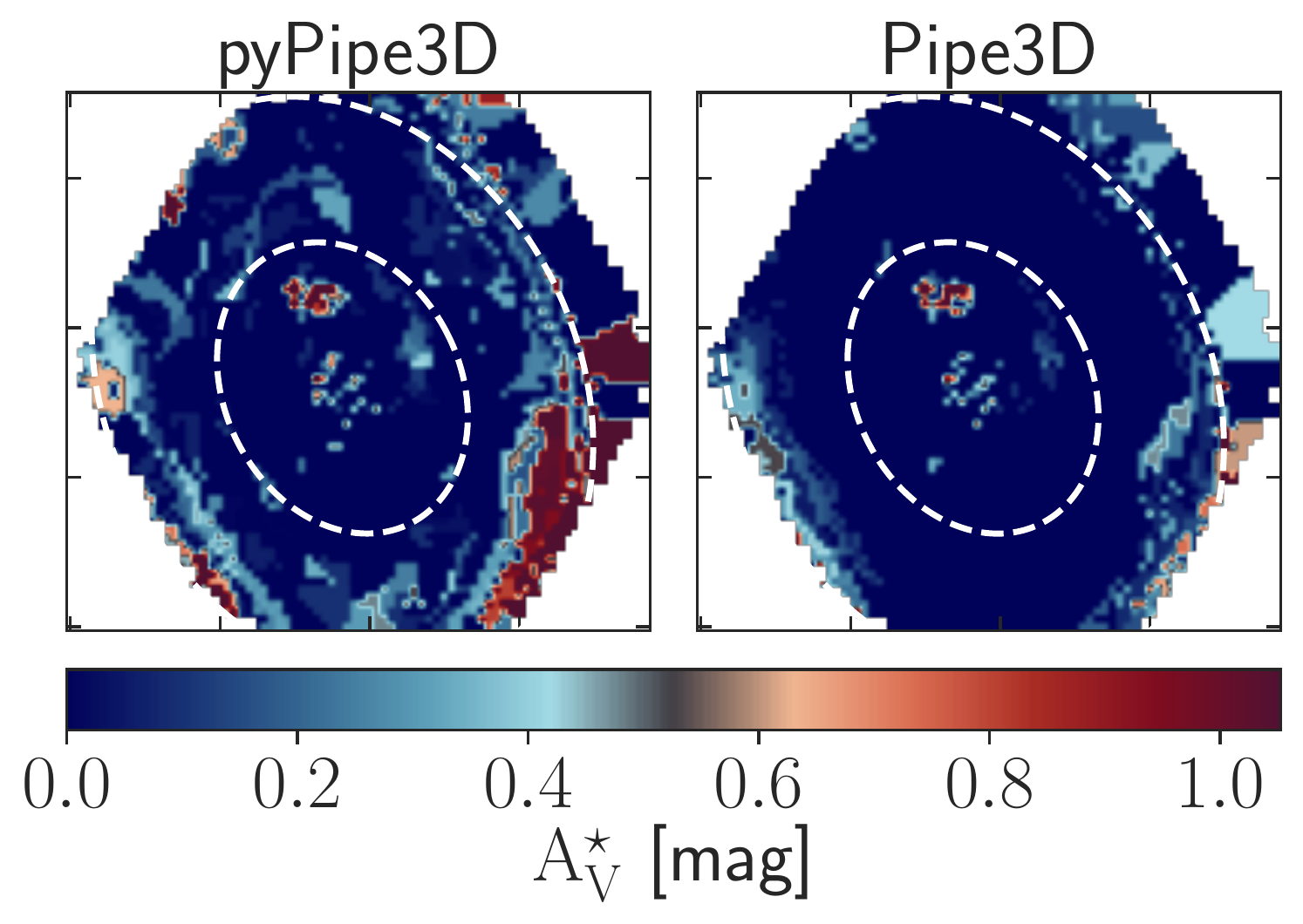}
    \includegraphics[width=0.49\columnwidth]{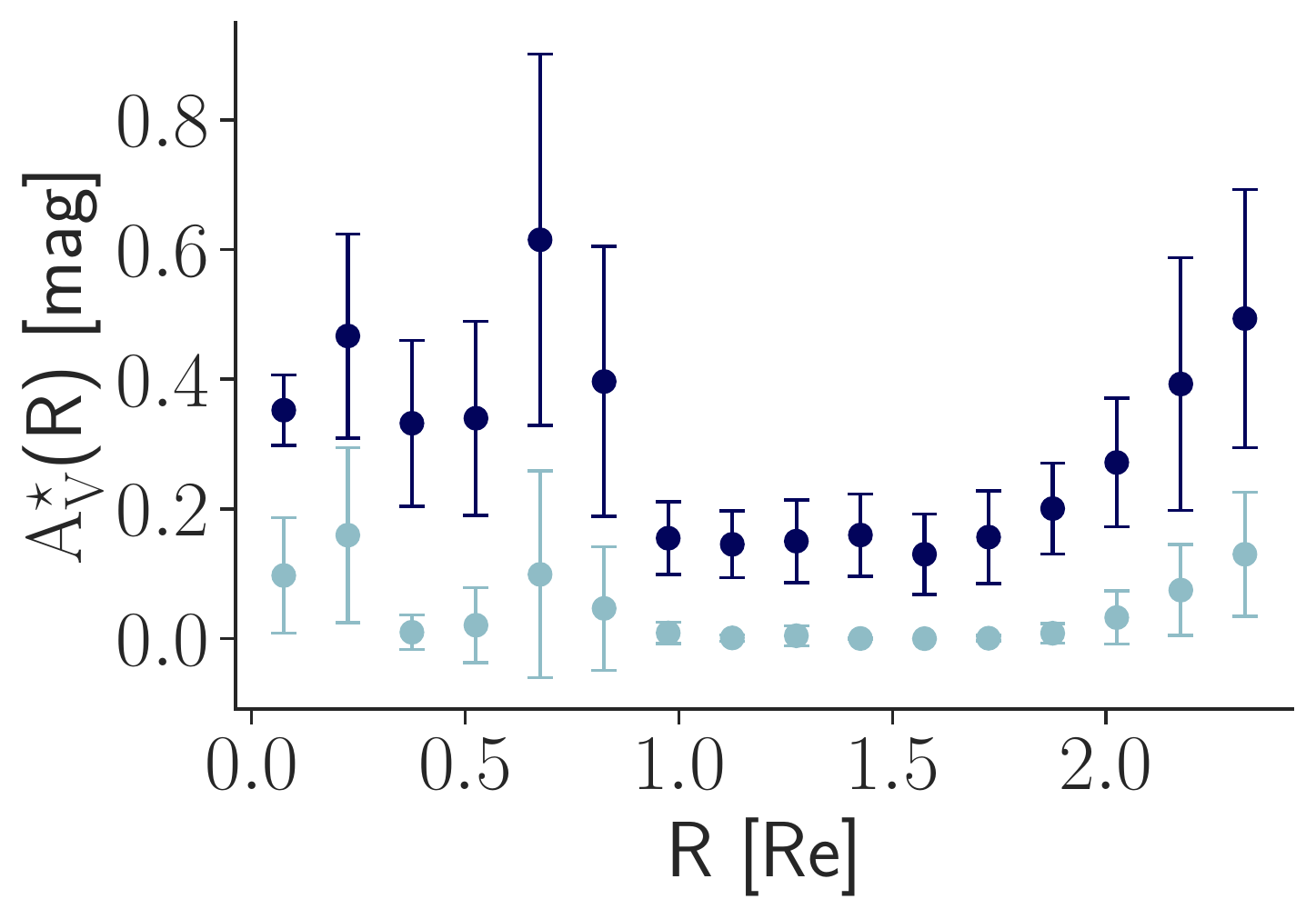} \\
    \caption{Comparison of the spatial distribution (left panels) and radial profiles (right panels) for the non-linear parameters derived by \pyp and Pipe3D for the  NGC\,2916 galaxy observed by CALIFA:  $v_{sys}$ (top panels, in absolute values), $\sigma_\star$ (middle panels), ${\rm A}_{\rm V}^\star$ (bottom panels).
    The white ellipses in the maps represents the location of one and two effective radius (Re). The error-bars in the left panels represent the mean and 1$\sigma$ interval of the azimuthally averaged parameters.
    }
    \label{fig:compare_R_syn_nlprop_NGC2916}
\end{figure}

\subsubsection{Accuracy of the estimated uncertainties of the emission line properties}
\label{sec:accur:eml:uncert}

We repeat here the test of the hypothesis that the errors estimated by \pyf could represent the real uncertainty in the recovery of each parameter (e.g., Sec. \ref{sec:accur:stpop:uncert}, Figure \ref{fig:DS_star}), this time for the emission line analysis.
Like in the case of the stellar populations the comparison between the uncertainties and the errors should be done in an statistical way. In Appendix \ref{app:rel} we discuss the individual comparison between both parameters. Figure \ref{fig:DS_eml} shows the distribution of $\Delta(par)/\epsilon(par)$ for the explored properties of the emission lines (flux, velocity dispersion and systemic velocity) based on the simulations described in the previous section (i.e.,  the equivalent to Figure \ref{fig:DS_star}, but for the emission line properties).
We first notice that there is no significant bias (offset) between the recovered and simulated values for each of the three explored parameters, in agreement with the results shown in Figure \ref{fig:elines_pairplot}.
Due to that the observed distributions are well center in zero. Like in the case of the errors estimated for the stellar population properties we introduce a correction factor ($f$), to match the observed distribution with the expected one (a Gaussian function centered in zero with a $\sigma=$1). As a result we appreciate that the estimated errors are of the same order of the real uncertainties for the emission line fluxes ($f\sim$1), however, for the kinematics parameters there is a clear underestimation of the errors ($f\sim$6-7). The observed and expected distributions are very similar once applied this correction factor. Therefore the corrected errors can be used as good estimations of the real uncertainties.



Finally, we repeat this experiment using the two methods implemented in \pyf to explore the emission lines, the {\bf RND method}, already present in FIT3D, and the new {\bf RND+LN method}.  Figure \ref{fig:elines_residuals_histograms} shows the distribution of $\Delta(par)/\epsilon(par)$ for two different fitting rounds (one for each method) over a set of 10,000 simulated spectra including the emission line system described before (\Niip\,+\,\Ha). 
There is a remarkable good agreement between the distributions of the real and estimated errors for the second method, with the distribution being well centered in -0.04 dex with a standard deviation of 1.05. It is clear that the concomitant work of both methods produces a better result than using the {\bf RND method} only. For this previous method the distribution presents a bias of -0.21 dex and narrower distribution ($\sigma=$0.85), what suggest that there is a bias in the recovered values and an overestimation of the errors. In summary, combining the {\bf RND+LM methods} improves the precision and accuracy of the code.

\begin{figure}
    \centering
    \includegraphics[width=0.49\columnwidth]{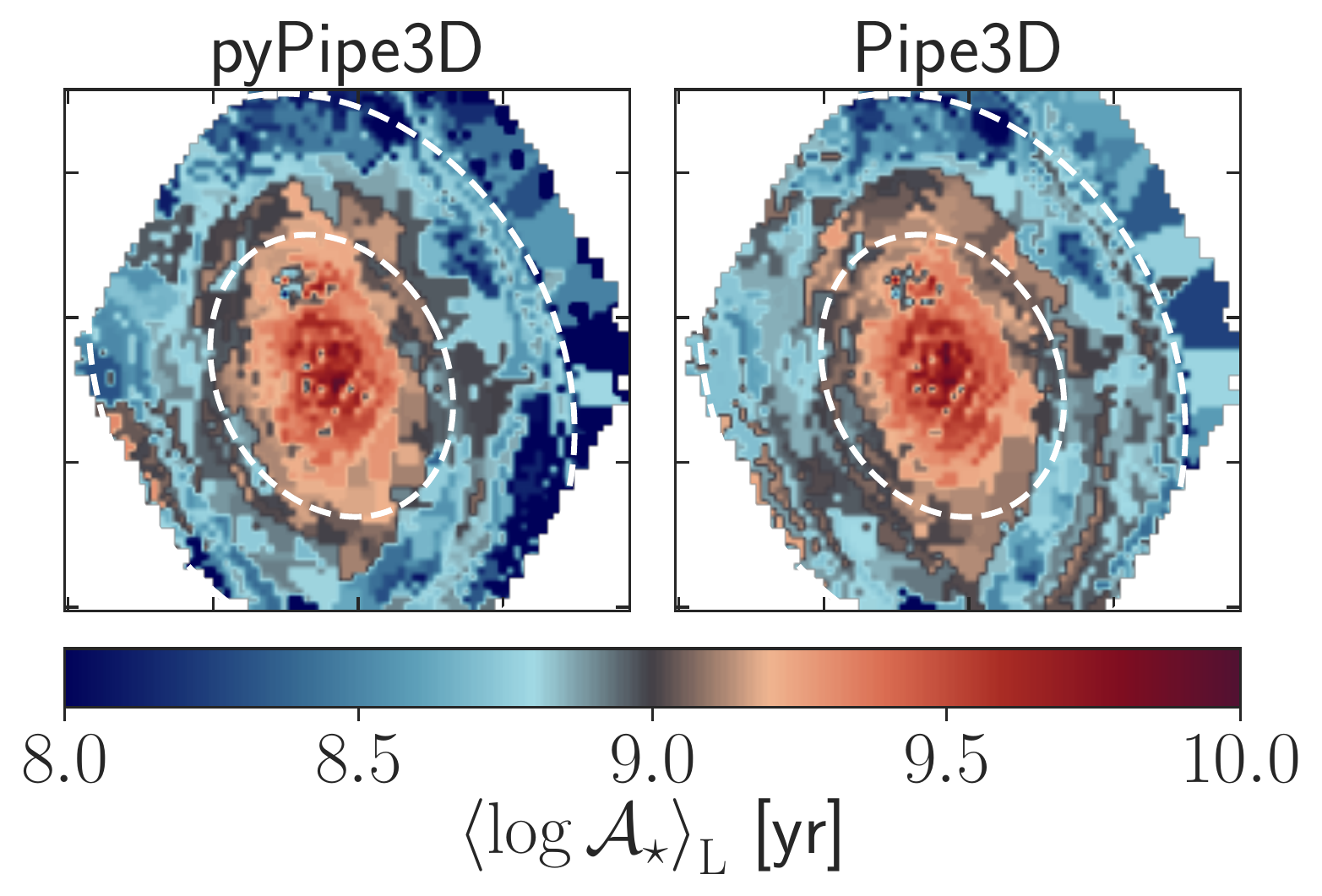}
    \includegraphics[width=0.49\columnwidth]{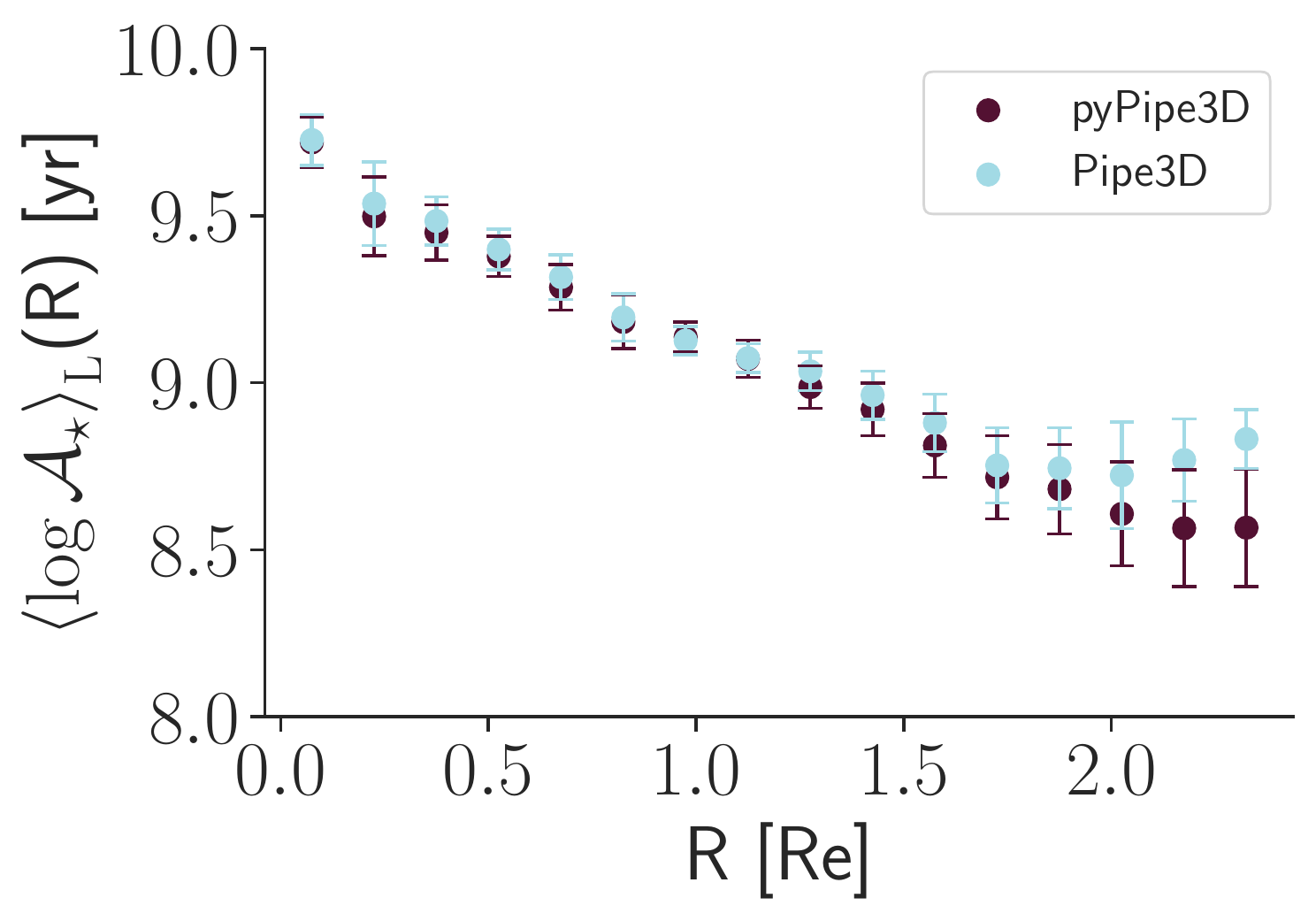} \\
    \includegraphics[width=0.49\columnwidth]{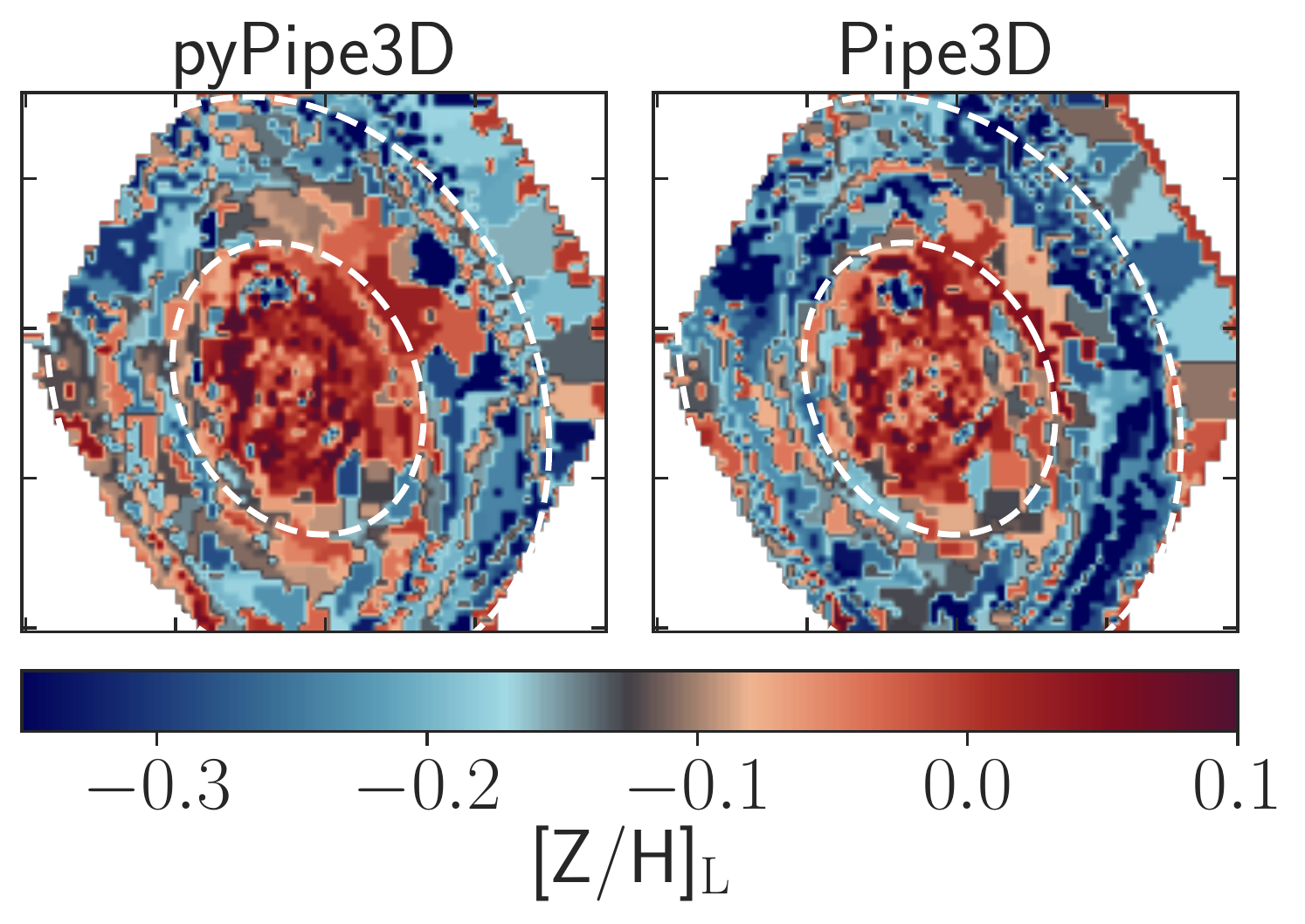}
    \includegraphics[width=0.49\columnwidth]{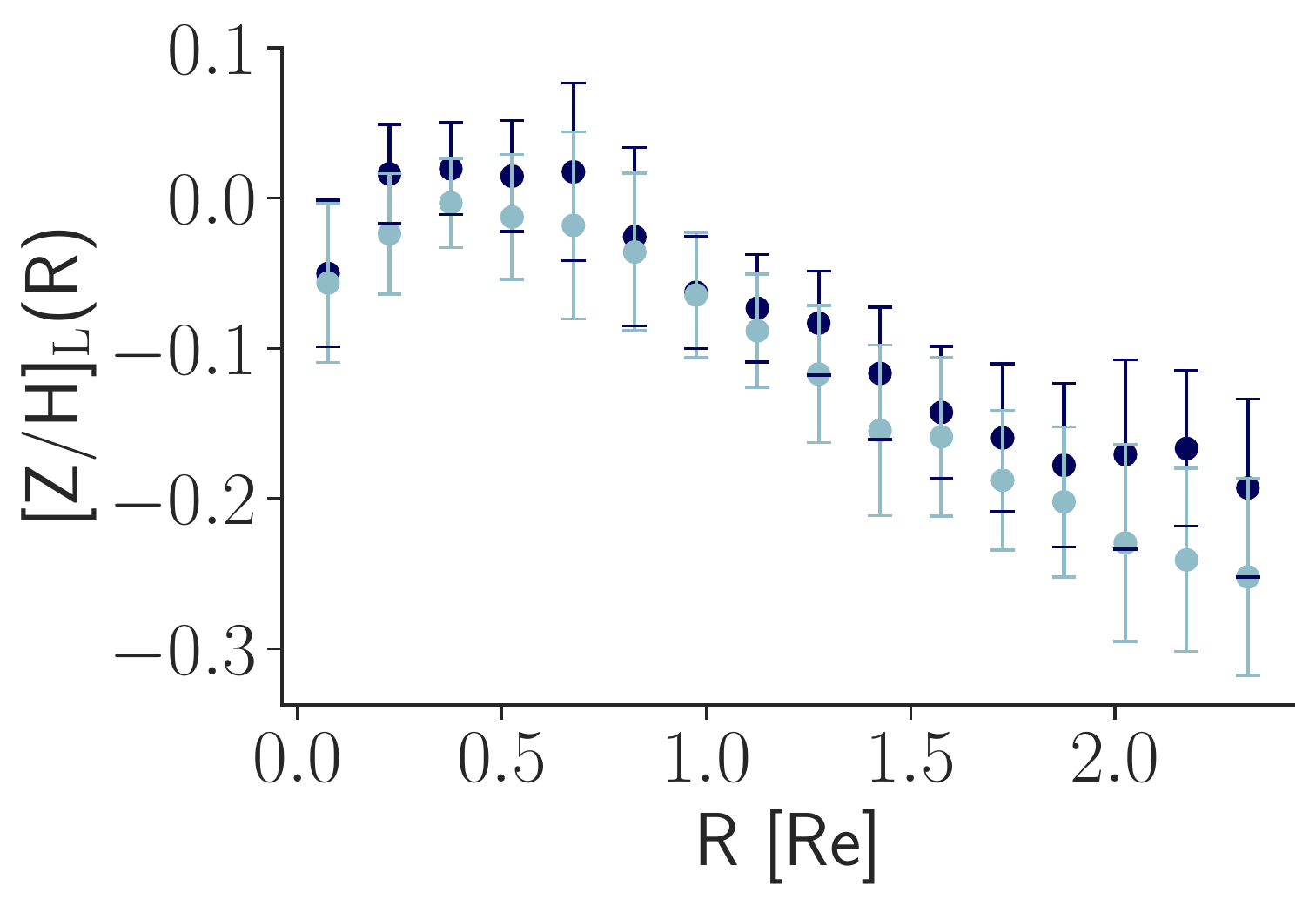}
    \caption{Similar comparison as the one shown in Figure \ref{fig:compare_R_syn_nlprop_NGC2916} for the \stageL and \stmetL, using the same nomenclature.
    }
    \label{fig:compare_R_syn_stprop_NGC2916}
\end{figure}

\section{\pyp applied on real data}
\label{sec:accur:realdata}

The experiments in the previous sections, using simulated data, with well controlled inputs allowed us to demonstrate that \pyf yield reliable results, characterizing the uncertainties associated to the derived parameters. However, it does not exemplify the full applicability of \pyp,  what is only possible when using it over real IFS data.
As a showcase of the use of the pipeline (Sec. \ref{sec:new_ver:pipeupd}) we reanalyze the V500-setup data for the galaxy NGC\,2916, provided by the CALIFA survey. This is a
an Sbc galaxy with $\log ($M$_\star/$M$_\odot) \sim 10.8$ and redshift $\sim 0.0122$ ($v_\star^{\rm cen} = 3669$ km/s), which comprises a wide range of stellar populations and emission line properties within its optical extension. For this reason was selected as the showcase example for the presentation of the previous version of the code (S16b). This way we will be able to compare the results provided by the two versions of the code. Considering that the previous version has been extensively used with several datasets, confronted with hydrodynamical simulations and compared with other fitting tool (Sec. \ref{sec:intro}), this comparison will allow us to estimate the quality of the analysis performed by the new version too.

Furthermore it will allow us to compare the speed of the two processes. Using the same CPU (Intel Xeon Gold 6130 (64) @ 2.101GHz), for a single-core, for a computer with 128Gb of RAM, the full analysis of the NGC\,2916 datacube takes a total of 12602s ($\sim$3.5h) for \pyp. The same process took 29859s ($\sim$8.3h) to Pipe3D. Thus, the new procedure, even including a second round in the analysis of the emission lines (i.e., the {\bf RND+LM} method), is $\sim$2.4 times faster. Similar numbers are derived when fitting a single spectrum (i.e., when running the \pyf routine): using the same hardware configuration the new code requires 15s while the old one consumes 34s, thus, $\sim$2.3 times faster.

\begin{figure}
    \centering
    \includegraphics[width=\columnwidth]{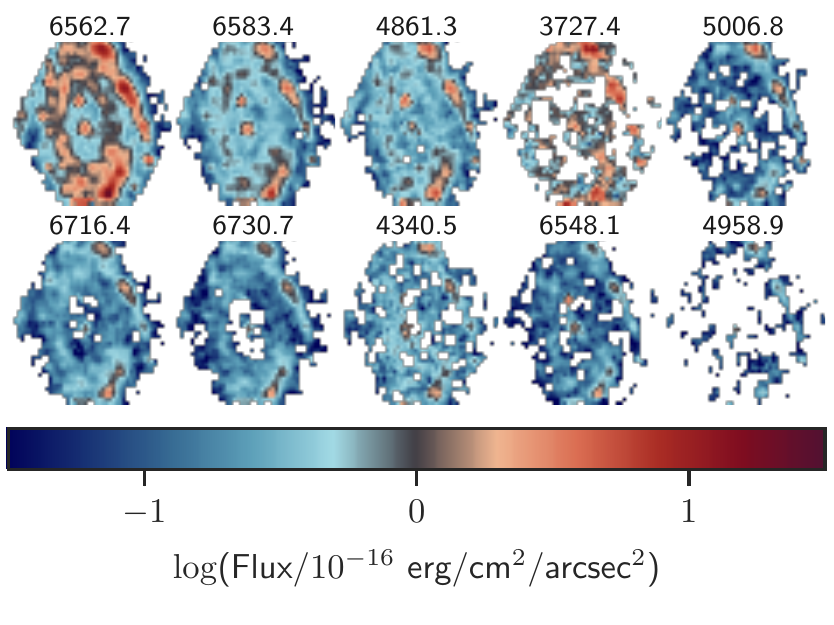}
    \caption{
    Flux intensity maps for the ten strongest emission lines observed in the galaxy NGC\,2916, sorted by the mean flux.
    The spiral shape of the galaxy is very clearly traced by these emission line intensity maps.
    As the mean flux decreases, the noise is more evident as a patchy pattern in the images. }
    \label{fig:flux_elines_NGC2916}
\end{figure}

\subsection{Stellar population analysis}
\label{sec:accur:realdata:stpop}

Figure \ref{fig:compare_R_syn_nlprop_NGC2916} shows the spatial distributions and radial profiles (in units of the effective radius, Re) of the properties estimated during the non-linear analysis (Sec. \ref{sec:new_ver:fitupd}), using both \pyp and Pipe3D, for the considered dataset. The radial profiles were calculated up to 2.5 Re, adopting an azimuthal average of the properties inside elliptical concentric bins of 0.15 Re width.
For practical reasons, we choose to show the velocity map instead of the redshift (i.e., $v_\star = c z_\star$), although both parameters are interchangeable.
No spatial masking was applied to the data since our goal is to compare the output produced by \pyp with that from Pipe3D regardless the quality of potential issues with the original data. For this reason the location of a foreground star at about 8$\arcsec$ North-East of the center of the galaxy is clearly shown as a location of anomalous values in all maps.
The left map is produced by first while the right one by latter.

In general, the new version of the code presents maps and profiles qualitatively similar to those of the previous one, and also with those previously published by other studies using different techniques \citep[e.g.][]{CF.etal.2013, pipe3d_ii}. Regarding the velocity profile, the old distribution (and radial profile) presents a non realistic drop in the outer regions (above 1.5Re). Despite the fact that this region already corresponds to an area of low S/N, it is evident that this effect is not seen in the velocity maps provided by new version of the code. Thus, we consider that it provides with a better determination of the velocity in the outskirts of the galaxy. Furthermore, the spatial distribution of $\sigma_\star$ is clearly smoother, more realistic, with a slightly steeper profile towards the center. However, the differences clearly seen along the map are not propagated to the radial profile, which is very similar for both versions of the code. The only difference is a that the new version of the code provides somehow lower values of the velocity dispersion in the outer regions (from $\sim$60 km/s to $\sim$30km/s in the last radial bins). The main quantitative difference is found in the clear systematic offset in ${\rm A}_{\rm V}^\star$. For this parameter the values derived by \pyp are in general $\sim$1.2 larger than the ones derived by the old version of the code. This difference is most probably due to the different implementation of the currently adopted extinction curve. Pipe3D implemented our own coded algorithm that reproduces the polynomials published by \citet{Cardelli.etal.1989}, while the new version uses the {\sc extinction} module implemented in {\sc python}. In addition we found an issue with the normalization wavelength at which the dust attenuation was derived for FIT3D in the re-codding of \pyf. We consider that the two combined effects explain this difference.



The maps and profiles of the light-weighted average properties of the stellar populations, age and metallicity, are shown in Figure \ref{fig:compare_R_syn_stprop_NGC2916}. NGC\,2916 is an Sbc galaxy and it follows the same age and metallicity profiles reported for galaxies with the same morphology and stellar mass \citep{CF.etal.2013, GonzalezDelgado.etal.2015, MejiaNarvaez.etal.2020, SFS.2020}. Both properties present negative gradients along the radius with the metallicity suffering a flattening or drop towards the inner regions (R $<$ 0.5\,Re). For these parameters the two versions of the code provide with very similar spatial and radial distributions of the explored parameters, despite of the mild/small differences reported in the non-linear parameters discussed before.


\subsection{Emission lines analysis}
\label{sec:accur:realdata:gas}

For the emission lines properties \pyp implements two different procedures, as discussed in Section \ref{sec:new_ver}. Figure \ref{fig:flux_elines_NGC2916} show the maps of the flux intensity for the ten strongest emission lines, ordered by the mean flux intensity of the map, calculated by the moment analysis (see \ref{sec:new_ver:fitupd:momana}). Since the analysis implemented in \pyp is not restricted to a wavelength range by construction, this list of emission lines is part of a configuration file and can be altered by the final user. Actually, the pipeline code has been tested including the derivation of the parameters of more than 200 emission lines with no significant increase over the total time of the analysis. As expected, the maps of the weakest emission lines present a noisy pattern due to low S/N, on top of the patchy structure intrinsic to emission line maps (due to the nature of the ionization, that it is not homogeneous across the optical extension of galaxies).


\begin{figure}[htb!]
    \centering
    \includegraphics[width=0.49\columnwidth]{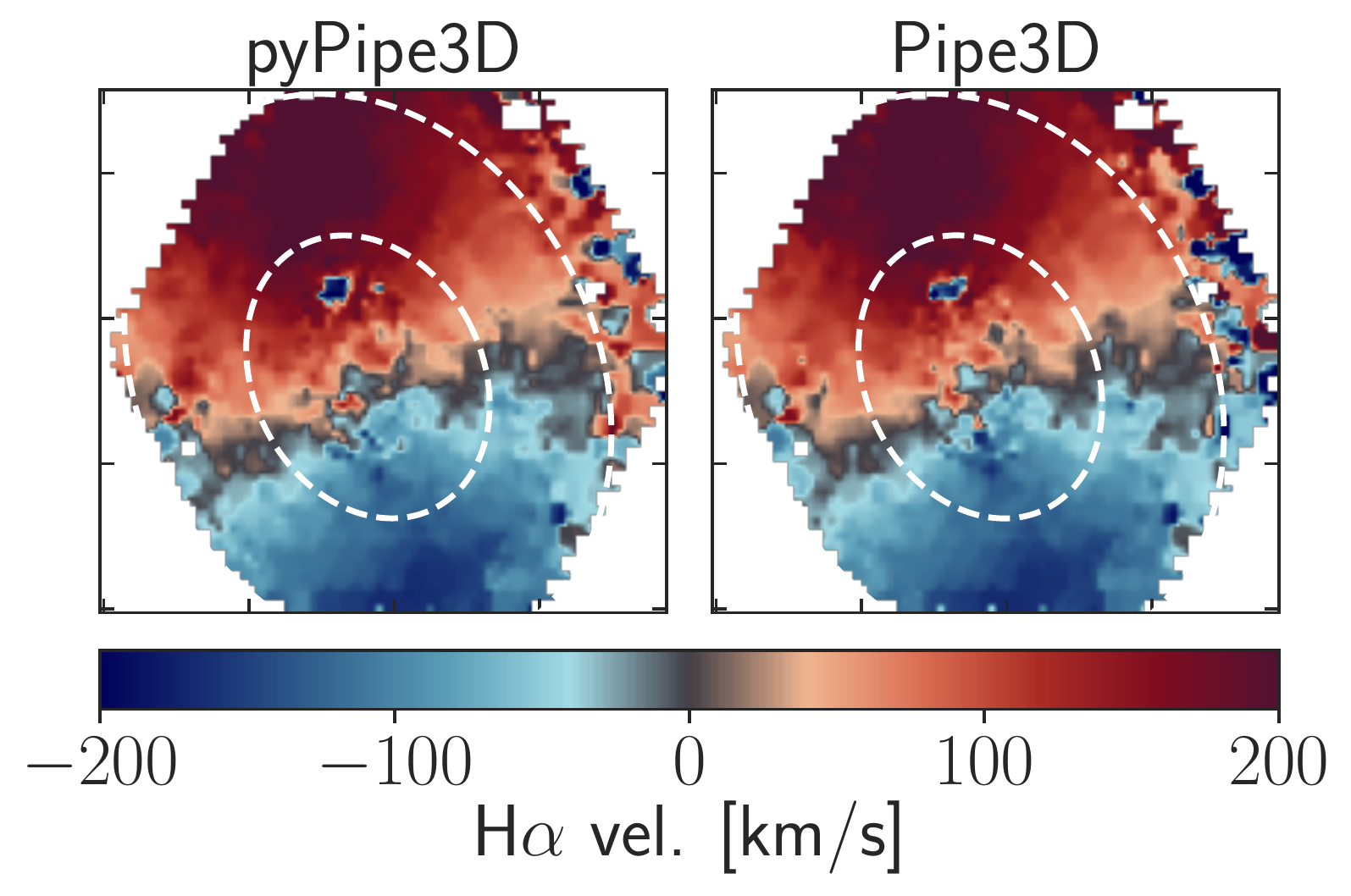}
    \includegraphics[width=0.49\columnwidth]{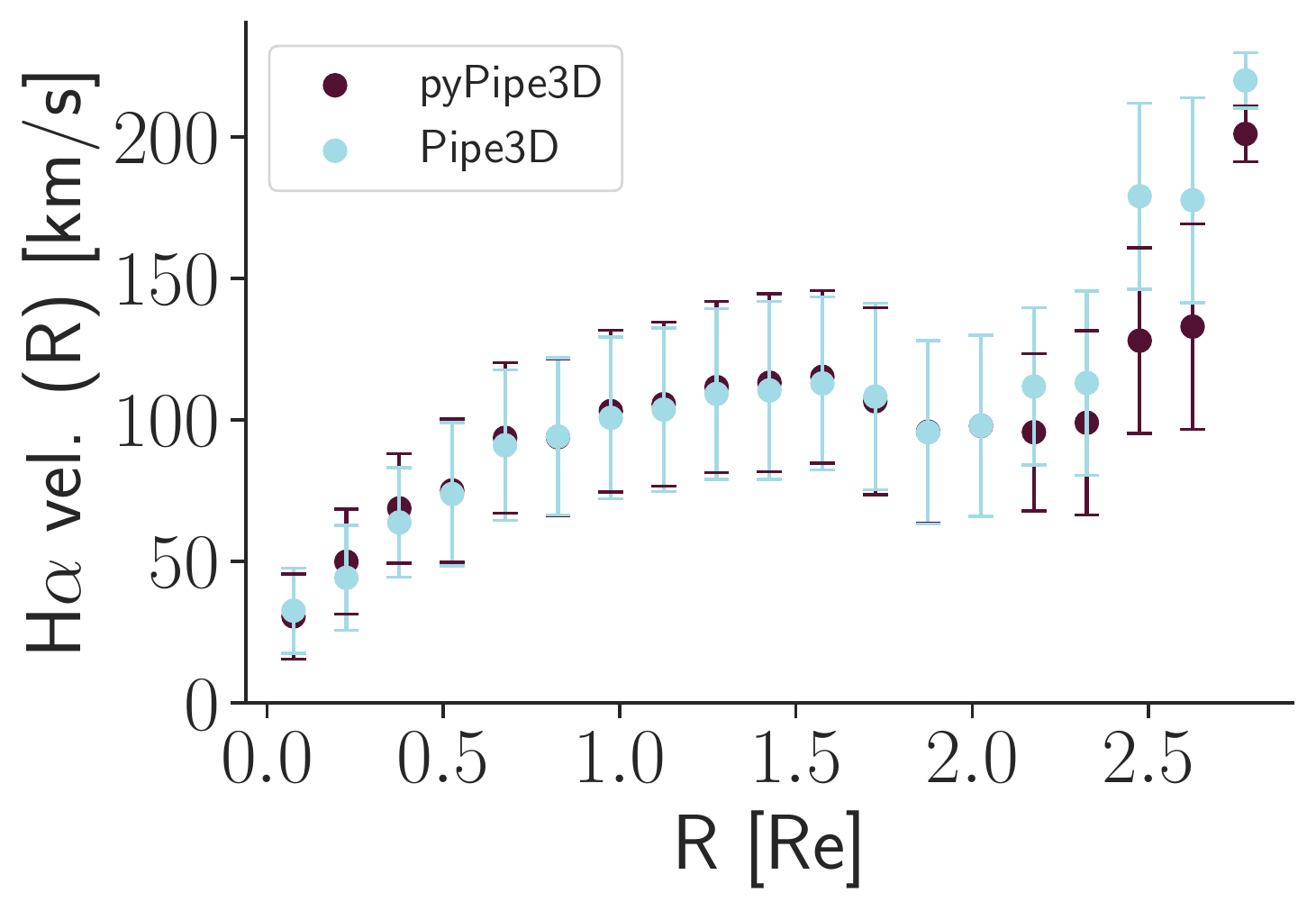} \\
    \includegraphics[width=0.49\columnwidth]{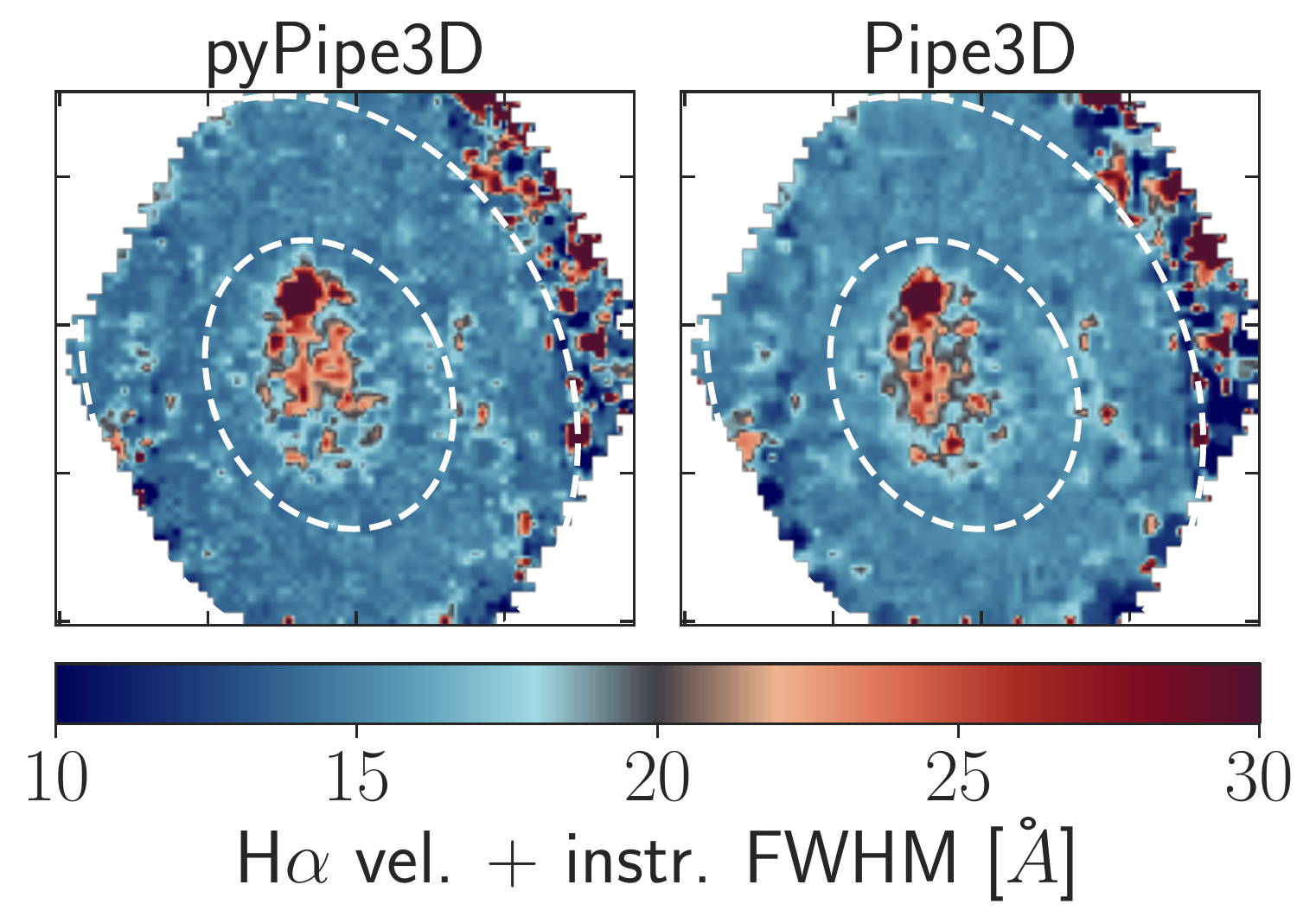}
    \includegraphics[width=0.49\columnwidth]{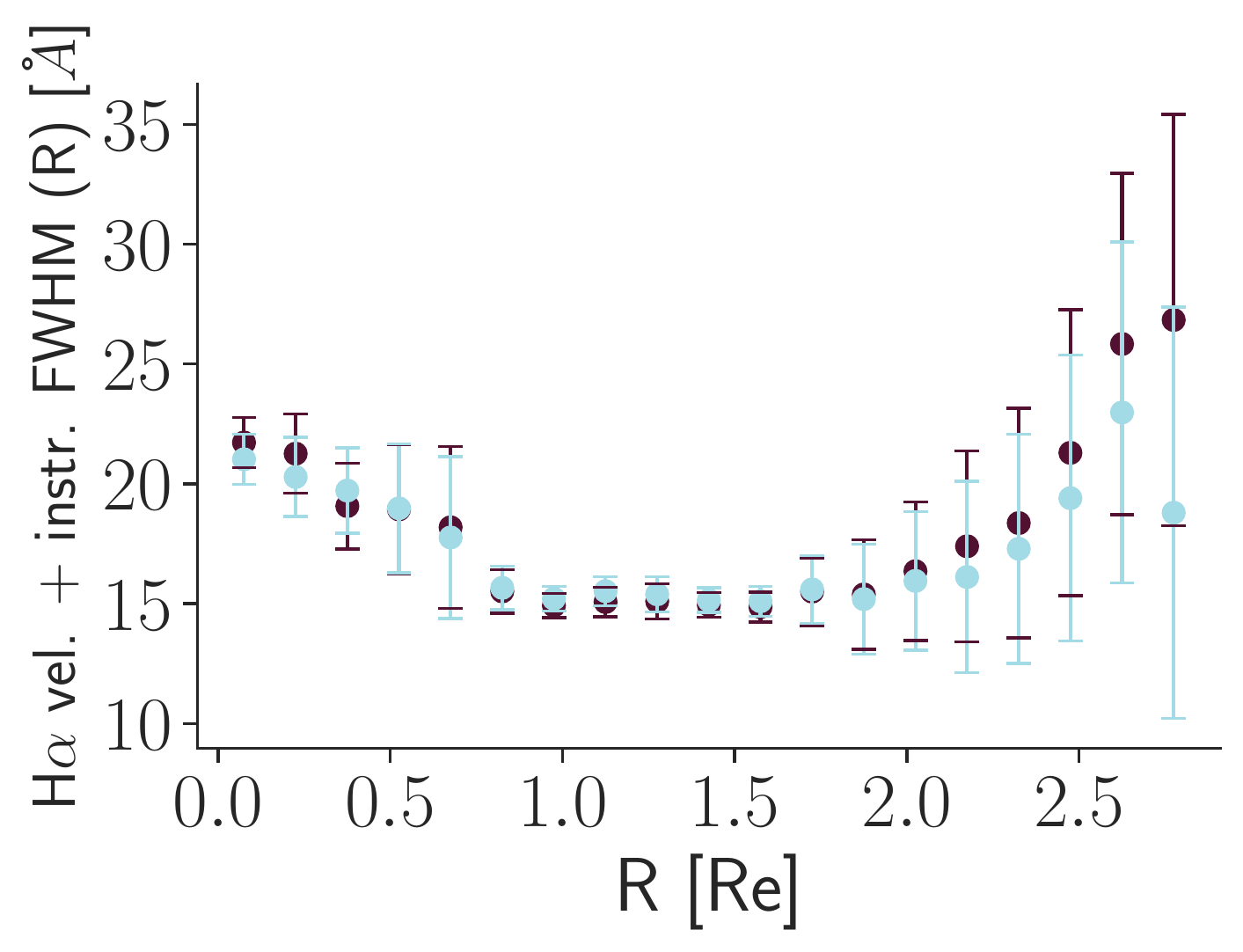} \\
    \includegraphics[width=0.49\columnwidth]{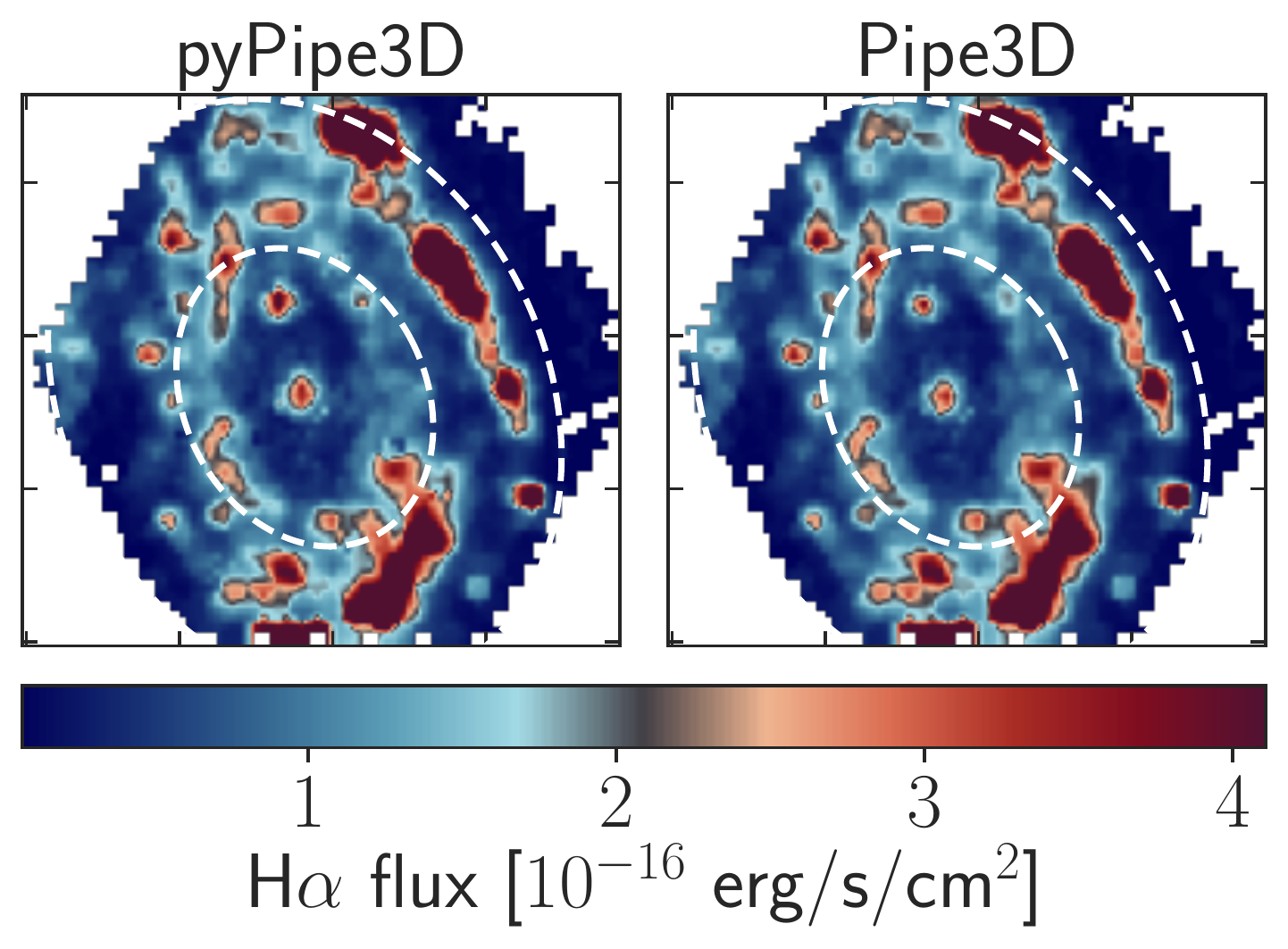}
    \includegraphics[width=0.49\columnwidth]{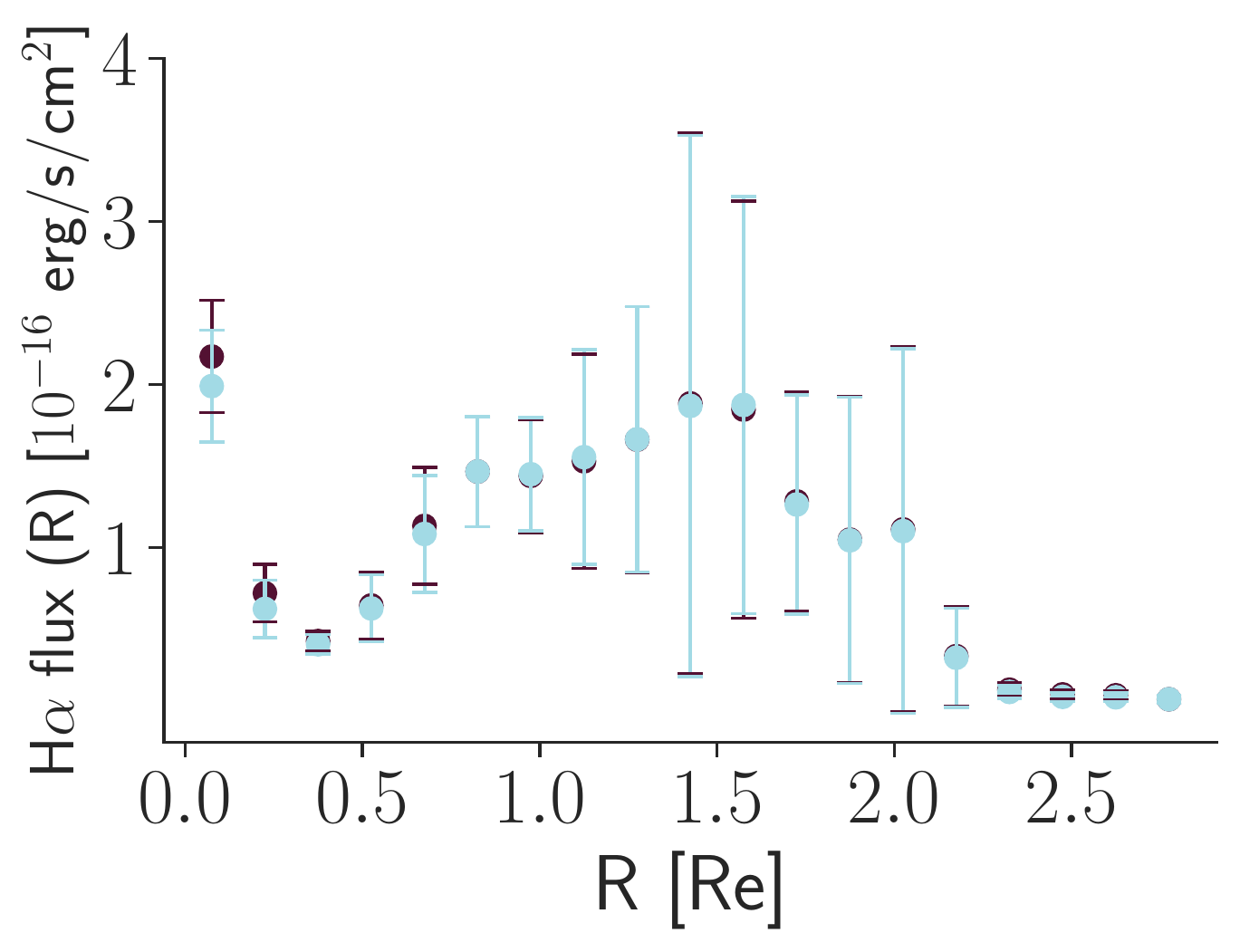}
    \caption{Comparison of the spatial distribution (left panels) and radial profiles (right panels) for the parameters derived by the moment analysis for the \Ha\ emission line by  \pyp and Pipe3D for the  NGC\,2916 galaxy observed by CALIFA: velocity (1st row panels, in absolute values), velocity dispersion (2nd row panels), flux intensity (3rd row panels), and EW(\Ha) (4rd row panels). White ellipses in the maps, symbols and error-bars in the radial plots follow the same scheme adopted in Figure \ref{fig:compare_R_syn_nlprop_NGC2916}.
    }
    \label{fig:comp_elines_NGC2916}
\end{figure}

Like in the case of the stellar populations we rely on the well proved abilities of the former version of the code, Pipe3D, to extract reliable properties for the emission lines. Therefore we compare the results provided by \pyp with those provided by Pipe3D in order to estimate the quality of the derived quantities. Figure \ref{fig:comp_elines_NGC2916} shows the comparison between the spatial distribution and radial profiles of the velocity, velocity dispersion, flux intensity and EW for the \Ha\ emission line derived by both codes using the moment analysis, following the same schemes adopted in Figure \ref{fig:compare_R_syn_nlprop_NGC2916} and \ref{fig:compare_R_syn_stprop_NGC2916}. Like in those cases no masking is applied to highlight the behavior of both codes even under non optical conditions (low-S/N, presence of foreground sources). There is a remarkable agreement, with even less differences than in the case of the properties of the stellar populations. In summary, the new code provides with a total consistent derivation of the properties of the emission lines for the explored procedure.

\begin{figure*}
    \includegraphics[width=0.33\textwidth]{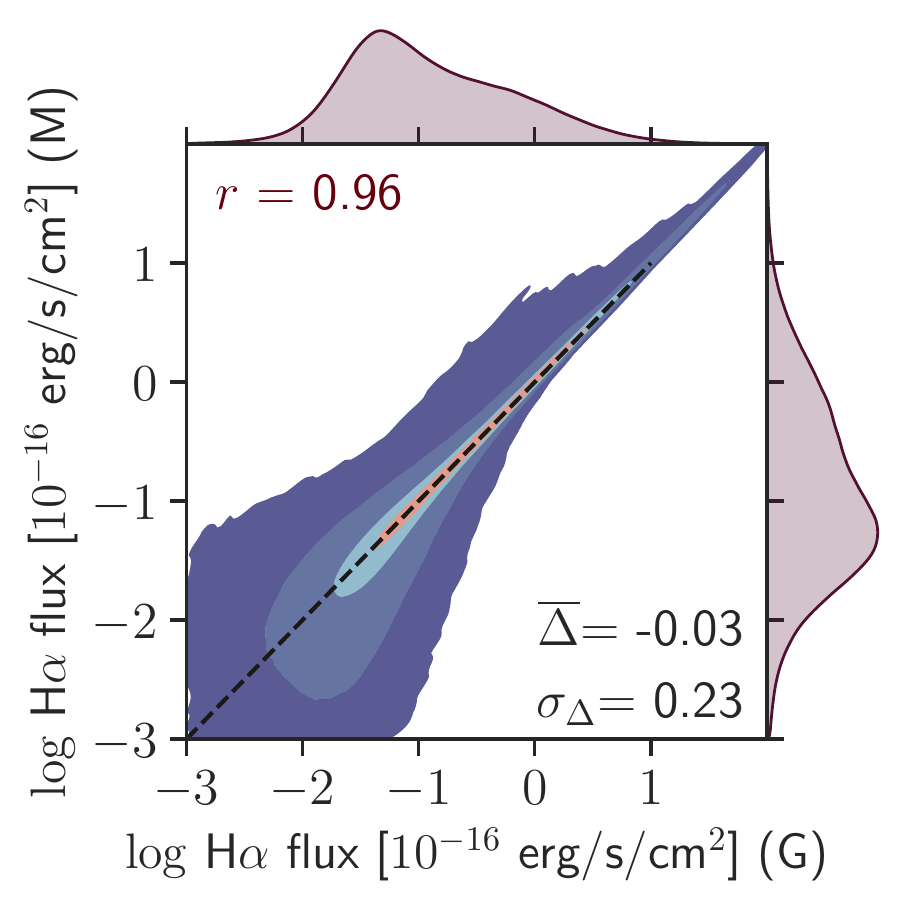}
    \includegraphics[width=0.33\textwidth]{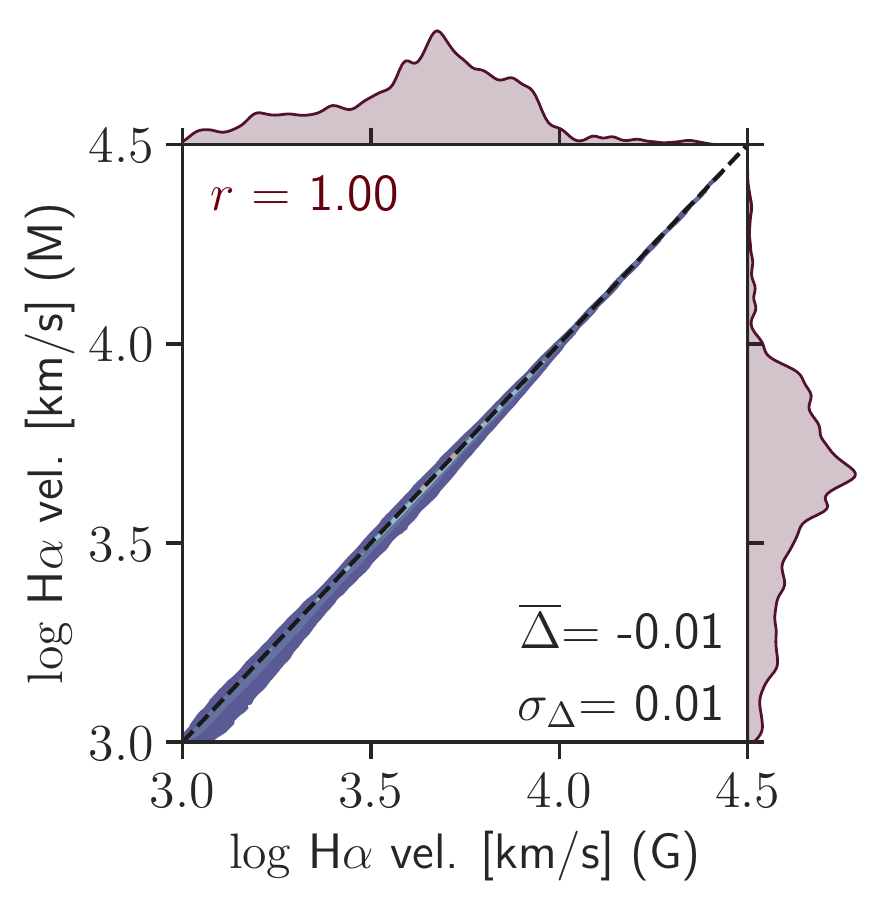}
    \includegraphics[width=0.33\textwidth]{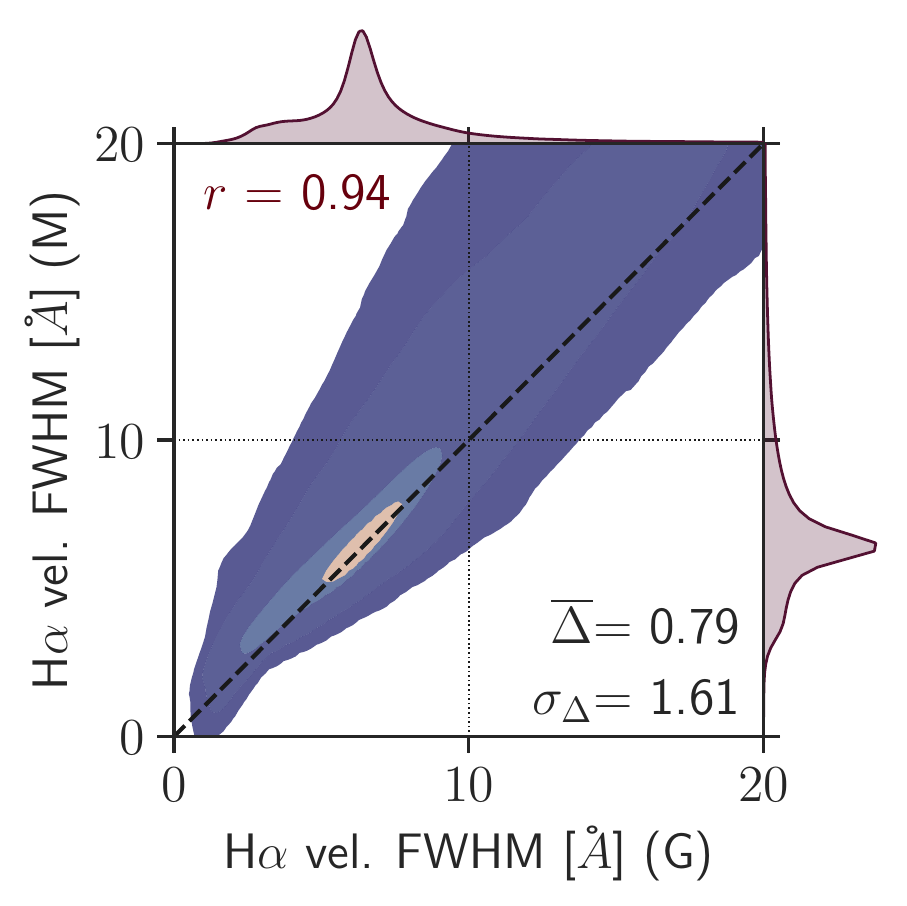}
    \caption{Comparison of the estimated parameters (flux intensity, velocity and velocity dispersion) of the \Ha emission line derived by both methods, Gaussian models (G) versus weighted-moment analysis (M), implemented in \pyp, for the full sample of eCALIFA 867 datacubes \citep{Lacerda.etal.2020}.
    Both dashed lines in the rightmost panel corresponds to a FWHM of 10\,\AA\, a value below which the \Ha emission line is not significantly blended with the \Niip doublet (for the spectral resolution of this data).
    Contours and density histograms in each panel follow the same scheme presented on Figure \ref{fig:stpop_resid_simul}.
    }
    \label{fig:G_M_comp}
\end{figure*}

As mentioned before, \pyp also derives the properties of the strongest emission lines adopting a Gaussian parametrization of the emission line, using the {\bf RND+LM method} (Sec. \ref{sec:new_ver:fitupd:elfit}). Figure  \ref{fig:G_M_comp} shows the comparison between the parameters derived for the \Ha emission line using both procedures, moment analysis and Gaussian parametrization, including the flux intensity, velocity and velocity dispersion (in this case, converted to FWHM). 
To increase the number statistics as much as possible we use the full dataset provided by the eCALIFA survey \citep[867 galaxies][]{Lacerda.etal.2020}, with comprises more than 4 million analyzed spaxels/spectra in total.
The same Figure was presented in S16b just of one galaxy (NGC\,2916), using the former version of the code (i.e. Pipe3D). The conclusions can be directly extrapolated to the case of the entire survey analysis. The flux intensity is the parameter with presents the lowest offset between both methods of analysis ($\Delta {\rm F}_{\Ha}=-0.04 \pm 0.55\,10^{-16}$ erg/s/cm$^2$). The kinematic parameters present a non-negligible bias ($\Delta v_{\Ha} = 46.4 \pm 59.8$\,km/s and $\Delta \sigma_{\Ha} = -0.78 \pm 1.61$\,\AA). A priori it is expected that the weighted-moment algorithm works very well for emission lines which are not blended, while the Gaussian parametrization should recover the parameters better in those cases (if the S/N is high enough). We included in the last panel two lines delimiting the maximal dispersion where the \Ha lines are not blended with \Niip doublet (corresponding to
a FWHM of 10\,\AA). Selecting only those spaxels/spectra under this limit in the velocity dispersion, the offset and the dispersion for $\Delta \sigma_{\Ha}$  decreases significantly to a value of $-0.38 \pm 0.65$\,\AA.


\section{Summary and conclusions}
\label{sec:summary}

Along this article we have presented the new implementation of Pipe3D IFS analysis pipeline and FIT3D, the fitting tools adopted by the pipeline, fully transcribed from \texttt{perl} to \texttt{python}.
The new pipeline conserves all the implemented tools and functionality from previous versions with a complete documentation available online\footnote{\url{http://ifs.astro.unam.mx/pyPipe3D}}. Moreover, the entire project comprises a new documentation, with examples, of how to use it, to facilitate the final user to understand how the analysis is made and also to build pieces of analysis themselves. 

This article is based on the assurance of the quality of the results already yielded by the Pipe3D pipeline over the last five years \citep[e.g.][]{CALIFADR3, CanoDiaz.etal.2016, BarreraBallesteros.etal.2016, SanchezMenguiano.etal.2017, sanchez18, LopezCoba.etal.2019, SFS.2020}.
For this reason we maintained the main fitting sequence and the basic philosophy of the algorithms. We described the new coding philosophy and the main steps of the \pyf fitting procedure, aimed to decouple the emission by stellar populations and the ionized gas and analyze both of them. We explain how this procedure is performed in three main steps: (i) the analysis of the non-linear parameters of the stellar population (i.e., $z_\star$, $\sigma_\star$ and A$_{\rm V}^\star$), (ii) the analysis and removal of the emission lines and (iii) the modeling of the stellar spectrum by a multi-SSP decomposition. In addition we describe further explorations that can be performed on both the stellar component (indices analysis) and the emission lines (moment analysis).

All these different pieces of analysis were integrated into the final \pyp pipeline, aimed to analyze automatically IFS observations on a single galaxy, extracting once more the properties of the stellar populations and the ionized gas. We described along this article the different procedures performed by this pipeline, which algorithms from \pyf are adopted (highlighting the differences with respect to the previous version of the code), and described briefly the main dataproducts delivered. In order to determine the accuracy and precision in the derivation of the parameters, and to compare with previous results, we applied our new code to an extensive set of simulated spectra and to real data.

The main conclusions of this study are the following ones:

\begin{itemize}
    \item The new version of the code implements all the algorithms and analysis tools of the previous version, respecting the input and output format of the data and products.
    However, the modules and tools under which the entire package is built are designed to be reused, facilitating also the work under exploration environments such as IPython \citep{ipython}. Furthermore, taking advantage of the reusability of the code, the main scripts of \pyf and \pyp are distributed with new implementations, accepting new input arguments and output logs.
    \item Although the fitting philosophy adopted by \pyf has not change, the code runs significantly faster allowing us to include a second round on the determination of the properties of the emission lines \pyf. In addition, this change in the implementation of the code improves the accuracy and the precision of the estimated parameters and uncertainties, as by simulations.
    \item In general the code present a good accuracy and precision in the recovery of the explored parameters. The best recovered stellar parameters are the redshift ($\Delta {\rm v}\sim$10 km/s), and the worst ones are the velocity dispersion ($\Delta \sigma_\star\sim$22 km/s) and the metallicity ($\Delta \stmetL[]$0.6 dex). The remaining stellar parameters are recovered with a precision better than 0.05 dex, and an even better accuracy.
    \item Our results confirm that the derivation of the different stellar parameters is not fully independent one-each-other. The strongest correlations are found between the uncertainties of the age and those of the dust, M/L and metallicity, and between the metallicity and the velocity dispersion, but not among themselves. Of them, the only clearly strong correlations are two first ones (age-dust and age-M/L, $r>$0.5), with the remaining ones present a much lower correlation ($r\sim$0.3). These correlations are inherent to the method, emerging from the intrinsic uncertainties of the stellar population models. These results are qualitatively and quantitatively similar to those found in the literature.
    \item The uncertainties estimated by \pyf for the properties of the stellar populations are of the same order of the real errors, with mild underestimation. However, there is no one-to-one correspondence between the estimated and real uncertainties.
    \item Regarding the emission lines, we found that the new implemented {\bf RND+LM} method recovers the simulated parameters with a much better precision ($\sim$1\% ) and accuracy ($\sim$3-9\%) than the previously implemented {\bf RND}, when the S/N$>$3. Furthermore, we have not found any significant interdependence in the determination of the different parameters. Finally, the estimated errors are of the order of the real ones, with a slight overestimation of a $\sim$5\%\ in the error of the flux, and a 1$\sigma$ agreement for the errors in the kinematics parameters.
    \item The comparison of the results derived from the current version of the code and those obtained with its previous implementation on real data shows that there is a qualitative and quantitative agreement in the derived parameters for both the stellar populations and the emission lines. The main difference is found in the derivation of ${\rm A}_{\rm V}^\star$, what we attribute to a bug discovered in the normalization wavelength in the previous version.
\end{itemize}

In summary, the new implementation of the pipeline could reproduce any of the studies already performed along the years using the previous version of the code and it is prepared for the analysis of IFS data from any of the present surveys and for those planned to be released over the next years. Since the new code is substantially faster, studies that require explorations using different input libraries, larger libraries or just over larger datasets are now feasible. Future upgrades already under-development will include (i) the modifications required to explore of the [$\alpha$/Fe], (ii) a differential treatment of the dust attenuation depending on the properties of the stellar populations, (iii) a spectro-kinematic decomposition of the different stellar component, and (iv) the inclusion of more complex models for the emission lines (e.g., Voigt models or models including asymmetries in the profiles).


\section*{Acknowledgements}
We acknowledge the support of the PAPIIT-DGAPA IN100519 and IG100622 projects. J.B-B acknowledges support from the grant IA101522 (DGAPA-PAPIIT, UNAM) and funding from the CONACYT grant CF19-39578.
EADL and SFS thank to Laura S\'{a}nchez-Menguiano for her valuable commentaries and time spent analyzing the products of beta versions of \pyp and \pyf improving the present versions. EADL thanks to Carlos Henrique Mantovani Sammartin de Abreu for the art design of the \pyp logo.
This study uses data provided by the extended Calar Alto Legacy Integral Field Area (eCALIFA) survey (\url{http://califa.caha.es/}).
This research has not been possible without the support from the UNAM and the DGAPA Post Doctoral Scholarship Program.

\appendix

\section{FIT3D format for the SSP templates}
\label{app:ssp}

As indicated before \pyf (and \pyp) can use, in principle, any stellar or SSP library to fit the stellar component of the analyzed spectra. However, it requires that these library adopts a certain format. The actual format is a row-stacked spectra (RSS) FITS file, in which each row corresponds to one of the components to be fitted (i.e., each $j$ component $f_{\lambda,j}^{\rm ssp}$ in Equations \ref{eq:nlfit_model} and \ref{eq:spec_mod}). So far the code is prepared to analyze spectra which wavelength solution is linear, i.e., each spectral pixel (X-axis of the RSS file) corresponds to a certain wavelength with a fixed wavelength interval in \AA between adjacent pixels. It also expects that the SSP templates are in this format. Therefore, the SSP FITS file should include in the header the reference pixel at which the starting wavelength is defined (CRPIX1), the value of this wavelength in \AA (CRVAL1), and the constant step in wavelength in the spectral axis (CDELT1). The SSP templates should be normalized to one at a certain wavelength withing the interval covered by the spectra (WAVENORM). The program requires the flux in solar units at this wavelength to be stored in the header with an entry for each individual SSP (NORM$j$), in order to provide with a reliable \mlL and stellar mass. In addition, to provide with reliable estimations of the \stageL and \stmetL (and \stageM and \stmetM, too), it is required an entry in the header for each individual SSP (NAME$j$) following the format {\tt spec\_ssp\_{\bf AGE}Gyr\_z{\bf MET}.spec}, where {\bf AGE} corresponds to the age in Gyr of the SSP and {\bf MET} corresponds to the decimals of the metallicity (i.e., $Z=0.1${\bf MET}). For instance {\tt spec\_ssp\_06.0000Gyr\_z0021.spec} corresponds to an SSP with an age of 6 Gyr and a metallicity of 0.0021\footnote{A few examples of the format have been included in the following webpage: \url{http://ifs.astroscu.unam.mx/pyPipe3D/templates/}}

\section{Relation between errors and uncertainties}
\label{app:rel}

\begin{figure*}[t]
    \includegraphics[width=0.333\textwidth]{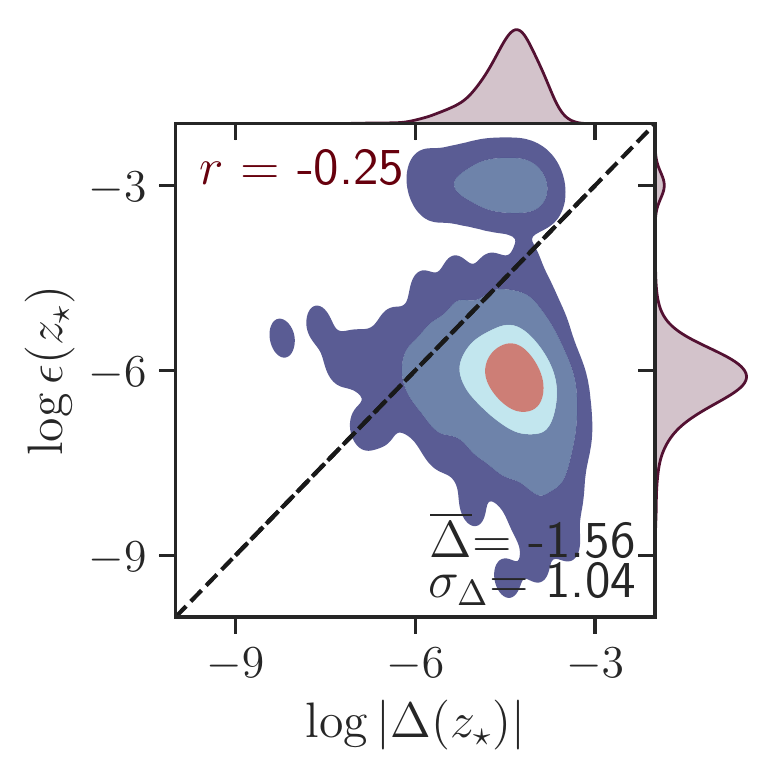}\includegraphics[width=0.333\textwidth]{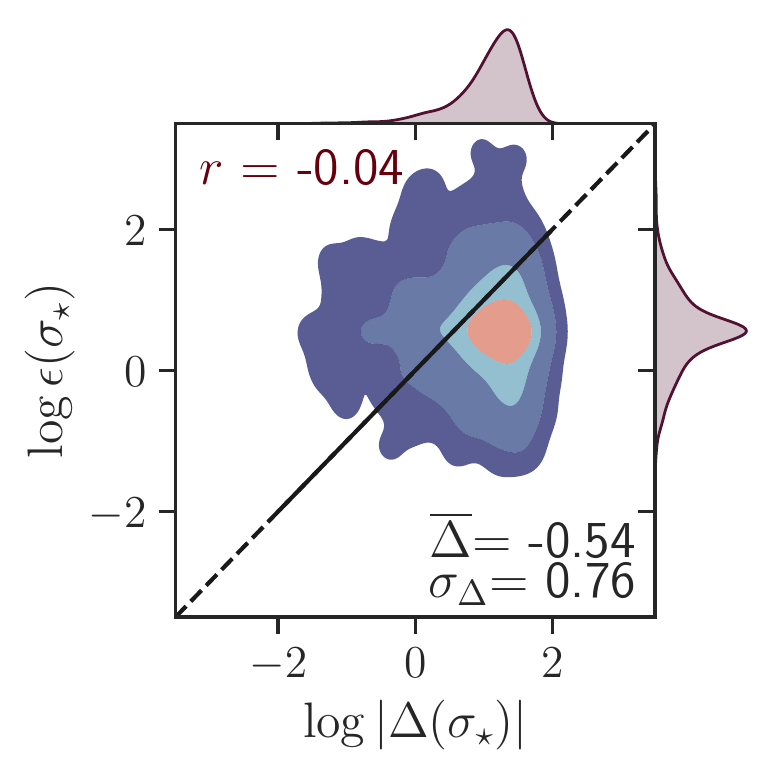}\includegraphics[width=0.333\textwidth]{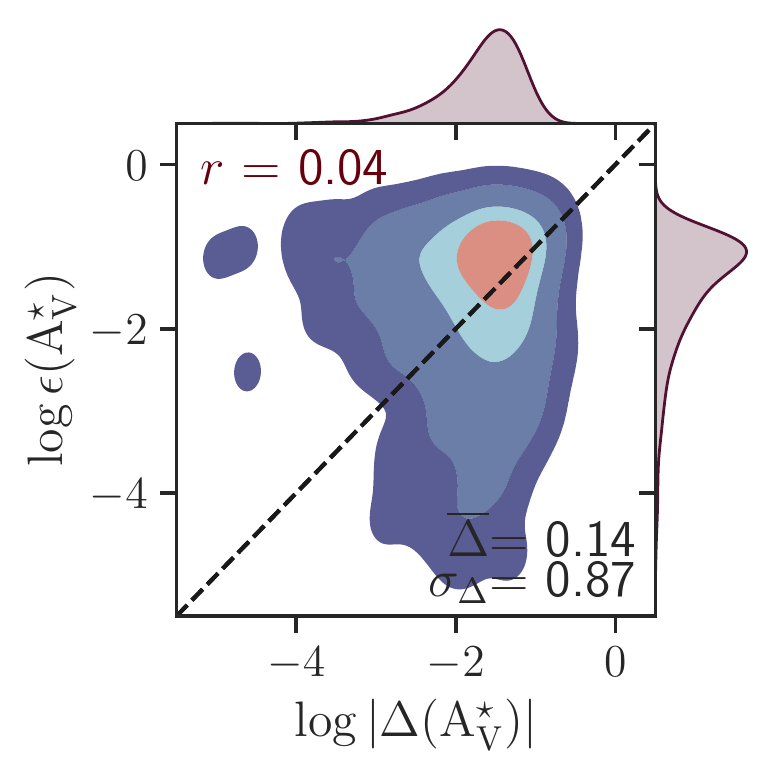}
    \includegraphics[width=0.333\textwidth]{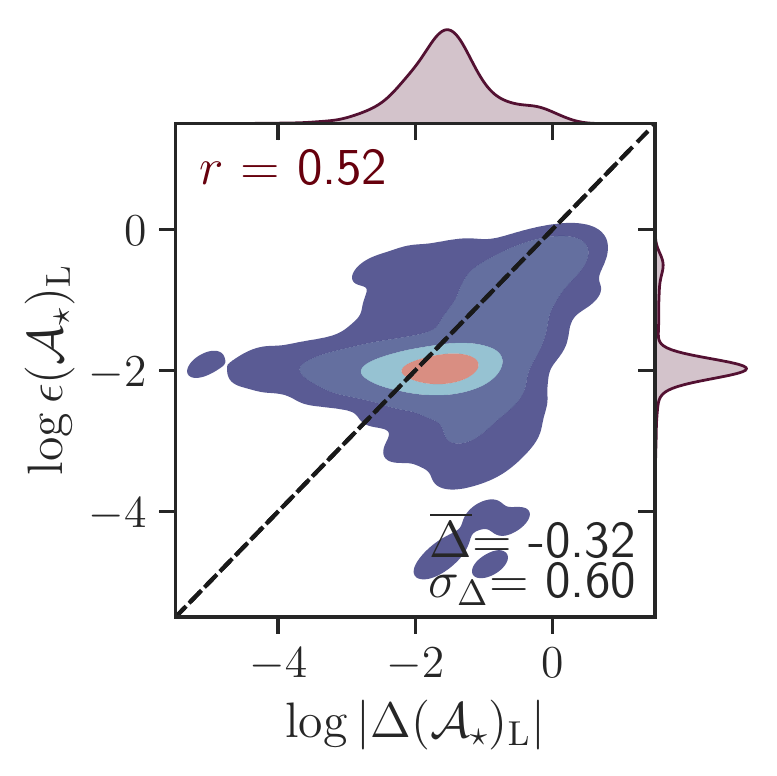}\includegraphics[width=0.333\textwidth]{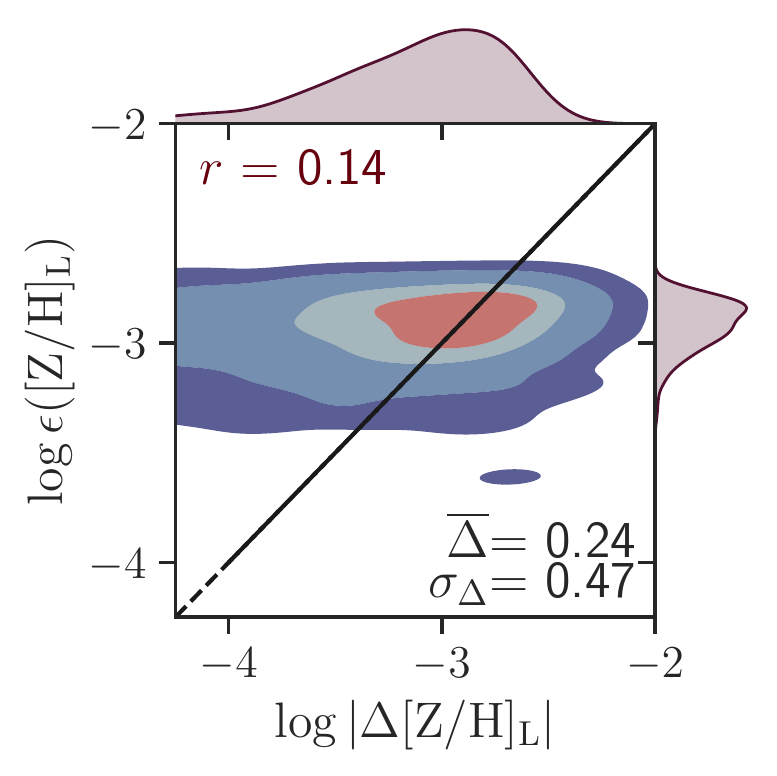}\includegraphics[width=0.333\textwidth]{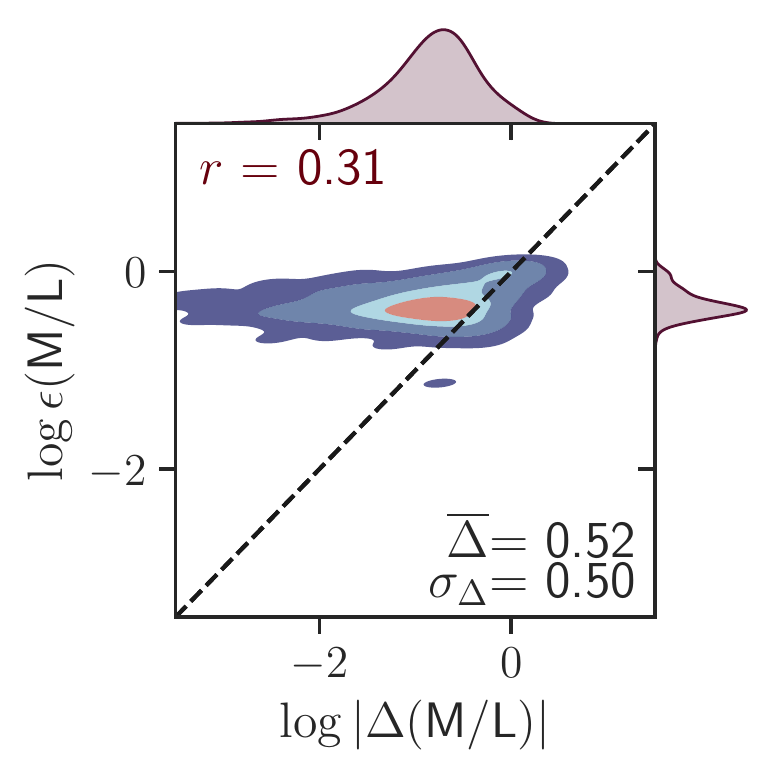}
    \caption{
    Distribution of the estimated error, $\epsilon$par, versus the absolute value of the difference between the estimated and input value, $\Delta par$, for each of the stellar parameters shown in Figure \ref{fig:stpop_simul}. Contours and density histograms in each panel follow the same scheme presented on Figure \ref{fig:stpop_resid_simul}. Each panel includes the correlation coefficient between the represented parameters (up-left legend), together with the average and standard deviation of the difference between them (bottom-right legend).}
    \label{fig:stpop_dppepp}
\end{figure*}

Figure \ref{fig:stpop_dppepp} shows the distribution of the estimated errors for the stellar populations,$\epsilon$(par), as a function of the absolute value of the uncertainty  $|\Delta({\rm par})|$, for properties discussed in Section \ref{sec:accur:stpop}. Despite of the good statistical correspondence between both parameters once applied the correction factor discussed in Sec. \ref{sec:accur:stpop:uncert}, there is no significant correlation between both parameters. Only the errors for the redshift and \stageL present some degree of correlation.

In many cases the distribution is near or on-top of the one-to-one relation (A$_{\rm V}^\star$, \stageL, \stmetL and \mlL), what indicates that the average errors estimated by \pyf traced the real ones. However, already discussed in Sec. \ref{sec:accur:stpop:uncert}, and shown in Figure \ref{fig:DS_eml} the errors are clearly underestimated for different properties. In the case of the kinematics ones both kinematic, the underestimation of is considerably larger than in the case of the velocity dispersion. In addition, the redshift presents a tiny bimodality as a consequence of the cases when it does not find a better solution than the input {\it guess}. In these cases \pyf assumes the {\it guess} as the redshift value and the input {\it interval} as the error. Those cases correspond to less than a 1\% of the total number of simulations. The same happens with the velocity dispersion but the substantial scatter on $\epsilon(\sigma_\star)$ does not allow us to visualize it in the Figure \ref{fig:stpop_dppepp}. On the other hand, for the dust attenuation, the center of the $\epsilon \times |\Delta|$ distribution shows a noteworthy relation. However, the estimated errors cover a wide range of values, and therefore, as indicated before, there is no significant correlation between both parameters. The errors of \stageL present a clear underestimation, covering a narrower range of values than the real ones. However, as we described before, this is the parameter for which the estimated errors present the strongest correlation with the real uncertainties. This is not the case for the \stmetL and \mlL, where the estimated errors cover a much narrower range of parameters than the real ones. In both cases there is a clear over-estimation of the errors. In summary,
there is no direct correspondence between the real uncertainties and the errors estimated by the code for each individual simulation, and the agreement between both parameters is valid on an statistical basis, as discussed in Sec. \ref{sec:accur:stpop:uncert}.

\begin{figure*}[t]
    \includegraphics[width=0.33\textwidth]{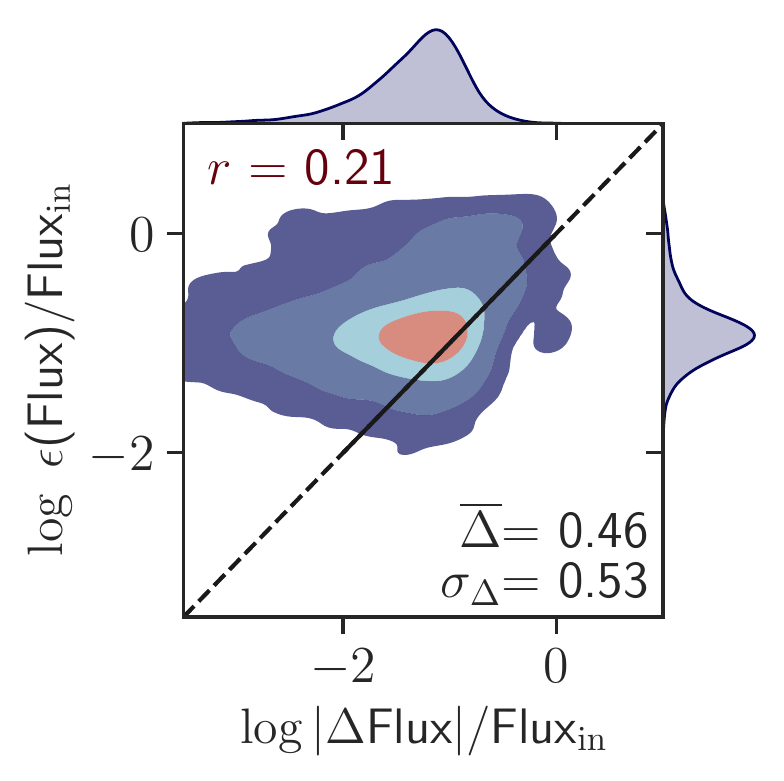}
    \includegraphics[width=0.33\textwidth]{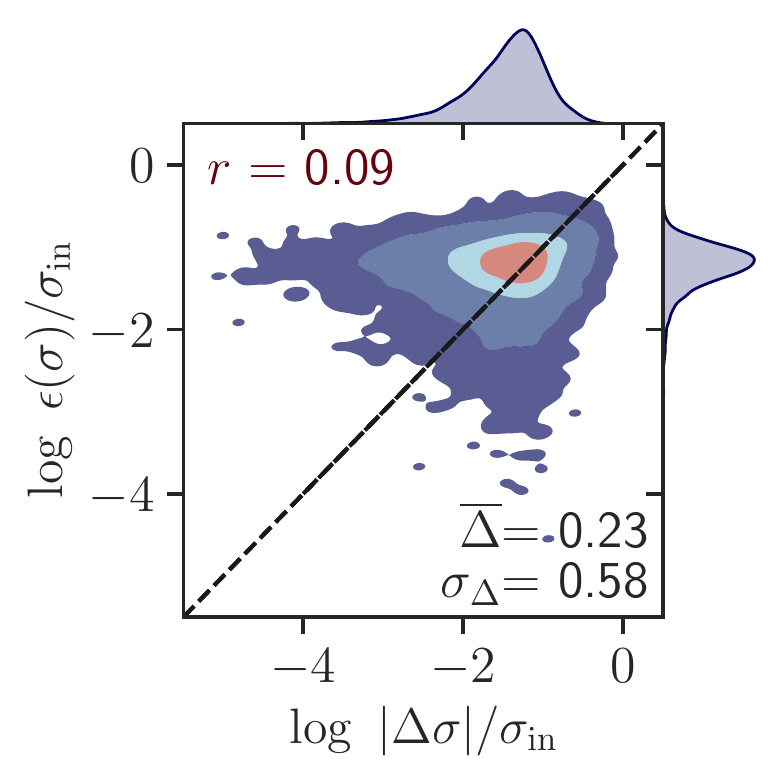}
    \includegraphics[width=0.33\textwidth]{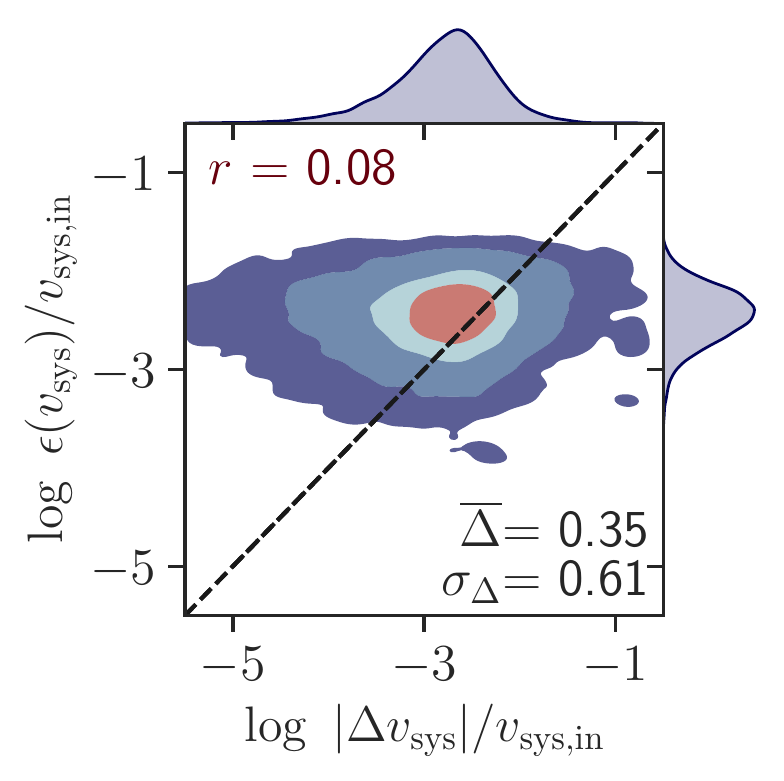}
    \caption{
    Distribution of the errors estimated by \pyf as a function of the absolute value of the relative difference between the derived and input values for each of the explored parameters of the simulated emission lines: flux intensity (left), velocity dispersion (middle) and systemic velocity (right).
    Contours and density histograms in each panel follow the same scheme presented on Figure \ref{fig:stpop_resid_simul}.
    }
    \label{fig:elines_dppepp}
\end{figure*}

Figure \ref{fig:elines_dppepp} shows the distribution of the relative value of the estimated errors, $\epsilon$(par)/par$_{\rm in}$, as a function of this tracer of the uncertainties, i.e. $|\Delta$par$|$/par$_{\rm in}$, for the simulated emission lines discussed in Sec. \ref{sec:accur:eml}.
As already discussed in Sec. \ref{sec:accur:eml:uncert} the three parameters present a mild underestimation of the errors, what results in $\epsilon$(par)/par$_{\rm in}$ covering a much narrower range of values than $|\Delta$par$|$/par$_{\rm in}$. As expected, there is no clear correlation between the uncertainties and errors ($r<$0.3 in all cases), as both parameters should be compared only in an statistical sense.



\section{Parameters adopted for the showcase fitting example}
\label{appendix:showcase_config}


For the description of the fitting procedure of \pyf in Section \ref{sec:new_ver:fitupd} we analyze the central spectrum ($5\arcsec$ aperture) of the galaxy NGC\,5947, extracted from the data provided by the CALIFA V500 IFS datacube as a showcase example. The chosen input values for the non-linear parameters are (guess/step/min/max):
\begin{itemize}
    \item $z_\star$: 0.01/0.001/0.003/0.05;
    \item $\sigma_\star$: 30/10/75/375 km/s;
    \item ${\rm A}_{\rm V}^\star$: 0.15/0.05/0/1.8 mag.
\end{itemize}
\noindent All domains are chosen covering up the real observed intervals for CALIFA data. The kinematic stellar parameters are estimated using the wavelength interval from 3800 to 4700\,\AA\  because this spectral region is dominated by stronger stellar absorption features. All subsequent analysis use the wavelength interval from 3800 to 7000\,\AA.

For the emission lines analysis, we choose to fit five different {\it systems} with ten different emission lines: \Hd; \Hg; \Oiiip\,+\,\Hb ; \Niip\,+\,\Ha and \Siip, with the procedure fitting one {\it system} at a time. The kinematic is selected to be the same for all emission lines, assuming a physical scenery where they are formed at same conditions. We should note that this is not a limitation of the method, just the choice done for the current analysis. For the \Oiiip\,+\,\Hb and emission line {\it system} we tie the integrated flux of each \Oiiip line as $_{\Oiii} = (1/3)$F$_{\OOiii}$. Following the same hypothesis, for the \Niip\,+\,\Ha\ {\it system} we link the integrated flux of the \Niip line as F$_{\Nii} = (1/3)$F$_{\NNii}$. An example of a simple configuration file of an emission lines {\it system} can be found in the \pyp webpage\footnotetext{\url{http://ifs.astroscu.unam.mx/pyPipe3D/pipeline\_example\_README.html}}. The {\it RND method} is configured to proceed 20 MC loops for the search of the kinematic parameters and the integrated flux. This procedure
is repeated five times, with the parameters intervals narrowed around the best fitted parameters, as described in Sec. \ref{sec:new_ver:fitupd:elfit}.

The adopted SSP template library was the \texttt{gsd156}, precisely described in \citep{CF.etal.2013}. It includes templates extracted from two projects: (a) synthetic stellar spectra from the GRANADA library \citep{Martins.etal.2005} for the ages $<$ 65 Myr and (b) the SSP library of MILES \citep{SanchezBlazquez.etal.2006, Vazdekis.etal.2010, FalconBarroso.etal.2011}. The full spectral library is composed of 39 stellar ages (from 1 Myr to 13 Gyr) and 4 metallicities (Z/Z$_\odot$ = 0.2, 0.4, 1 and 1.5). Again, this is not a limitation of the procedure, and indeed we have tested the code using several different SSP libraries. The results of that comparison will be presented elsewhere. For the non-linear fit of the showcase example we use a limited version of the \texttt{gsd156} library including only 12 SSPs (\texttt{gsd12}: 3 ages, t$_\star$ = 0.001, 0.5 and 14.1 Gyr; and the same 4 metallicities) for the reasons mentioned in Sec. \ref{sec:new_ver:fitupd:nlfit}. The actual configuration file required to run the \pyf script on this spectrum can be found again in the \pyp webpage.

\bibliography{pyPipe3D_biblio}{}

%
%

\end{document}